\documentclass[fleqn,usenatbib]{mnras}

\usepackage{newtxtext,newtxmath}
\usepackage[T1]{fontenc}
\usepackage{ae,aecompl}
\usepackage{graphicx}	% Including figure files
\usepackage{amsmath}	% Advanced maths commands
\usepackage{subfigure}  % Subfigures
\usepackage{placeins}   % FloatBarrier
\usepackage{tabularx}   % Table with fixed width
\usepackage{mathrsfs}   % for \mathscr
\usepackage{appendix}   % for appendix
\usepackage{dsfont}     % fuer ganze Zahlen Z
\usepackage{enumitem}   % for enumerate functionality

\newcommand{\vect}[1]{\mathbf{#1}}
\newcommand*\diff{\mathop{}\!\mathrm{d}}

\newcommand{\appropto}{\mathrel{\vcenter{
  \offinterlineskip\halign{\hfil$##$\cr
    \propto\cr\noalign{\kern2pt}\sim\cr\noalign{\kern-2pt}}}}}
\usepackage[dvipsnames]{xcolor}
\defcitealias{2019MNRAS.484.3291T}{T19}	

\title[The Galactic bar in action space]{Identifying resonances of the Galactic bar in \emph{Gaia} DR2:\\I. Clues from action space}

\author[W.~H.~Trick et al.]{Wilma H.~Trick$^{1}$\thanks{E-mail: trick@mpa-garching.mpg.de}, Francesca Fragkoudi$^{1}$, Jason A.~S.~Hunt$^{2,3}$, J.~Ted Mackereth$^{4}$,
\newauthor 
and Simon D.~M.~White$^{1}$
\\
$^{1}$Max-Planck-Insitut f\"ur Astrophysik, Karl-Schwarzschild-Str. 1, D-85748 Garching b. M\"unchen, Germany\\
$^{2}$Dunlap Institute for Astronomy and Astrophysics, University of Toronto, 50 St. George Street, Toronto, Ontario, M5S 3H4, Canada\\
$^{3}$Center for Computational Astrophysics, Flatiron Institute, 162 5th Av., New York City, NY 10010, USA\\
$^{4}$School of Astronomy and Astrophysics, University of Birmingham, Edgbaston, Birmingham, B15 2TT, UK
}

\date{Accepted XXX. Received YYY; in original form ZZZ}

\pubyear{2020}

\begin{document}
\label{firstpage}
\pagerange{\pageref{firstpage}--\pageref{lastpage}}
\maketitle

\begin{abstract}
Action space synthesizes the orbital information of stars and is well-suited to analyse the rich kinematic substructure of the disc in the \emph{Gaia} DR2 radial velocity sample (RVS). We revisit the strong perturbation induced in the Milky Way (MW) disc by an $m=2$ bar, using test particle simulations and the actions $(J_R,L_z,J_z)$ estimated in an axisymmetric potential. These make three useful diagnostics cleanly visible. (1.) We use the well-known characteristic flip from outward to inward motion at the Outer Lindblad Resonance (OLR, $l=+1,m=2$), which occurs along the axisymmetric resonance line (ARL) in $(L_z,J_R)$, to identify in the \emph{Gaia} action data three candidates for the bar's OLR and pattern speed $\Omega_\text{bar}$: $1.85\Omega_0$, $1.20\Omega_0$, and $1.63\Omega_0$ (with $\sim0.1\Omega_0$ systematic uncertainty). The \emph{Gaia} data is therefore consistent with both slow and fast bar models in the literature, but disagrees with recent measurements of $\sim1.45\Omega_0$. (2.) For the first time, we demonstrate that bar resonances---especially the OLR---cause a gradient in vertical action $\langle J_z \rangle$ with $L_z$ around the ARL  via ``$J_z$-sorting'' of stars. This could contribute to the observed coupling of $\langle v_R \rangle$ and $\langle | v_z | \rangle$ in the Galactic disc. (3.) We confirm prior results that the behaviour of resonant orbits is well approximated by scattering and oscillation in $(L_z,J_R)$ along a slope $\Delta J_R/\Delta L_z = l/m$ centered on the $l$:$m$ ARL. Overall, we demonstrate that axisymmetrically estimated actions are a powerful diagnostic tool even in non-axisymmetric systems.
\end{abstract}


\section{Introduction} \label{sec:introduction}

\subsection{Moving groups and bar resonances} \label{sec:intro_bar}

In the local Solar neighbourhood ($\sim 200~\text{pc}$) of the pre-\emph{Gaia} era, stellar moving groups have been identified in both the stars' velocities (e.g., \citealt{1996AJ....112.1595E,1998AJ....115.2384D,2005A&A...430..165F}) and orbit space (e.g., \citealt{2006A&A...449..533A,2008ApJ...685..261K}). The amount of kinematic substructure in the in-plane motions of the Galactic disc stars discovered by the \citet{2018A&A...616A..11G} in the radial velocity sample (RVS) of the second \emph{Gaia} data release (DR2) \citep{2016A&A...595A...1G,2018A&A...616A...1G} was still surprising (c.f. \citealt{2009ApJ...700.1794B}). At least seven arches or ridges are visible in the $(U,V)$ velocity space within $\sim1.5-4~\text{kpc}$ from the Sun \citep{2018A&A...616A..11G}, in the $(R,v_T)$ plane \citep{2018Natur.561..360A,2018MNRAS.479L.108K} or orbital action space (\citealt{2019MNRAS.484.3291T}, \citetalias{2019MNRAS.484.3291T} hereafter).

Their origin is still largely unexplained. Early studies suggested a star cluster origin \citep{1996AJ....112.1595E,1998A&A...340..384C}, which is however not supported by age-abundance measurements \citep{2004A&A...418..989N,2007ApJ...655L..89B,2007A&A...461..957F,2008A&A...483..453F}. Several dynamic processes have been proposed that could cause these kinematic ridges: Spiral arms \citep{1991dodg.conf..323K,2003AJ....125..785Q,2004MNRAS.350..627D,2005AJ....130..576Q,2012ApJ...751...44S,2018MNRAS.480.3132Q,2018MNRAS.479L.108K,2019MNRAS.484.3154S,2020A&A...634L...8K} and bar resonances (e.g., \citealt{2000AJ....119..800D,2001A&A...373..511F,2019MNRAS.488.3324F,2020MNRAS.494.5936F}), secular evolution of the disc in general (e.g., \citealt{2012ApJ...751...44S,2015A&A...581A.139F,2015A&A...584A.129F}; going back to \citealt{1981seng.proc..111T}) and transient processes like phase-mixing caused by transient spiral structure \citep{2015MNRAS.449.1982F,2018MNRAS.481.3794H} and satellite interactions \citep{2009MNRAS.396L..56M,2012MNRAS.419.2163G,2018Natur.561..360A,2019MNRAS.485.3134L,2019MNRAS.489.4962K}.

Resonance effects in particular have been studied in depth in the literature (see \citet{2016AN....337..703M} for a paedagogical introduction). A star's orbit experiences lasting changes if its fundamental frequencies are commensurate with the pattern speed of the periodic perturber, $\Omega_\text{bar}$. In other words, the radial frequency with which the star oscillates in the radial direction, $\Omega_{R,\text{true}}$, and the circular frequency around the Galactic center, $\Omega_{\phi,\text{true}}$, \citep[\S3.2.3 ]{2008gady.book.....B} are related with the pattern speed of the bar by
\begin{equation}
    m \cdot \left(\Omega_\text{bar} - \Omega_{\phi,\text{true}}\right) - l \cdot \Omega_{R,\text{true}} = 0, \label{eq:res_condition_real_freqs}
\end{equation}
with $m,l \in \mathds{Z}$. At $l=0$, the star is in co-rotation resonance (CR) with the bar. $l>0$ describes resonances outside, and $l<0$ resonances inside of CR in the Galactic disc. Depending on the mass distribution of the perturber, the resonances at different $m$ have different strength. The $m=2$ Fourier component of a typical galaxy bar is dominant \citep{2006AJ....132.1859B}, so its $l=\pm1$, $m=2$ resonances, the Outer and Inner Lindblad Resonances (OLR and ILR), are expected to have a strong effect on the Galactic disc.

In the pre-\emph{Gaia} era, many studies focused on explaining the Hercules stream with bar resonances. The \emph{short fast bar} model with $\Omega_\text{bar}\sim51~\text{km/s/kpc}$ associates Hercules with the OLR signature of the bar \citep{2000AJ....119..800D,2007A&A...467..145C,2007ApJ...664L..31M,2014AA...563A..60A,2017MNRAS.466L.113M,2017MNRAS.465.1443M,2010MNRAS.407.2122M,2019MNRAS.488.3324F}. Recent evidence from gas and stellar structures in the inner Galaxy converges, however, on $\Omega_\text{bar}\sim40~\text{km/s/kpc}$ (e.g. \citealt{2008A&A...489..115R,2013MNRAS.428.3478L,2015MNRAS.454.1818S,2017MNRAS.465.1621P,2019MNRAS.488.4552S,2019MNRAS.489.3519C,2019MNRAS.490.4740B}). Depending on the Milky Way (MW) rotation curve, this explains the Hercules stream either by CR (the \emph{long slow bar}; \citealt{2017ApJ...840L...2P,2019AA...626A..41M,2020ApJ...890..117D}) or the 1:4 resonance (\emph{slightly faster slow bar}; \citealt{2018MNRAS.477.3945H}). (More details in Section \ref{sec:discussion_literature_comparison}.)

Action space has proven to be especially powerful to study resonance effects (e.g. \citealt{1972MNRAS.157....1L,1978mmcm.book.....A,1981CeMec..25...93M,1996NewA....1..149R,2012ApJ...751...44S}). Strategies include the calculation of the so-called slow and fast actions for a given, individual resonance region (e.g., \citealt{1979MNRAS.187..101L,2015MNRAS.449.1982F,2017MNRAS.471.4314M}), or the perturbation of action-based distribution functions (e.g., \citealt{1989MNRAS.240..991S,2015MNRAS.449.1967F,2015MNRAS.449.1982F,2015ApJ...806..117F,2016MNRAS.457.2569M,2017MNRAS.465.1443M}), or the perturbation of the orbital tori themselves \citep{2018MNRAS.474.2706B,2020MNRAS.495..886B}.

\subsection{Axisymmetric action and frequency estimates as a diagnostic tool} \label{sec:intro_actions}

The \emph{true actions} are true integrals of motion. For general gravitational potentials, it is however not always known how to calculate them, or (all three) actions might not even exist.  The \emph{axisymmetric actions} $\vect{J} \equiv (J_R,J_\phi,J_z)$ are integrals of motion in some static, axisymmetric potentials, but not conserved in more general galaxy potentials with bars and/or spiral arms. Using a best-fit model for the background axisymmetric potential of the MW (e.g. \citealt{2011MNRAS.414.2446M,2015ApJS..216...29B,2019ApJ...871..120E}), we can still calculate instantaneous \emph{axisymmetric action estimates} $\vect{J}$ from observed stellar positions and velocities, with the azimuthal action $J_\phi = R \cdot v_T \equiv L_z$ being the z-component of the angular momentum.

In an axisymmetric potential $\Phi_\text{axi}$, an orbit has the fundamental frequencies 
\begin{equation}
\Omega_{i,\text{axi}}(\vect{J}) \equiv \frac{\partial \mathscr{H}_\text{axi}(\vect{J})}{\partial J_i} \label{eq:frequencies_Hamiltonian_definition}
\end{equation} 
with $i \in [R,\phi,z]$ denoting the Galactocentric cylindrical coordinates, where $\mathscr{H}_\text{axi}$ is the Hamiltonian of the axisymmetric system (see, e.g., \citealt[\S3.5]{2008gady.book.....B}). These frequencies describe the oscillation a star on a given orbit $\vect{J}$ experiences in the three coordinate directions. An in-plane rosette-like disc orbit, for example, can be considered as a superposition of (i) a circular orbit with the (guiding-center) radius $R_g(L_z|\Phi_\text{axi})$ and (ii) an epicycle. The guiding-center of the epicycle moves along the circular orbit with frequency $\Omega_{\phi,\text{axi}}$. In the axisymmetric limit, $\vect{\Omega}_\text{axi} = \vect{\Omega}_\text{true}$. The star moves around the epicycle with frequency $\Omega_{R,\text{axi}}$. The larger the radial action $J_R$, the more extended is the radial oscillation amplitude of the orbit's rosette. Even though the axisymmetric disc model does not contain a bar, we set up a resonance condition\footnote{At many time steps along a bar-affected orbit, $\vect{\Omega}_\text{true} \neq \vect{\Omega}_\text{axi}$. A star can be either a resonantly trapped orbit and \emph{always} satisfy Equation \eqref{eq:res_condition_real_freqs}, or a non-resonant, circulating orbit that \emph{never} satisfies Equation \eqref{eq:res_condition_real_freqs}. The former \emph{will} sometimes satisfy Equation \eqref{eq:res_condition_axisym_freqs}, i.e., when it just crosses the ARL. The latter \emph{may} sometimes cross the ARL and thus temporarily satisfy Equation \eqref{eq:res_condition_axisym_freqs}, depending on how close the star lives on average to the resonance, and how much it oscillates in action space. This is a consequence of $\vect{\Omega}_\text{axi}$ being a 
`wrong' local estimate. The above described behaviour in action space will become clear later on in Section \ref{sec:background} and Appendix \ref{app:oscillation}.} analogous to Equation \eqref{eq:res_condition_real_freqs} using these axisymmetric frequencies:
\begin{equation}
    m \cdot \left(\Omega_\text{bar} - \Omega_{\phi,\text{axi}}\right) - l \cdot \Omega_{R,\text{axi}} = 0 \label{eq:res_condition_axisym_freqs}
\end{equation}
For a given $(l,m)$ and $\Omega_\text{bar}$, we can identify those stars with zero vertical excursions, $J_z=0$, that satisfy this condition. In action space $(L_z,J_R)$, these stars lie along a line. We call this resonance line the \emph{axisymmetric resonance line} (ARL) for short as it is calculated on the basis of axisymmetric action and frequency estimates. It was first plotted by \citet{2010MNRAS.409..145S} and also used in, e.g., \citet{2018MNRAS.474.2706B}. To construct ARLs, we do not require any knowledge about or the existence of a perturber---except for an assumed value for $\Omega_\text{bar}$. In practice, we find the ARL by fitting a linear line to simulation particles with $J_R < 0.15L_{z,0}$ and $J_z < 0.001L_{z,0}$ for which Equation \eqref{eq:res_condition_axisym_freqs} is satisfied.

ARLs have a negative slope in the $(L_z,J_R)$ plane \citep{2010MNRAS.409..145S}. For studies of resonances in position-velocity coordinates this means that the exact Galactocentric radius $R$ (related to $L_z$) at which the resonance condition is satisfied depends on the eccentricity of the orbit (related to $J_R$). A resonance spans therefore a whole region in $R$, the ``Lindblad zone'', as shown by \citet{2015MNRAS.450.2217S}.

In the pre-\emph{Gaia} era, dynamic effects could only be studied locally ($<200~\text{pc}$ from the Sun) where velocities stand in for orbit labels. In the extended RVS sample of \emph{Gaia} DR2, larger parts of an orbit are captured; velocities change along the orbit while integrals of motion stay (roughly) constant. They are therefore better suited to characterize different orbits across the Galaxy. The classical integrals---the total energy $E$ together with $L_z$---are for example often used to find orbit substructure in the Galactic halo (e.g. \citealt{1999Natur.402...53H,2017A&A...598A..58H,2010MNRAS.408..935G,2018ApJ...860L..11K,2018MNRAS.478.5449M}). The main advantage of the action integrals is that they are conserved during adiabatic changes of the (axisymmetric) gravitational potential. Their intuitive meaning and the convenient properties of the canonical action-angle coordinate space make them our integrals of choice.

In this work, we investigate if axisymmetric action and frequency estimates can be informative about the unknown non-axisymmetries in the MW, as also suggested in earlier work by \citet{2018MNRAS.474.2706B}, \citet{2010MNRAS.409..145S}, and others.

\begin{figure*}
    \centering
    \includegraphics[width=\textwidth]{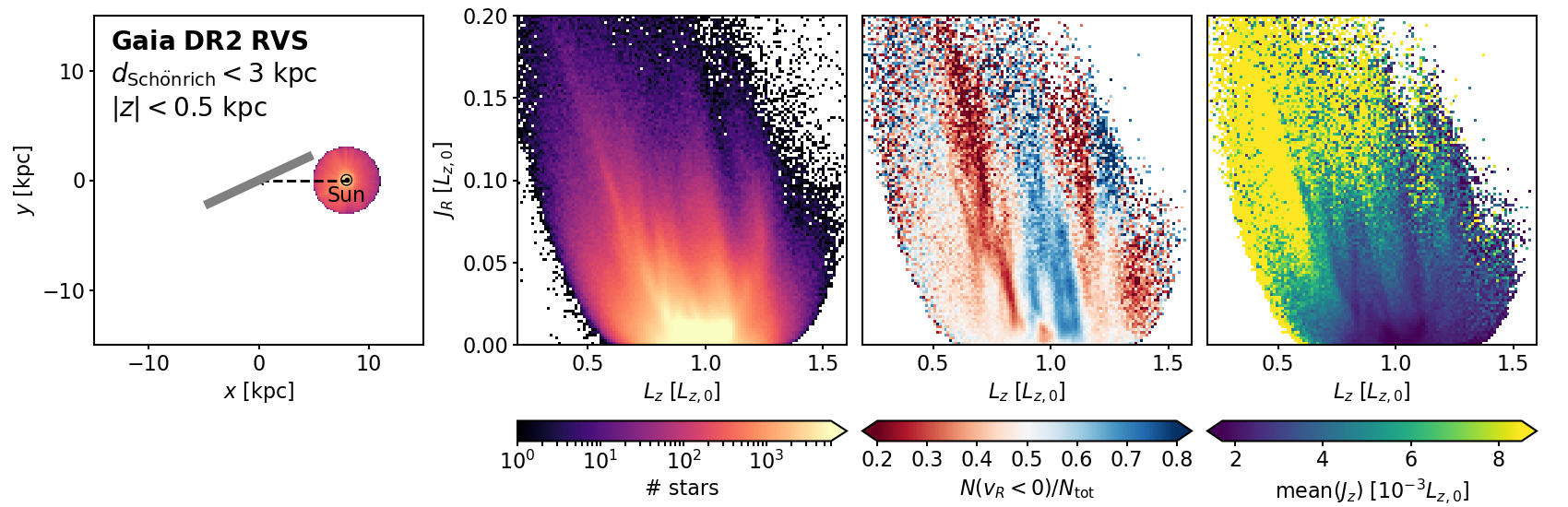}
    \caption{\emph{Gaia} DR2 RVS data in axisymmetric action space estimated in the \texttt{MWPotential2014}. To summarize the main findings in \citetalias{2019MNRAS.484.3291T}, we show the stellar density, the $v_R$-asymmetry and the mean vertical action $\langle J_z \rangle$ as a function of the in-plane actions $(L_z,J_R)$. As opposed to \citetalias{2019MNRAS.484.3291T}, we use the distances by \citet{2019MNRAS.487.3568S} and show the action distribution within the larger volume of $d_\text{Sch\"{o}nrich}<3~\text{kpc}$. The actions are given in units of $L_{z,0} = 8~\text{kpc} \times 220~\text{km/s}$. Overdensity ridges are related to stripes of predominantly inward- (blue) or outward (red) motions, and on average low $J_z$.}
    \label{fig:Gaia_actions_Solar_neighbourhood}
\end{figure*}

One goal of Galactic Dynamics is to perform \emph{quantitative} dynamical modelling of the \emph{Gaia} data, including bar and spiral arms. Several authors (e.g., \citealt{2008gady.book.....B,2013A&ARv..21...61R}) advocate to first strive for an axisymmetric dynamical model of the MW (e.g. as in \citealt{2017ApJ...839...61T}), in which the effect of non-axisymmetries can subsequently be included using quasi-linear perturbation theory \citep{1971ApJ...166..275K,2001MNRAS.328..311W,2015ApJ...806..117F,2016MNRAS.457.2569M,2017MNRAS.471.4314M,2018MNRAS.474.2706B}. However, the sheer amount of substructure in the \emph{Gaia} data requires that we first \emph{qualitatively} disentangle the mechanisms creating the individual ridges.

It has been shown that very different bar and spiral arm models can be tuned to look like the local \emph{Gaia} data \citep{2019MNRAS.490.1026H} or convincingly explain all observed features at once (e.g. \citealt{2019AA...626A..41M,2020A&A...634L...8K,2019arXiv191204304C}). Complex models with many free parameters face therefore---at least at this point in time---the M\"{u}nchhausen trilemma.\footnote{Any model that reproduces all the substructure in the \emph{Gaia} data can only be considered as truth, if also the model assumptions (e.g. about the many mechanisms involved, the bar, spiral arms, and satellite interactions) themselves are close to the truth.} Different strategies can mitigate this problem. Firstly, external information can be employed as a prior to constrain some degrees of freedom, e.g., by using existing bar models based on Galactic center data \citep{2017MNRAS.465.1621P} to model local disk data \citep{2019AA...626A..41M}. Secondly, covering a larger region in the 6D phase-space, with observations and in the modelling, increases the information content available in the data; \citet{2019AA...626A..41M} and \citet{2019MNRAS.490.1026H} study, for example, different 2D and 3D projections of the 4D in-plane phase-space. 

The chosen approach in this work inverts the first strategy. We present a detailed exploration of \emph{just one} strong perturbing mechanism in the Galactic disc: the $m=2$ resonances of the bar with a constant pattern speed. By building intuition about its signatures in the space of axisymmetric actions, the \emph{Gaia} data \emph{alone} might reveal candidates for the bar's true OLR. These `uninformed' results are subsequently compared to external information from other studies. Agreement should then make the identification of the remaining features observed in the Galactic disk and their respective perturbers easier. Discrepancies should be used to inform us with respect to which parts of the OLR signature the simple bar model requires modification and more complexity.

Concerning the second strategy, we use \emph{Gaia} data within 3 kpc from the Sun and study them in three in-plane dimensions, $(J_R,L_z)$ and the radial phase proxy $v_R$, and---for the first time---also in one out-of-plane dimension, the vertical action $J_z$. In this work, we do not cover the fourth in-plane dimension, i.e., the Galactocentric azimuth $\phi$ or the related orbital tangential phase-angle $\theta_\phi$. A companion study, investigating the resonance signatures in the space of orbital phase-angles, is currently in preparation.

This paper is organized as follows. In Sections \ref{sec:data} and \ref{sec:method}, we present the \emph{Gaia} DR2 action data and the test particle simulation of a barred galaxy. Section \ref{sec:background} recapitulates the background theory of bar resonances by means of numerically integrated orbits in action space. Readers who are already familiar with resonant phenomena in action space are encouraged to skip to Section \ref{sec:results} where we lay out our main results: In Section \ref{sec:OLR_signature_and_pattern_speed}, we demonstrate how the OLR signature in action space can be used to estimate the bar's pattern speed from the \emph{Gaia} data; in Section \ref{sec:results_vertical_action}, we show how resonances affect the distribution of the vertical action $J_z$. In Section \ref{sec:discussion}, we investigate the proposed bar pattern speeds more closely and discuss them with respect to existing models in the literature. We summarize and conclude in Section \ref{sec:conclusion}.

\section{Data} \label{sec:data}

We use the \emph{Gaia} DR2 action data introduced in \citetalias{2019MNRAS.484.3291T}. For distances larger than $\sim1~\text{kpc}$, the inverse parallax as employed by \citetalias{2019MNRAS.484.3291T} is a poor distance measurement. In this work, we therefore use instead the $(\alpha,\delta,\mu_\alpha^*,\mu_\delta,v_\text{los})$ from \emph{Gaia} DR2 RVS \citep{2016A&A...595A...1G,2018A&A...616A...1G,2019A&A...622A.205K} together with the Bayesian distance estimates by \citet{2019MNRAS.487.3568S} that include a systematic parallax offset of $0.048~\text{mas}$. This allows us to use the \emph{Gaia} action data out to $d=3~\text{kpc}$ from the Sun. In addition, we restrict the data to within $|z|=500~\text{pc}$ from the Galactic plane and do not apply any quality cuts, as discussed in \citetalias{2019MNRAS.484.3291T}. This data set includes $\sim4.8$ Million stars.

\begin{figure*}
    \centering
    \includegraphics[width=\textwidth]{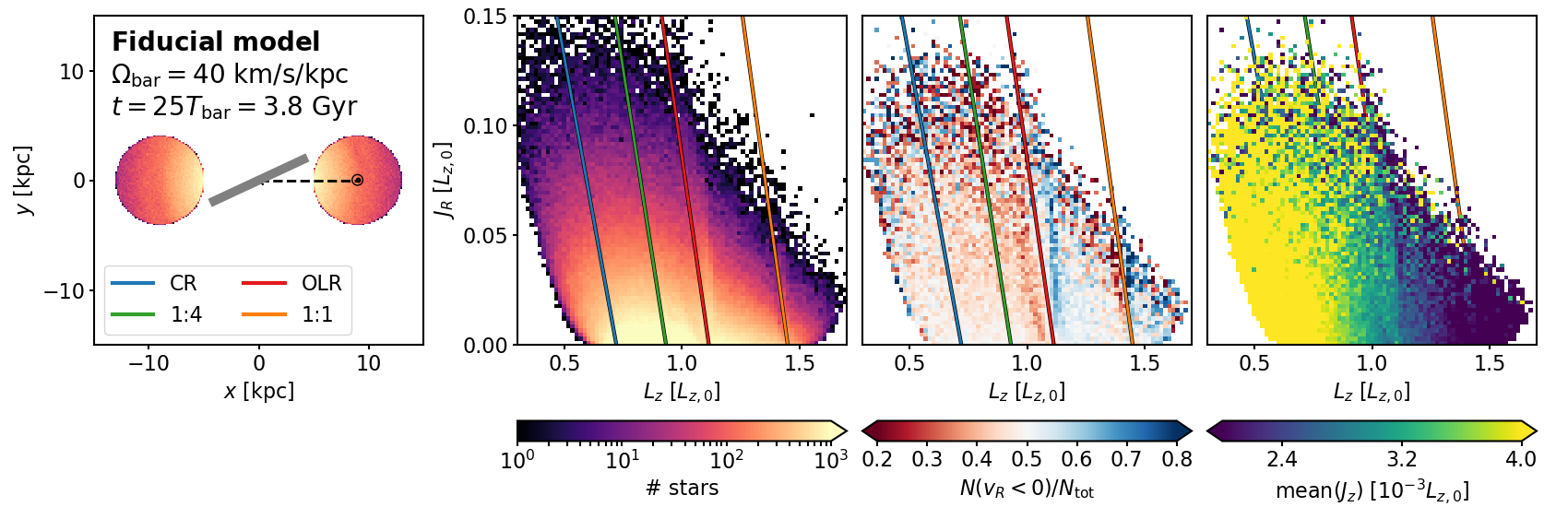}
    \caption{Axisymmetric action space for the test particle simulation in the \texttt{Fiducial} bar model in a \emph{Gaia}-like survey volume (cylinder with radius $4~\text{kpc}$ and $|z| < 500~\text{pc}$ around a solar-like position at $R=9~\text{kpc}$). This figure shows the location of the survey volume with respect to the bar (after $t=25T_\text{bar}=3.8~\text{Gyr}$ of orbit integration), the stellar number density in the $(L_z,J_R)$ plane, as well as action space colour-coded by the number of inward moving stars ($v_R < 0$) to the total number of stars $(N_\text{tot})$ per bin and the mean vertical action $\langle J_z \rangle$. We overplot the ARLs for $\Omega_\text{bar} = 40~\text{km/s/kpc}$. The signatures caused by the OLR and 1:1 resonance resemble qualitatively some of the features observed in the \emph{Gaia} data in Figure \ref{fig:Gaia_actions_Solar_neighbourhood}.}
    \label{fig:Fiducial_bar_model_actions}
\end{figure*}

In Figure \ref{fig:Gaia_actions_Solar_neighbourhood}, we show the corresponding axisymmetric action estimates in the \texttt{MWPotential2014} model by \citet{2015ApJS..216...29B}, with the $(L_z,J_R)$ distribution colour-coded by stellar density, $v_R$-asymmetry, and $\langle J_z \rangle$, thus summarizing the three main findings by \citetalias{2019MNRAS.484.3291T}:

(i) The \emph{stellar overdensities} in the in-plane velocity space are the local manifestation of an extended system of orbit structures in the Galactic disc which reach consistently out to (at least) $\sim 3~\text{kpc}$ from the Sun and lie along ridges of slightly negative slopes $\Delta J_R / \Delta L_z$ around $L_z\sim\text{const}$.

(ii) Action space colour-coded by the fraction of inward-moving stars, reveals a strong \emph{$v_R$-asymmetry pattern} of predominantly inward- or outward motion along the ridges. This corresponds to asymmetric numbers of stars at $+v_R$ vs. $-v_R$ in the well-known $(U,V) \sim \left(-v_R,v_T-v_\text{circ}\right)$ plane of the Solar neighbourhood. In some recent papers, the same features were shown by \citet{2019MNRAS.488.3324F} and \citet{2019MNRAS.485.3134L}, who colour-coded the $(R,v_T)$ plane by $\langle v_R \rangle$, and by \citet{2019MNRAS.490.5414F}, who plotted $-\langle v_R \rangle$ as a function of $L_z$ and in the $(L_z,\phi)$ plane.

(iii) Another property of action space found in \citetalias{2019MNRAS.484.3291T} was that the overdensity ridges were related to on average \emph{low vertical action $J_z$}. \citet{2019MNRAS.489.4962K} also noted that the ridges in $(R,v_T)$ live mostly at low $|z|$.

Any proposed creation mechanism for the ridges needs to be able to explain all three of these properties.

\section{Simulation}  \label{sec:method}

\subsection{Test particle simulation} \label{sec:method_sim}

\emph{Simulation setup.---}To investigate the effect of the bar on individual orbits in action space, we set up an idealized MW-like galaxy test particle simulation. The initially axisymmetric galaxy uses the \texttt{MWPotential2014} as the gravitational background potential (with $R_0 \equiv 8~\text{kpc}$, $v_0 \equiv v_\text{circ}(R_0) = 220~\text{km/s}$, $L_{z,0}\equiv R_0\times v_0=1760~\text{kpc km/s}$, and $\Omega_0\equiv v_0/R_0 = 27.5~\text{km/s/kpc}$). We generate a stellar disc with 5 Million massless test particles from the quasi-isothermal distribution function (DF) by \citet{2011MNRAS.413.1889B}. Then, we add the 3D quadrupole ($m=2$) bar model by \citet{2000AJ....119..800D} and \citet{2016MNRAS.461.3835M} to the potential, orientated at an azimuth of $25~\text{degrees}$ with respect to and ahead of the Sun \citep{2019MNRAS.490.4740B}, and rotate it with a pattern speed of $\Omega_\text{bar}=40~\text{km/s/kpc} = 1.45\Omega_0$. Our \texttt{Fiducial} bar model has a weak bar strength. Its pattern speed was chosen for illustration purposes and is only by coincidence close to the \emph{(slightly faster) slow bar} pattern speed in the literature. We integrate the orbits of all mock stars in the barred potential for up to 25-50 bar periods using \texttt{galpy}\footnote{The Python package for Galactic dynamics \texttt{galpy} by \citet{2015ApJS..216...29B} can be found at \url{http://github.com/jobovy/galpy}.}. Their final phase space coordinates $(\vect{x},\vect{v})$ are then used to estimate the actions $\vect{J}$ and frequencies $\vect{\Omega}_\text{axi}$ in the axisymmetric background \texttt{MWPotential2014} using the \emph{St\"{a}ckel Fudge} algorithm by \citet{2012MNRAS.426.1324B} \citep{2013ApJ...779..115B,2016MNRAS.457.2107S}. The model parameters of the \texttt{Fiducial} bar model are summarized in Table \ref{tab:bar_models}. In Appendix \ref{sec:app_simulation_details}, we present more details about the simulation setup, and illustrate it in Figure \ref{fig:Fiducial_model_intro}.

\emph{Methodological context to related and recent studies.---}A similar study using the idealized test particle simulation approach was performed by \citet{2003A&A...401..975M}. They investigated bar resonances in the velocity moments as a function of $R$ at $z=0$. We are interested in action space $(L_z,J_R)$ and use a 3D bar to study also signatures in $J_z$. \citet{2018MNRAS.474.2706B} investigated the DF in $(L_z,J_R)$ evolving under the influence of a bar by applying perturbation theory. To be able to interpret action-angle space of Galactic surveys, correcting for selection effects is crucial \citep{2011MNRAS.418.1565M}. Combining selection functions with perturbed action-angle-based DFs is non-trivial, so we resort to test particle simulations. \citet{2019AA...626A..41M}, \citet{2019MNRAS.490.1026H}, and \citet{2019MNRAS.484.3154S} studied the signatures of bars and/or spiral arms in action space at fixed positions in the Galaxy using the backwards-integration method by \citet{2000AJ....119..800D}. This method finely resolves the ridges, but becomes noisy when integrating for more than $\sim 10$ bar periods. \citet{2019MNRAS.488.3324F} and \citet{2019MNRAS.490.1026H} studied resonance signatures in self-consistent N-body and test particle simulations with more complex bar models, respectively. To complement these studies, we focus here on isolating and describing the effect of the $m=2$ bar only.

\emph{Parameter space exploration.---}We have run test particle simulations with different (a) pattern speeds, (b) integration times, (c) bar strengths, (d) either slowly introducing the bar over several bar periods or switching it on instantaneously, (e) including different $m=4$ bar components. The behaviour of tests (a)-(d) was well-behaved for weak bars, as expected when comparing to our \texttt{Fiducial} bar model, and left the results in this work qualitatively unchanged. We therefore focus on the \texttt{Fiducial} model only. Adding the $m=4$ bar component in tests (e) confirmed the findings by \citet{2018MNRAS.477.3945H} and \citet{2019AA...626A..41M} in some respects (see Section \ref{sec:discussion_Gaia_vs_sim}), and disagreed in others. It increased the space of bar parameters to explore and introduced complex behaviour. The signatures of the $m=2$ component, however, remained very similar. A detailed discussion of the effect of the $m=4$ components is therefore beyond the scope of this paper.

\subsection{Action space of the \texttt{Fiducial} model around the Sun} \label{sec:method_fiducial_solar_neighbourhood}

In Figure \ref{fig:Fiducial_bar_model_actions}, we show the action distribution analogously to the \emph{Gaia} data for the \texttt{Fiducial} bar model for test particle stars within the two cylinders of radius $4~\text{kpc}$ centered on an observer's position at $(x,y,z)_\odot = (\pm9,0,0)~\text{kpc}$. By using the frequencies ($\Omega_R,\Omega_\phi)_\text{axi}$, we overplot the ARLs for the known $\Omega_\text{bar}$ of the system.

In the stellar density distribution, we note (i) an high-$J_R$ overdensity ridge to the right (and an underdensity region to the left) of the OLR and 1:1 ARL, as expected (see Section \ref{sec:scattering_theory}), (ii) an otherwise smooth distribution similar to an axisymmetric distribution (c.f. Figure \ref{fig:Fiducial_model_intro}(b)), which is---except for the parabolic lower envelope\footnote{The parabolic envelope at low $J_R$ in the action distribution is due to the cylindrical selection of the data. See \S 2.3.2 in \citetalias{2019MNRAS.484.3291T} for an explanation of this selection effect.}---(iii) very similar to the action space without any spatial subselection (c.f. Figure \ref{fig:Fiducial_model_intro}(e)). The latter is one of the advantages of using action space.

The $v_R$-asymmetry panel in Figure \ref{fig:Fiducial_bar_model_actions} shows that the ARL of the OLR (and also the 1:1 ARL) separates an outward-moving (red) stripe from an inward-moving (blue) stripe. In addition, there is a weak trend towards outward-moving (red) stars for $L_z$ smaller than the OLR. We will investigate this OLR signature in more detail in Section \ref{sec:OLR_signature_and_pattern_speed}.

The mean vertical action in Figure \ref{fig:Fiducial_bar_model_actions} shows the expected trend of decreasing $\langle J_z \rangle$ with $L_z$. The reason is that orbits with the same $z_\text{max}$ (i.e., maximum height above the plane that can be reached), have higher $J_z$ if they live in the in the inner disk because of the higher surface-mass density. Interestingly, this trend is broken around the OLR: The underdensity region to the left of the OLR has higher $\langle J_z \rangle$ while the overdensity ridge to the right of the OLR resonance has lower $\langle J_z \rangle$. This is surprising, as the bar potential model depends on $z$ only very weakly in the Solar vicinity, and the $J_z$ of the individual stars did not change significantly during orbit integration; the mean change is just $\langle |J_{z,\text{end}} - J_{z,\text{start}}|\rangle/L_{z,0}\sim4\cdot10^{-5}$. We will investigate this further in Section \ref{sec:results_vertical_action}.

Overall, Figure \ref{fig:Fiducial_bar_model_actions} illustrates that bar resonances, in particular the OLR, can give rise to signatures in the space of axisymmetric actions qualitatively similar to some of those observed in the \emph{Gaia} data.

\section{Background} \label{sec:background}

\begin{figure}
    \centering
     \subfigure[Action evolution of bar-affected orbits versus the corresponding orbit in the axisymmetric potential for an integration time of $5T_\text{bar} = 0.8$ Gyr. \label{fig:example_orbits_T5}]{\includegraphics[width=\columnwidth]{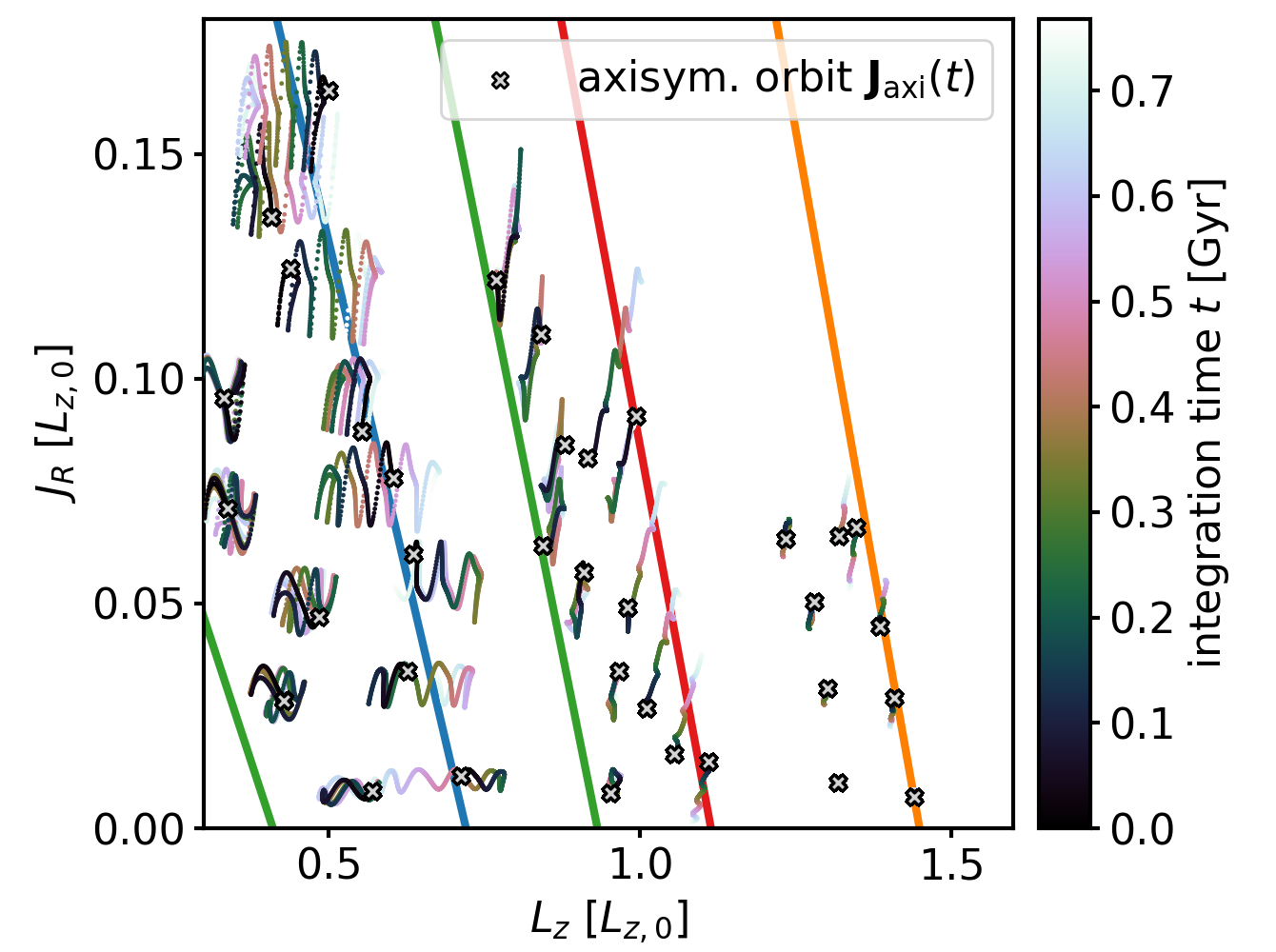}} \hfill
    \subfigure[For an integration time of $50T_\text{bar} = 7.7$ Gyr, all stars reveal an oscillation in axisymmetric action space. Resonant orbits oscillate around their ARL. The colour-coding according to peri-/apocenter illustrates the orbit libration.\label{fig:example_orbits_T50}]{\includegraphics[width=\columnwidth]{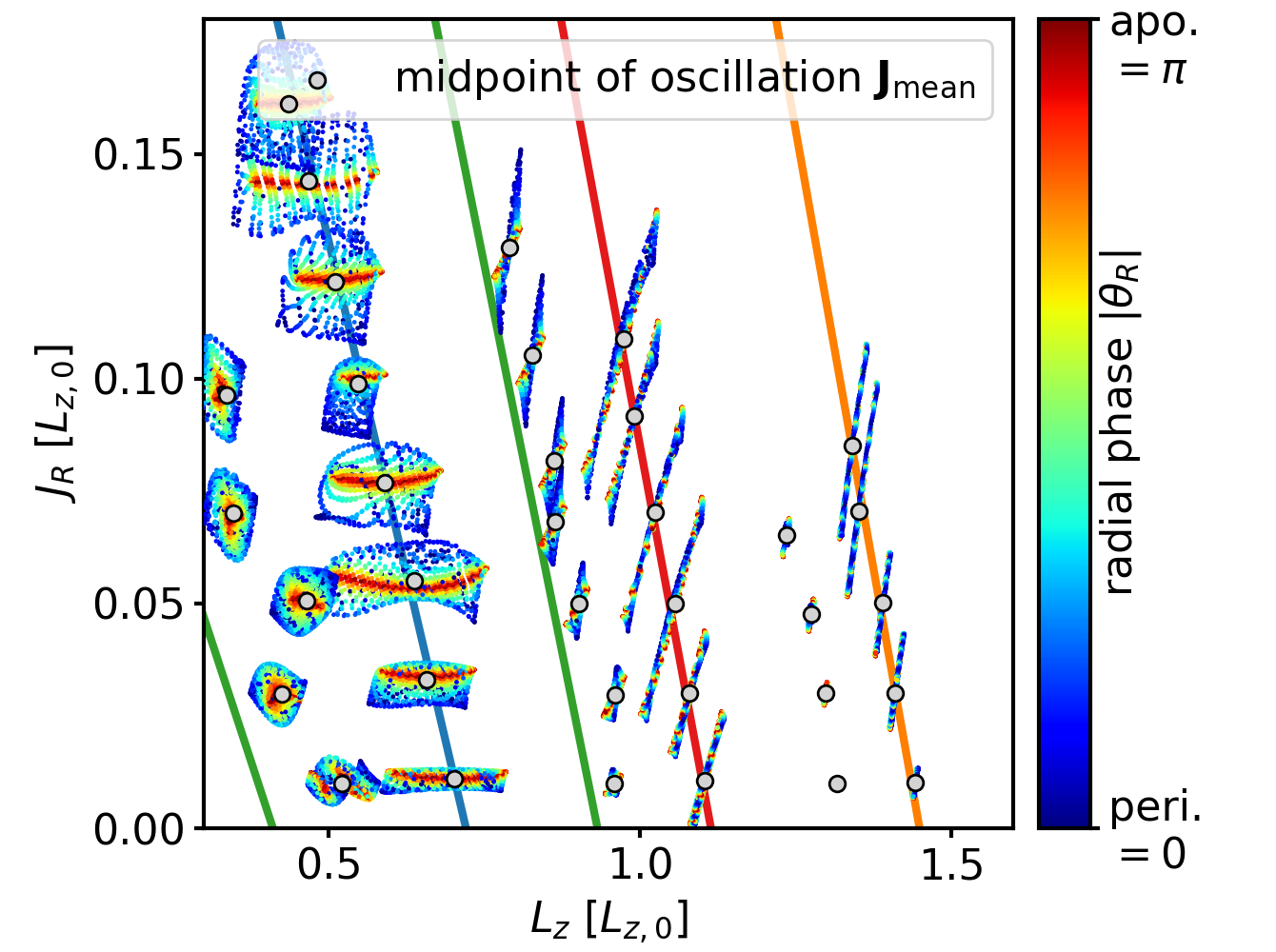}} %
    \subfigure[OLR orbits only, in action space (left) and in the $(x,y)$ frame (right) co-rotating with the bar (black). The librating orbits are shown in grey, their corresponding closed parent orbits are colour-coded by radial phase.  \label{fig:example_orbits_parents}]{\includegraphics[width=\columnwidth]{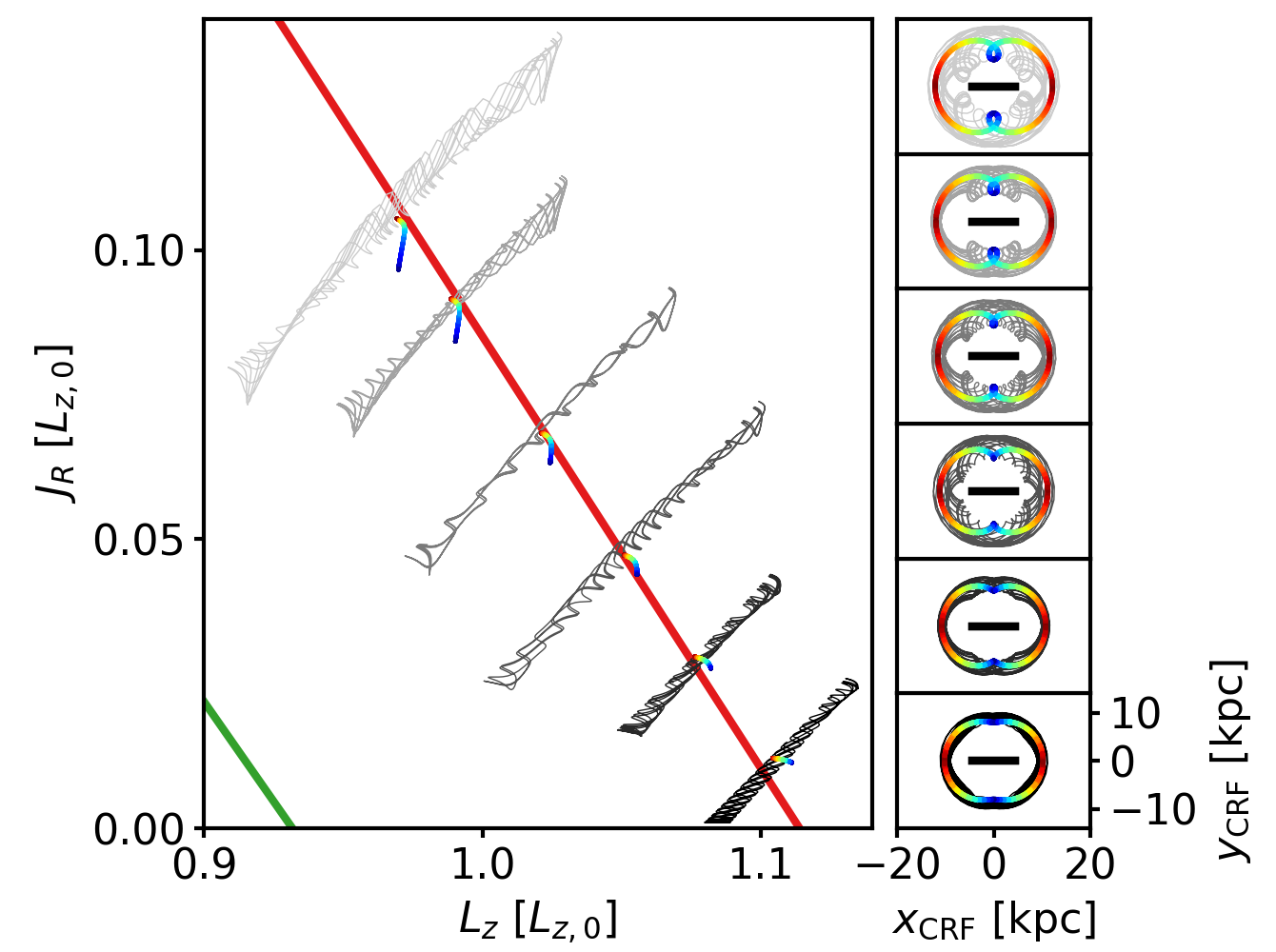}} %
    \caption{Example orbits integrated in the \texttt{Fiducial} bar potential and their time evolution in the axisymmetric $(L_z,J_R)$ action plane. The thick solid lines are the CR (blue), OLR (red), 1:1 (orange), $\pm$1:4 (green) ARLs. We show both resonant stars and stars in-between the ARLs.}
    \label{fig:example_orbits}
\end{figure}

\subsection{Numerical example orbits in action space}

The behaviour of bar-affected orbits in action space has often been studied on the basis of perturbation theory. (A recent and pedagogical explanation can be found, for example, in \citet{2019arXiv191204304C}.) To complement this, we show here numerically integrated orbits in the space of axisymmetric actions. We start with a few individual examples in Figure \ref{fig:example_orbits} to build intuition.

Figure \ref{fig:example_orbits_T5} shows (as grey crosses) the orbits integrated in the axisymmetric \texttt{MWPotential2014} for an integration time of $t=5T_\text{bar}=0.8~\text{Gyr}$. As the actions are conserved in this potential, the time evolution along each orbit corresponds to one single point $J_\text{axi}(t) = \text{constant}$. Inaccuracies in the orbit integration and action estimation do lead to small (unphysical) time variations in $\vect{J}_\text{axi}$, but they are smaller than the marker size in Figure \ref{fig:example_orbits_T5}.

If we integrate the stars with the same initial conditions in the potential with the rotating bar, they all move substantially in both $L_z$ and $J_R$ direction (colour-coded by time in Figure \ref{fig:example_orbits_T5}).

Figure \ref{fig:example_orbits_T50} shows the same orbits for a longer integration time of $t=50T_\text{bar}=7.7~\text{Gyr}$. All stars oscillate in a restricted area within $(L_z,J_R)$. We take the time-average to determine the central location around which the orbit oscillates (marked by a grey dot), i.e.,
\begin{equation}
    \vect{J}_\text{mean}\equiv\langle \vect{J}(t)\rangle_t.
\end{equation}

In the following, we discuss the resonant phenomena of \emph{scattering}, \emph{libration} around \emph{parent orbits}, \emph{oscillation}, and \emph{orbit orientation} by means of these numerical orbits in action space and knowledge from the literature. We explain how they lead to the signatures observed in Figure \ref{fig:Fiducial_bar_model_actions}.

\subsection{Scattering} \label{sec:scattering_theory}

\begin{figure}
    \centering
    \includegraphics[width=\columnwidth]{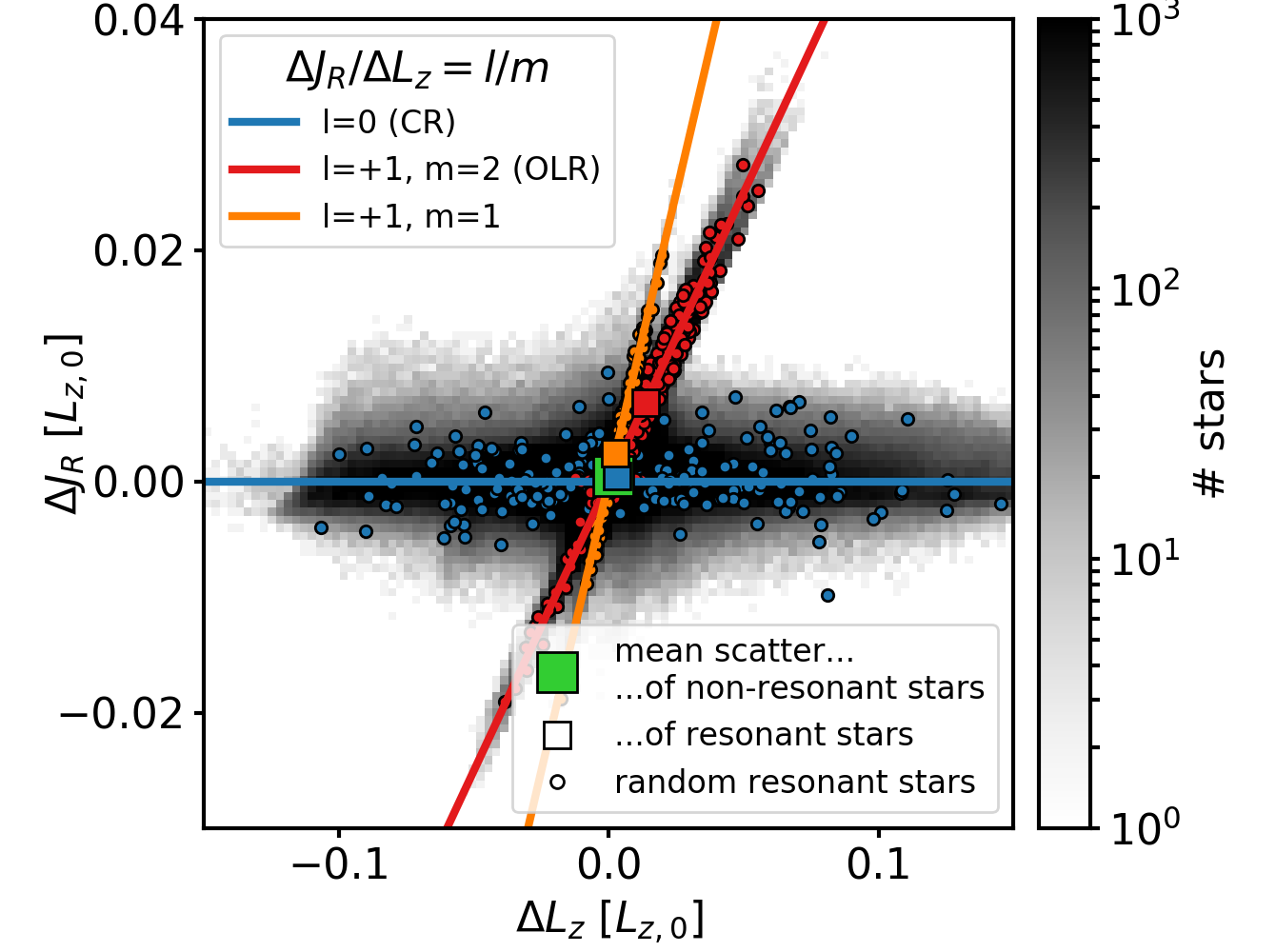}
    \caption{Scattering of all test particle stars with $R_\text{end}>5~\text{kpc}$ in the \texttt{Fiducial} bar simulation. ``Scattering'' is the lasting, average change in the axisymmetric action due to the bar (Equation \eqref{eq:midJ}). The majority of stars with significant scattering, $|\Delta \vect{J}| \gg 0$, are as expected stars in resonance with the bar. As predicted by $E_J$ conservation, resonant stars change their actions along $\Delta J_R/\Delta L_z \sim l/m$ \citep{2002MNRAS.336..785S}. A strong net change occurs at the OLR and 1:1 resonance (red and orange squares)}.
    \label{fig:scattering}
\end{figure}

It has long been known that \emph{resonant scattering} can change orbits substantially \citep{1972MNRAS.157....1L}. \citet{1989MNRAS.240..991S} for example showed that resonant scattering at spiral arms creates ridges in action space (their fig. 7). A rotating bar potential with a fixed pattern speed conserves the Jacobi integral \begin{equation}
E_J = E - \Omega_\text{bar} \times L_z \label{eq:Jacobi_integral}
\end{equation}
along a star's orbit \citep[\S 3.3.2]{2008gady.book.....B}. \citet{2002MNRAS.336..785S} demonstrated analytically that this implies the following relation:
\begin{equation}
    \Delta J_R = \frac{\Omega_\text{bar}-\Omega_\phi}{\Omega_R} \Delta L_z = \frac{l}{m}  \Delta L_z. \label{eq:scattering}
\end{equation}
This relation is valid in the epicyclic approximation of near circular orbits and still approximately true for more eccentric orbits \citep{1972MNRAS.157....1L,2002MNRAS.336..785S}. We illustrate this scattering process due to bar resonances, which changes a star's long-term average location $\vect{J}_\text{mean}$ with respect to its initial axisymmetric actions $\vect{J}_\text{axi}$, by plotting in Figure \ref{fig:scattering} the difference 
\begin{equation}
    \Delta \vect{J} \equiv \vect{J}_\text{mean} - \vect{J}_\text{axi}\label{eq:midJ}.
\end{equation}
$\Delta L_z$ describes an average, lasting change in $L_z$ due to the bar and contributes to the radial migration of stars within the disk. $\Delta J_R$ can be considered as a difference in the average amount of radial oscillation that the orbit experiences with respect to the unperturbed orbit. In Figure \ref{fig:scattering}, we overplot also...
\begin{enumerate}[leftmargin=*,topsep=0ex,label=(\roman*)]
\item ...stars that are according to Equation \eqref{eq:res_condition_real_freqs} truly in resonance with the bar.\footnote{We determine the real fundamental frequencies of the orbits, $(\Omega_\phi,\Omega_R)_\text{true}$, from a Fourier analysis of $\vect{x}(t)$ analogous to \citet{2019MNRAS.488.3324F} (see also \citealt{1982ApJ...252..308B,1993CeMDA..56..191L}).} It is therefore the stars in main resonances that experience substantial scattering (the extended wings with $|\Delta \vect{J}| \gg 0$ in Figure \ref{fig:scattering}).
\item ...the analytic scattering relation \eqref{eq:scattering}. As expected, the resonant scattering wings lie along slopes of 
\begin{equation}
\Delta J_R / \Delta L_z \sim l/m.
\end{equation}
The exact scattering direction depends on the star's instantaneous phase angles at the time the bar is switched on. If the bar in the simulation is slowly grown, the resonant stars follow the scattering relation more closely. In a test particle simulation with a stronger bar ($\alpha_{m=2}=0.015$), stars can get scattered further.
\item ...separately the average net change for all stars at a given resonance, and also for all non-resonant stars. We find that only at the OLR resonance occurs a significant \emph{net} change in both $L_z$ and $J_R$. As can also be seen in Figure \ref{fig:example_orbits_T5}, more stars at the OLR get scattered towards higher $L_z$ than towards lower $L_z$.
\end{enumerate}

The latter is a consequence of relation \eqref{eq:scattering}, which couples at the OLR and 1:1 resonance the direction of $\Delta L_z$ to the change in $J_R$. Stars with the same $\vect{J}_\text{mean}$ got scattered from their initial $\vect{J}_\text{axi}$: either from higher $J_{R,\text{axi}}$ on the right of the ARL downwards, $\Delta L_z = 2 \Delta J_R < 0$, or from lower $J_{R,\text{axi}}$ on the left upwards, $\Delta L_z = 2 \Delta J_R > 0$. In the overall disk population, the stellar density decreases steeply with increasing $J_R$, i.e., many more stars stars live initially at lower $J_{R,\text{axi}}$. Consequently, scattering in the $\Delta L_z = 2 \Delta J_R > 0$ direction occurs more often than in the opposite direction. In addition, near-circular orbits with $J_{R,\text{axi}}\sim0$ can only increase their radial oscillations and therefore $L_z$. This is also illustrated in the $(L_z,E)$ plane by fig. 1 in \citet{2002MNRAS.336..785S}. Overall, this process leads to the underdensity vs. overdensity signature around the OLR and 1:1 resonances in Figure \ref{fig:Fiducial_bar_model_actions}.

The strongest radial migration is, as expected, observed for individual CR stars. The CR's weak effect on $J_R$ and the absence of a preferred scattering direction in $L_z$ explains the absence of obvious substructure around the CR ARL in Figure \ref{fig:Fiducial_bar_model_actions} (see also Appendix \ref{app:oscillation}).\footnote{The scattering relation in Equation \eqref{eq:scattering} is only satisfied in potential models with a fixed bar pattern speed. A recent study by \citet{2018A&A...616A..86H} investigated the net change in angular momentum at CR. They found that in their galaxy simulations which consider a realistic, self-consistent bar formation process including growth and slow-down, the CR swipes through a large range of Galactocentric radii and pulls trapped stars along, causing a large $L_z$ net change. This radial migration of the stars due to the transient process is called ``churning''. For a fixed potential with constant pattern speed, they did not measure a $L_z$ net change, just periodic oscillations around CR, as in our Figure \ref{fig:example_orbits}.}

\subsection{Parent orbits}

In the axisymmetric system, rosette-shaped disk orbits can be grouped into families with the closed, circular orbit ($J_R=0$) as the parent orbit. Orbits with the same $L_z$ oscillate in the $(R,v_R)$ plane around it---the larger $J_R$, the larger the amplitude.

In the barred system, resonant orbit families are grouped by the same $E_J$. Their parent orbits close in the $(x_\text{CRF},y_\text{CRF})$ frame co-rotating with the bar and correspond to a point in the surface of section $(x_\text{CRF},v_{x,\text{CRF}}\mid y_\text{CRF}=0)$. In our mock simulation, the parent orbits do not get populated. We therefore use the algorithm by \citet{1993RPPh...56..173S} to determine the parent orbits belonging to the specific OLR example orbits in Figure \ref{fig:example_orbits}. Figure \ref{fig:example_orbits_parents} shows these parent orbits both in the co-rotating $(x_\text{CRF},y_\text{CRF})$ frame and in action space. The parent orbits oscillate between peri- and apocenter which correspond to slightly different locations in $(L_z,J_R)$ but constant $E_J$. This fast oscillation between the radial phases is in the literature described by the radial angle $\theta_R$, also called the \emph{fast angle} $\theta_f$ close to a resonance.

\subsection{Libration}

In perturbation theory studies, the orbit evolution is usually averaged over $\theta_f$. It follows from $E_J$ conservation that, on average, the quantity
\begin{equation}
    J_f \equiv J_R - \frac lm L_z \sim \text{const.}
\end{equation}
is close to an integral of motion. It is known in the literature as the \emph{fast action} of a given resonance (see also \citealt{1979MNRAS.187..101L,1994MNRAS.268.1041K,1994ApJ...420..597W,2017MNRAS.471.4314M}).

In our numeric study using axisymmetric estimates, we observed in Figure \ref{fig:example_orbits_T50} that the apocenter of each orbit changes with time along a line in $(L_z,J_R)$. This line is not perfectly linear. We have, however, checked in our simulation that the apocenters of true resonant stars at the main resonances (OLR, 1:1, CR, and also 1:4) move indeed along slopes 
\begin{equation}
    \delta J_R / \delta L_z \sim l/m,
\end{equation}
where $\delta J_i$ is the oscillation amplitude in the $J_i$-direction (Equation \eqref{eq:oscillation_amplitude}; see also fig. 4 in \citealt{2018MNRAS.474.2706B}). (From Figure \ref{fig:example_orbits_T50} it appears that the pericenters oscillate more strongly in action space.) The slow evolution along constant $J_f$ is called \emph{orbit libration} and the variation is described by the \emph{slow angle} $\theta_s$. Parent orbits do not librate (see Figure \ref{fig:example_orbits_parents}). In the $(x_\text{CRF},y_\text{CRF})$ frame co-rotating with the bar, the libration corresponds to the slow shift of the peri-/apocenter in azimuth $\phi_\text{CRF}$. The azimuthal range within which the peri-/apocenters librate is restricted---this is called the \emph{trapping} of the orbit at the resonance (see, e.g., the OLR example orbits in Figure \ref{fig:example_orbits_parents}, and fig. 1 and 6 in \citet{2017MNRAS.471.4314M}). 

The maximum libration amplitude possible at a resonance, depends on $J_{R,\text{mean}}$, and the strength of the bar. Beyond this boundary (called the \emph{separatrix}), orbits are circulating, i.e. the whole azimuthal range $\phi_\text{CRF} \in [-\pi,\pi]$ is available to the peri-/apocenters. More details can be found in, e.g., \citet{2018MNRAS.474.2706B,2019arXiv191204304C}.

\begin{figure}
    \centering
     \includegraphics[width=\columnwidth]{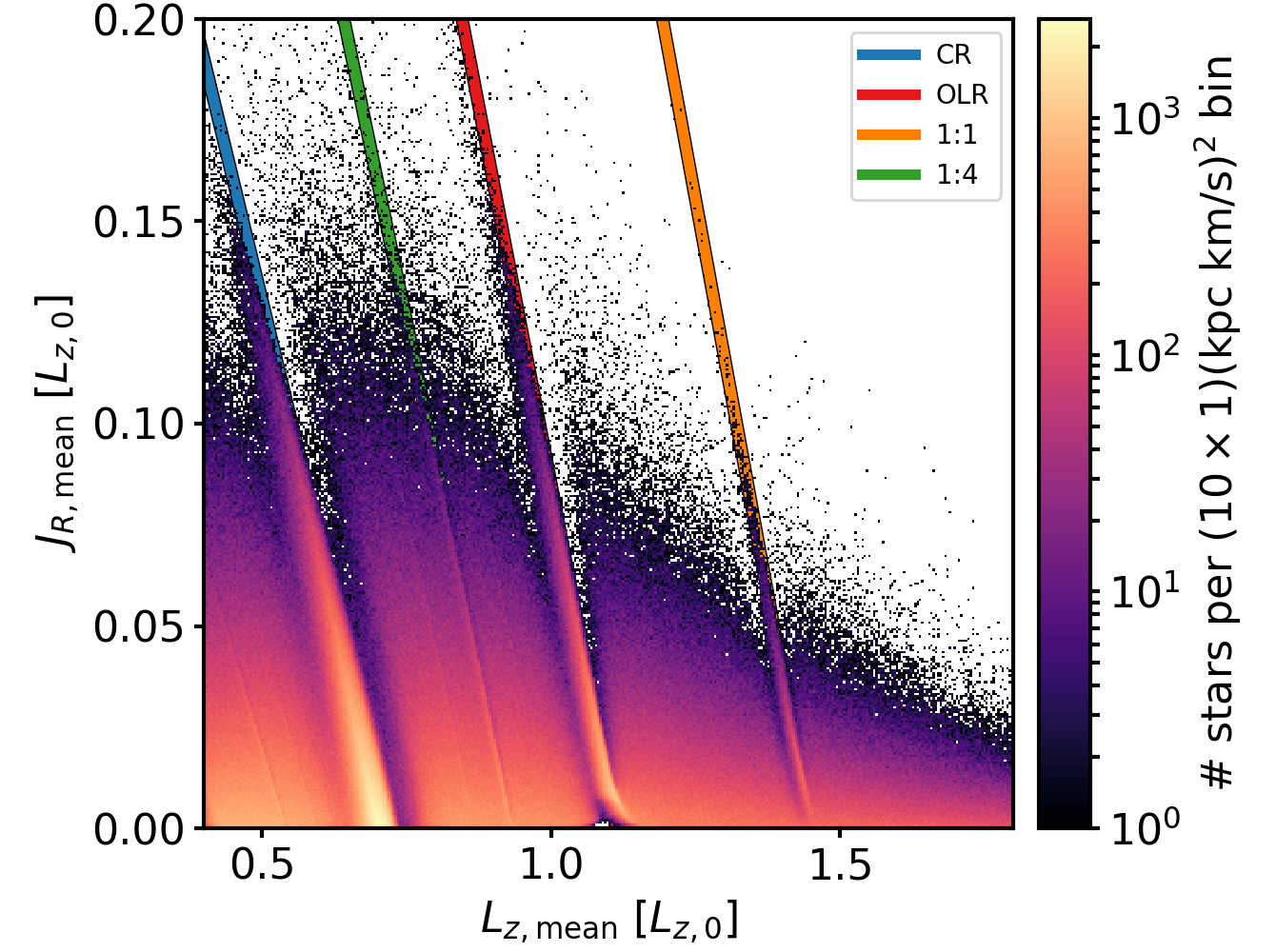}
    \caption{Oscillation midpoints $\vect{J}_\text{mean}$ in the \texttt{Fiducial} bar model. We define ``oscillation'' as the midpoint $\vect{J}_\text{mean}$ and the amplitudes $\delta J_R$ and $\delta L_z$ (see Figure \ref{fig:oscillation_amplitudes}) with which the stars oscillate around it in the plane of axisymmetric actions (see Figure \ref{fig:example_orbits_T50}). This figure illustrates a ``trapping'' of stars at the resonances: The distribution of oscillation midpoints is smooth in the actions $(L_z,J_R)$, only the vicinity of the ARLs gets depleted and stars accumulate along the $l:m$ ARLs.}
    \label{fig:LzJR_mid}
\end{figure}

\subsection{Oscillation} \label{sec:oscillation_theory}

To summarize, a resonant orbit `swings' fast between peri- and apocenter, and librates slowly along a line of average constant $J_f$. Our axisymmetric action estimates $L_z$ and $J_R$ oscillate in both cases---and also significantly along non-resonant orbits. In the following, we use the term \emph{oscillation} therefore to describe the general variation in $L_z$ or $J_R$, not just at the resonances. At the resonances, the idealized prediction in Equation \eqref{eq:scattering} is a good approximation to describe the overall behaviour of scattering and oscillation.

In Figure \ref{fig:LzJR_mid}, we show now the distribution of oscillation midpoints $(L_{z,\text{mean}},J_{R,\text{mean}})$ for all mock stars in the \texttt{Fiducial} model. We overplot the ARLs. Overall, the distribution looks smooth and similar to $(L_z,J_R)$, showing that most stars oscillate close to their $\vect{J}_\text{axi}$ (i.e. scattering $\Delta \vect{J} \sim 0$). Only in the vicinity of the ARLs, the resonance has depleted regions of $\vect{J}_\text{mean}$ and accumulated the stars' oscillation midpoints along the ARLs.\footnote{In Figure \ref{fig:LzJR_mid}, at high $J_R$, the stellar distribution in $\vect{J}_\text{mean}$ tilts away from our linear ARL fit to stars satisfying Equation \eqref{eq:res_condition_axisym_freqs}. This is because for $J_R>0.15L_{z,0}$, the ARL does actually not stay perfectly linear.}. The stars that are trapped at the resonances librate around the ARLs, was also one of the main findings by \citet{2018MNRAS.474.2706B} from the study of perturbed tori.

As laid out it Appendix \ref{app:oscillation}, all stars oscillate and, in general, the oscillation amplitude $\delta J_R$ increases with $J_R$, and $\delta L_z$ decreases with $L_z$. The resonances, in particular those with $l\neq0$, are locations of increased $J_R$ oscillation as expected (see Figure \ref{fig:oscillation_amplitudes}).

\begin{figure}
    \centering
    \includegraphics[width=\columnwidth]{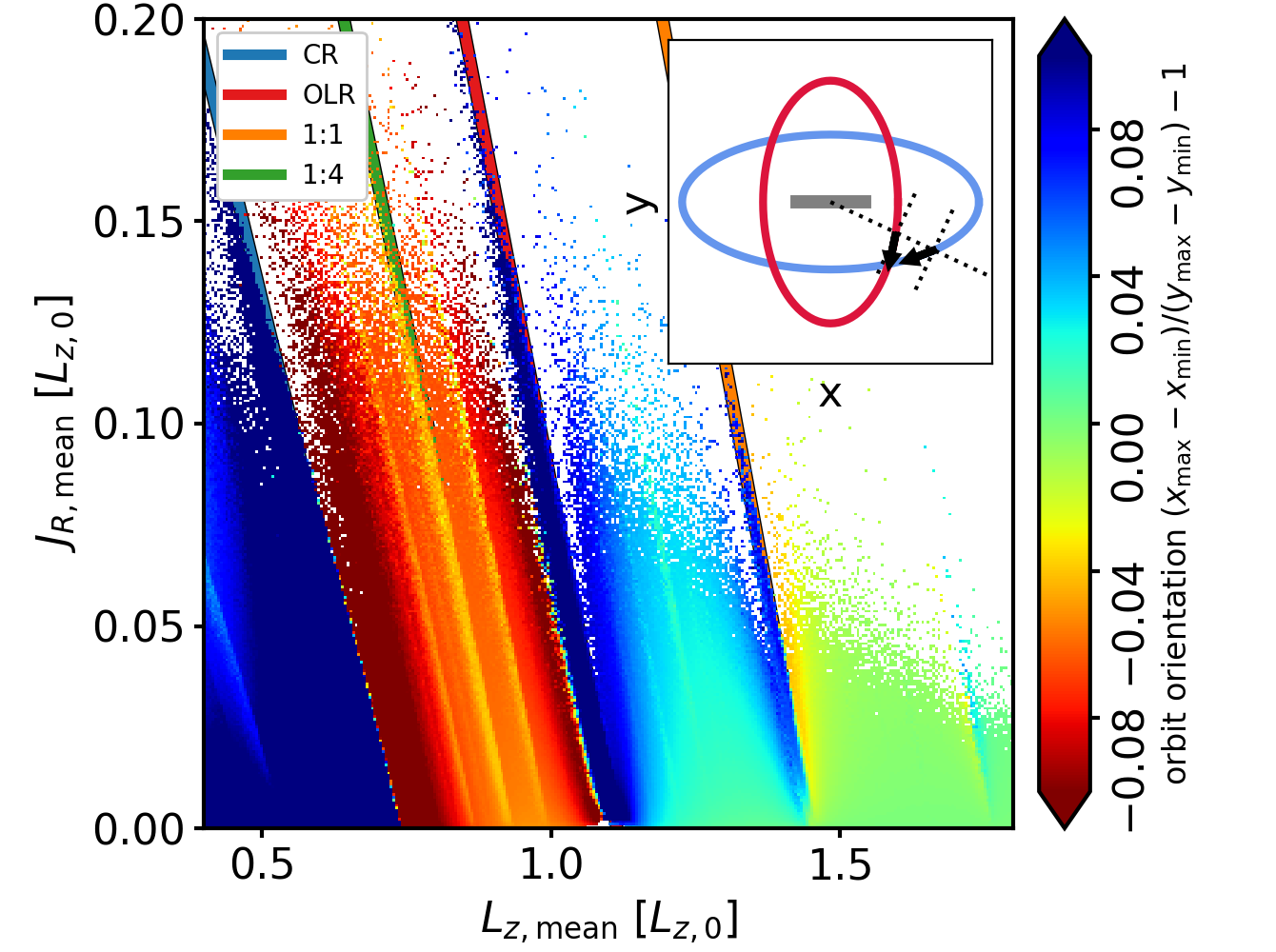}
    \caption{Orbit orientation in the frame co-rotating with the bar as a function of the oscillation midpoints in action space. For each orbit in the \texttt{Fiducial} simulation we get an estimate for the orbit elongation and its orientation as $q\equiv (x_\text{max}-x_\text{min})/(y_\text{max}-y_\text{min})-1$, where $x=0$ is always aligned with the bar (see inserted cartoon). For $q=0$, the orbit appears round, for $q \neq 0$ elongated. For $q>0$, the orbit is aligned with the bar. For $q<0$ the orbit is anti-aligned with the bar. This figure illustrates that, when averaging the axisymmetric action estimates over time, these orbit orientation flips occur cleanly along the ARLs in action space. This is a consequence of the well-known fact that librating resonant orbits follow the underlying pattern of their parent orbits. Around the OLR, the orbit orientation flip translates into a velocity flip in the Solar neighbourhood (in the cartoon at an angle of $\phi_\odot=25~\text{deg}$ behind the bar).}
    \label{fig:orbit_orientation}
\end{figure}

\subsection{The orbit orientation flip at the OLR} \label{sec:orbit_orientation_flip_theory}

\emph{Orbit orientation flips in the `time-averaged' action-space.}---It is well known that, in the $(x_\text{CRF},y_\text{CRF})$ frame co-rotating with the bar, the orientation of orbits changes its direction at the principle resonances (see, e.g., \citealt{1989A&ARv...1..261C,1993RPPh...56..173S}). At the OLR, the orbit orientation flips from anti-aligned inside of the OLR (the $x_1(2)$ orbit family) to being aligned with the bar outside of the OLR (the $x_1(1)$ orbit family). This was first discussed by \citet{1976ApJ...209...53S} for gas particle orbits, and also by \citet{1991dodg.conf..323K}. \citet{2000AJ....119..800D} illustrates this for stellar orbits.\footnote{See, e.g., fig. 8 in \citet{2000AJ....119..800D}, fig. 4 and 6 in \citet{2001A&A...373..511F}, and fig. 5 in \citet{2019MNRAS.488.3324F} for an illustration of resonant orbit types.}

We found that these orbit orientation flips can be illustrated especially well in the action plane $(L_{z,\text{mean}},J_{R,\text{mean}})$ of oscillation midpoints. The oscillation midpoints are quantities that we found as the time-average from integrating (part of) the whole orbit---in some sense they are therefore better ``integrals of motion'' or orbit labels than the instantaneous $L_z$ and $J_R$, because the variation due to libration and radial oscillation is averaged out. From Figures \ref{fig:example_orbits_T50}-\ref{fig:example_orbits_parents}, we see that the oscillation midpoints of librating stars are close to their parent orbits in action space. In Figure \ref{fig:orbit_orientation}, we plot for the \texttt{Fiducial} simulation in this plane the orbits' orientation and elongation. This shows that the orbit orientation flip occurs cleanly along the OLR ARL in action space.

\citet{2019MNRAS.488.3324F}, using a self-consistent $N$-body simulation, demonstrated that in the region inside of the OLR ($L_z = R \times v_T < \sim L_{z,\text{OLR}}$) the $x_1(2)$ orbit family overlaps with highly librating orbits from the $x_1(1)$ family. When comparing Figure \ref{fig:orbit_orientation} with the oscillation amplitude in Figure \ref{fig:oscillation_amplitudes}(a), we confirm this finding by \citet{2019MNRAS.488.3324F}: In the $(L_z,J_R)$ plane, the aligned OLR orbits ($q > 0$) librate strongly around the OLR ARL, while the anti-aligned orbits ($q < 0$) inside the ARL oscillate much less, leading to this orbit overlap inside the OLR ARL.

\emph{The OLR in velocity space.}---A well-known consequence of the orbit orientation flip around the OLR is that at the Solar azimuth stellar radial velocities switch from outward-moving to inward-moving (see \citealt{2000AJ....119..800D,2001A&A...373..511F,2003A&A...401..975M,2019MNRAS.488.3324F}, and the cartoon insert in Figure \ref{fig:orbit_orientation}). The bimodality of the pre-\emph{Gaia} $(U,V)$ velocity plane---the outward-moving Hercules stream around $(U,V) \sim (-30,-50)~\text{km/s}$ and the inward-moving Horn feature at $(U,V)\sim (50,-20)~\text{km/s}$ as, e.g., in fig. 22 of \citet{2018A&A...616A..11G}---has therefore been classically explained by the \emph{short fast bar's} OLR.

\begin{figure*}
    \centering
    \subfigure[The velocity plane of a small volume centered on $R_\text{OLR}$. \label{fig:OLR_UV_plane}]{\includegraphics[width=\columnwidth]{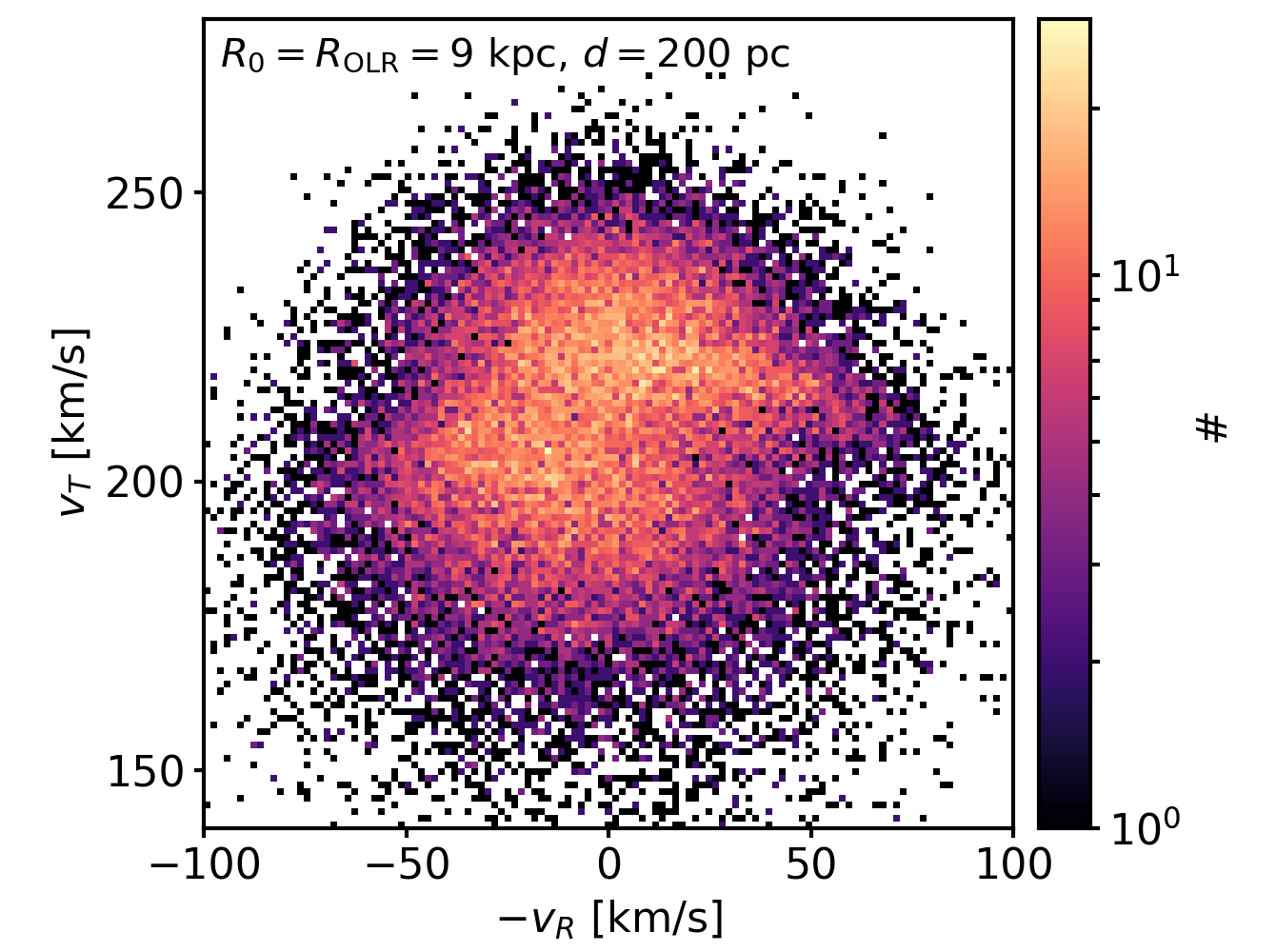}} %
    \subfigure[Action space colour-coded by the fraction of inward moving stars. \label{fig:OLR_LzJR_plane}]{\includegraphics[width=\columnwidth]{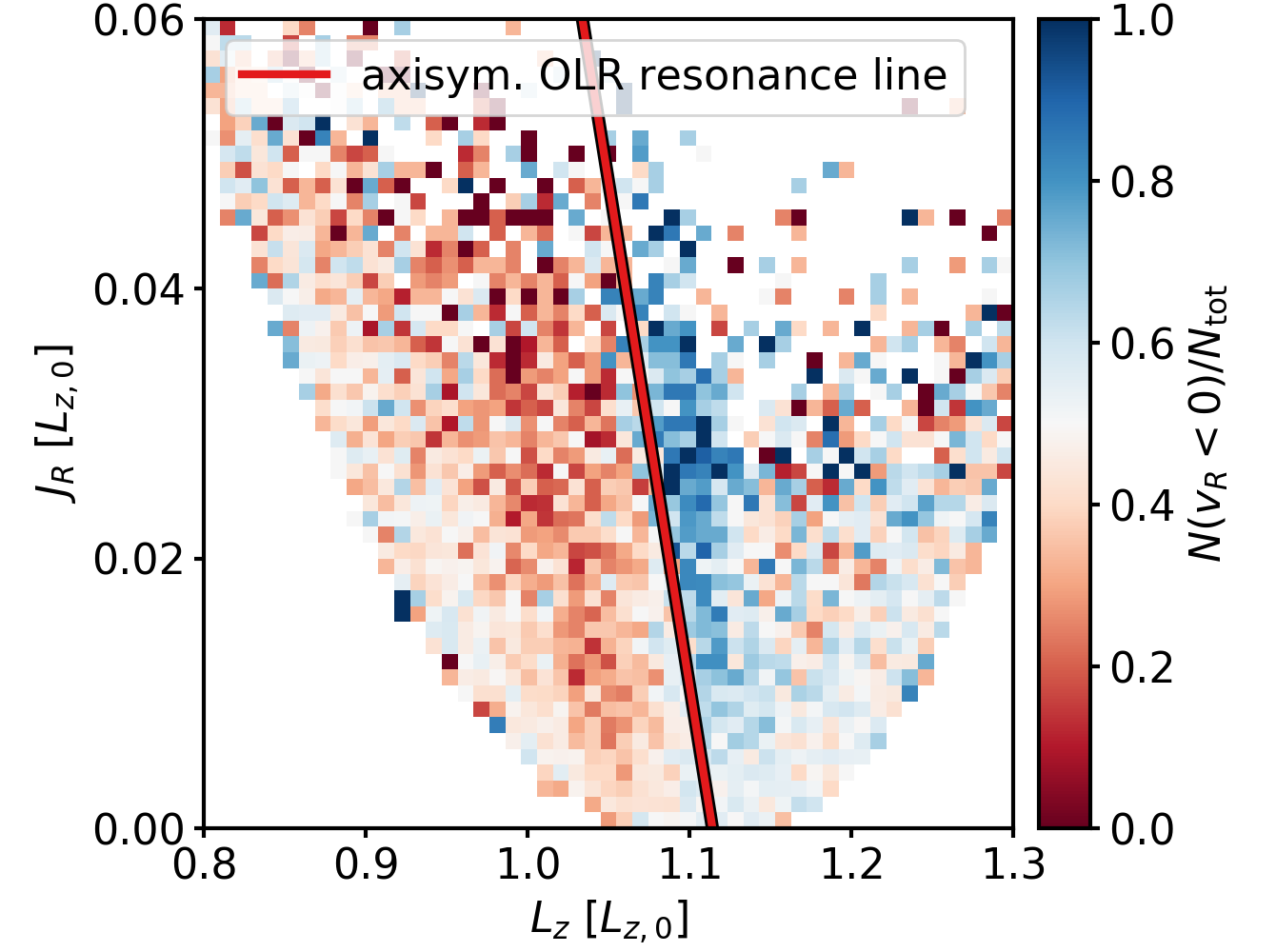}} %
    \caption{The outward/inward signature of the OLR in the \texttt{Fiducial} simulation. Panel \ref{fig:OLR_UV_plane} shows the classic velocity plane for a small volume ($d < 200~\text{pc}$) centered around $(R_\text{OLR},\phi_\text{bar}-\phi,z)=(9~\text{kpc},25~\text{deg},0)$. Panel \ref{fig:OLR_LzJR_plane} shows the corresponding action plane (note the parabolic envelope due to the small volume), colour-coded by the fraction of inward-moving stars, analogous to Figures \ref{fig:Gaia_actions_Solar_neighbourhood}-\ref{fig:Fiducial_bar_model_actions}. Overplotted is the axisymmetric OLR ARL(with the 
   ``Hercules''-analogue at $v_T \lesssim 210~\text{km/s}$ and the ``Horn''-analogue at $(v_T,-v_R) \sim (220,50)~\text{km/s}$), and in the action plane as a `red/blue' feature cleanly separated by the OLR ARL.}
    \label{fig:the_OLR_feature}
\end{figure*}

The $(v_R,v_T)\sim(-U,V+v_\text{circ})$ velocity plane that we show in Figure \ref{fig:OLR_UV_plane} stacks all snapshots of our \texttt{Fiducial} test particle simulation for which the bar was oriented at an angle of $25$ deg for stars located within 200 pc of $\vect{x}_\odot=(9~\text{kpc},0,0)$. This centers the survey volume on the OLR radius (see Table \ref{tab:bar_models}).

In velocity space, the orbit structure is more complicated than the simple `inside OLR $\longrightarrow$ anti-aligned $x_1(2)$ orbits $\longrightarrow$ outward-moving'. The `Hercules'-like OLR signature can contain also orbits from the $x_1(1)$: inward-moving parent orbits \citep{2000AJ....119..800D} and highly librating orbits exhibiting outward-movements at the Sun \citep{2019MNRAS.488.3324F}.

\emph{The OLR in 4D phase-space.}---The reason for the observed orbit overlap is the following: In the full 4D in-plane phase-space---$(J_R,L_z)$ together with their canonical conjugate angle coordinates $(\theta_R,\theta_\phi)$, or in the position-velocity space $(\vect{x},\vect{v})$---the two OLR orbit families are actually clearly distinct from each other. In 2D projection, it is only the angle space $(\theta_\phi,\theta_R)$ that reveals that the families do not overlap: $x_1(1)$ orbits have their apo- and pericenters at $\theta_\phi - \phi_\text{bar} = [0,\pm\pi]$ and $[\pm \pi/2]$, respectively; for the $x_1(2)$ orbits the opposite is true. In other 2D projections, the axisymmetric $(L_z,J_R)$, or velocity space as discussed above, or also in spatial positions, the two orbit families overlap in some regions. By adding a third dimension to action space, we can mitigate this.

\emph{The OLR in action-$v_R$ space.}---The same stars within 200 pc that were shown in velocity space in Figure \ref{fig:OLR_UV_plane}, are in Figure \ref{fig:OLR_LzJR_plane} shown in the action plane, colour-coded by the relative number of outward (red) and inward (blue) moving stars. As expected from Figure \ref{fig:orbit_orientation}, the OLR ARL also cleanly separates the red from the blue feature. (We colour-code action space by the stellar-number asymmetry in $\text{sign}(v_R)$ rather than $\langle v_R \rangle$ (as done in similar studies) to make the `red/blue' feature visible down to $J_R\longrightarrow0$ where $\langle v_R \rangle\longrightarrow0$.)

To summarize, the OLR signature at the Solar azimuth consists in the local (Figure \ref{fig:OLR_LzJR_plane}) and extended (Figure \ref{fig:Fiducial_bar_model_actions}) $(L_z,J_R)$ action plane of:
\begin{enumerate}[leftmargin=*,topsep=0ex,label=(\roman*)]
\item a tendency between CR and OLR to be outward-moving/red,
\item an outward-moving/red underdensity stripe to the low-$L_z$ side of the ARL, and
\item a sharp, inward-moving/blue, high-$J_R$ scattering ridge to the high-$L_z$ side of the ARL. The ridge is offset from the OLR ARL (e.g. \citet{2010MNRAS.409..145S,2019MNRAS.490.1026H}; see also Figure \ref{fig:Fiducial_bar_model_actions}).
\end{enumerate}
This signature is well-studied in different coordinate spaces, where the features correspond...
\begin{enumerate}[leftmargin=*,topsep=0ex,label=...,itemindent=*]
    \item locally, in the velocity plane, to (i) an extended ``Hercules''-like feature, (ii) an arch-shaped gap, (iii) a narrow ``Horn''-like feature (Figure \ref{fig:OLR_UV_plane}; see also, e.g., \citealt{2000AJ....119..800D,2001A&A...373..511F,2019MNRAS.488.3324F}).
    \item globally in the Galactic disk, as a function of Galactocentric $(R,\phi)$, to a wide outward-/inward-moving wiggle in $v_R$ at the same location as an underdensity/overdensity wiggle in stellar numbers (Figure \ref{fig:Fiducial_model_intro}(d); see also, e.g. \citealt{2003A&A...401..975M,2020ApJ...899L..14H}).
    \item globally, in the $(R,v_T)$ plane, to (i)-(ii) an outward-moving region and gap towards low $R$ and $v_T$ and (iii) the prominent, arch-like, inward-moving scattering ridge roughly near a line of constant $L_{z,\text{OLR}}$ (e.g., \citealt{2019MNRAS.488.3324F,2020MNRAS.494.5936F,2019MNRAS.490.1026H}). $J_R$ increases both towards higher and lower $v_T$, as well as towards smaller $R$ across this ridge.
    \item globally, in the $(L_z,\phi)$ plane, to thinner parallel (ii) red/underdense and (iii) blue/overdense stripes at constant $L_z$ (Figure \ref{fig:OLR_time_evolution} in Appendix \ref{app:OLR_time_evolution}; see also, e.g., \citealt{2019A&A...632A.107M,2019arXiv191204304C}).
\end{enumerate}

\section{Results}  \label{sec:results}

\begin{figure*}
    \centering
    \includegraphics[width=0.8\textwidth]{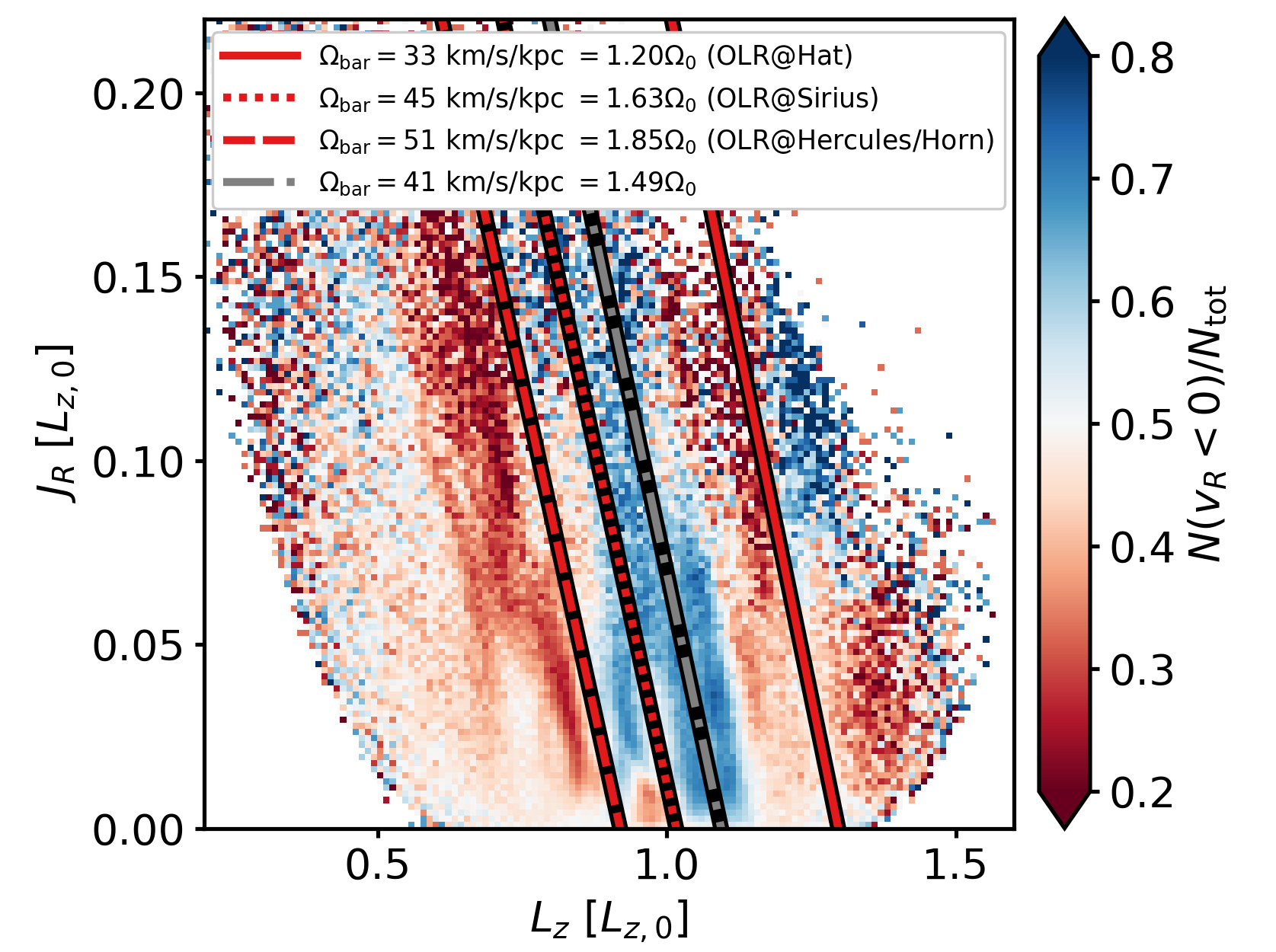}
    \caption{All candidates for the characteristic outward/inward (`red/blue') signature of the MW's bar OLR within 3 kpc from the Sun in \emph{Gaia} DR2 RVS. We show the $v_R$-asymmetry as a function of the axisymmetric actions estimated in the \texttt{MWPotential2014} ($\Omega_0 = 27.5~\text{km/s/kpc}$). The \emph{Gaia} action space reveals three prominent `red/blue' features at slopes similar to that of ARLs. We position (the red) OLR lines such that each separates an outward- from an inward-moving stripe, as observed in the simulation (see Figure \ref{fig:the_OLR_feature}). These OLR locations correspond to specific pattern speeds, thus allowing us to read off possible pattern speeds for the bar directly from the data. The corresponding pattern speeds, as well as the moving groups of the local Solar neighbourhood that could be related to this OLR, are indicated in the legend. In addition, we overplot (in grey) the OLR line for a pattern speed of $1.49\Omega_0=41~\text{km/s/kpc}$ \citep{2019MNRAS.488.4552S,2019MNRAS.490.4740B}, which does not separate a red from a blue feature (see also Figure \ref{fig:Hunt_actions} and Section \ref{sec:discussion_literature_comparison}).}
    \label{fig:Gaia_actions_and_OLR_candidates}
\end{figure*}

\begin{table*}
    \centering
    \begin{tabular}{c|ccc|c|cc|c|l}
         Pattern speed & \multicolumn{3}{c|}{Pattern speed} & OLR radius & \multicolumn{2}{c|}{Bar length} & Bar strength & Notes\\
          model name & derived from OLR-  & $\Omega_\text{bar}$  & $\Omega_\text{bar}$ & $R_\text{OLR}$ & $R_\text{bar}$ & $R_\text{CR}/R_\text{bar}$ & $\alpha_{m=2}$\\
          & like signature at & [km/s/kpc] & [$\Omega_0$] & [kpc] & [kpc] & & \\\hline
          \texttt{Fiducial}  & - & 40 & 1.45 & 9.0 & 4.5 & 1.25 & 0.01 & used only in the generic\\
          &&&&&&&&investigation of the bar's effect \\
          &&&&&&&&on the axisym. action space \\
          &&&&&&&&in Sections \ref{sec:method}-\ref{sec:results} \\\hline
         \texttt{Hercules} &  outward Hercules/ & 51 & 1.85 & 7.3 & 3.5 & 1.26 & 0.01 & c.f. \emph{short fast bar} in\\
         & inward Horn, & $52\dagger$ & $1.85\dagger$ & $7.5\dagger$ &&&& \citet{2000AJ....119..800D},\\
         & $L_z\sim0.9 L_{z,0}$&&&&&&& \citet{2014AA...563A..60A}\\
         &&&&&&&&\\
         \texttt{Sirius} & outward Hyades/ & 45 & 1.63 & 8.2 & 4 & 1.26 & 0.015 & - \\
         & inward Sirius, & $48\dagger$ & $1.70\dagger$ & $8.1\dagger$ &&&\\
         & $L_z\sim 1.0 L_{z,0}$ &&&&&&&\\
         &&&&&&&&\\
         \texttt{Hat} & outward/inward & 33 & 1.20 & 10.7 & 5 & 1.35 & 0.015 & c.f. \emph{long slow bar} in\\
          & at the Hat, & $36\dagger$ & $1.27\dagger$ & $10.5\dagger$ &&&& \citet{2017ApJ...840L...2P},\\
         & $L_z \sim 1.3 L_{z,0}$ &&&&&&& \citet{2019AA...626A..41M}\\\hline
         \texttt{S19B19} & - & $41\pm3$ & $1.49\pm0.1$ & $8.8\pm0.6$ & - & - & - & \emph{slightly faster slow bar}\\
         &&& $1.45\pm0.1\dagger$ & $9.3\pm0.7\dagger$&&&&with pattern speed taken from\\
         &&&&&&&& \citet{2019MNRAS.488.4552S},\\
         &&&&&&&& \citet{2019MNRAS.490.4740B}.\\
         &&&&&&&&Hercules/Horn related to the\\
         &&&&&&&&1:4 OLR \citep{2018MNRAS.477.3945H}.\\
         \end{tabular}
    \caption{Overview of the bar pattern speeds considered in this work. The pattern speeds of the \texttt{Hercules}, \texttt{Hat}, and \texttt{Sirius} bar models were derived from the \emph{Gaia} DR2 data (see text for details) for the \texttt{MWPotential2014} (\citealt{2015ApJS..216...29B}, with $\Omega_0= 27.5~\text{km/s/kpc}$). We have run test particle simulations for these pattern speeds and have assumed a bar orientation of $\phi_\text{bar}= 25~\text{deg}$ with respect to the Solar azimuth for a pure quadrupole bar model. The bar length was chosen to satisfy $R_\text{CR}/R_\text{bar}\sim 1.2$ \citep{1980A&A....81..198C,1992MNRAS.259..345A} and the bar strengths $\alpha_{m=2}$ (see Appendix \ref{app:method_bar}) were chosen to create OLR signatures of roughly similar strength. The pattern speeds marked with $\dagger$ were derived analogously from the \emph{Gaia} DR2 data, but for actions and frequencies estimated in the MW potential by \citet{2019ApJ...871..120E} with $\Omega_0=28.3~\text{km/s/kpc}$. The \texttt{S19B19} pattern speed is included as a comparison with another bar model in the literature.}
    \label{tab:bar_models}
\end{table*}

\subsection{Measuring the bar's pattern speed using the outward/inward feature of the OLR} \label{sec:OLR_signature_and_pattern_speed}

The outward/inward velocity flip created by the bar's OLR has often been used to identify the location of this resonance. Comparing Figure \ref{fig:OLR_LzJR_plane} ($<200~\text{pc})$ with Figure \ref{fig:Fiducial_bar_model_actions} ($<4~\text{kpc})$, shows that action space conserves the alignment of the OLR's outward/inward feature around the ARL when going from the local to the extended Solar neighbourhood. The \emph{Gaia} data allow therefore to search for the OLR beyond the local velocities. The actions in particular enable us for the first time to show all OLR candidates in one, clean overview plot in Figure \ref{fig:Gaia_actions_and_OLR_candidates}.

Every pair of outward/inward moving stripes separated by a line of the same slope as an ARL is a candidate for the signature of the bar OLR. Each OLR candidate corresponds to a specific $\Omega_\text{bar}$. In the \emph{Gaia} DR2 RVS actions, we count three prominent `red/blue' features (see Figures \ref{fig:Gaia_actions_Solar_neighbourhood} and fig. 7 in \citetalias{2019MNRAS.484.3291T}). A decrease in $\Omega_\text{bar}$ shifts the OLR ARL toward larger $L_z$ and makes it steeper. We read off the value for $\Omega_\text{bar}$ whenever the OLR ARL separates a red stripe on the left from a blue stripe on the right, as illustrated in Figure \ref{fig:Gaia_actions_and_OLR_candidates}. These three direct $\Omega_\text{bar}$ measurements assume the \texttt{MWPotential2014} potential model, and are $\Omega_\text{bar}\sim1.85~\Omega_0$, $\Omega_\text{bar}\sim1.2~\Omega_0$, and $\Omega_\text{bar}\sim1.63~\Omega_0$.

We also implemented the more recent MW potential model by \citet{2019ApJ...871..120E} (with $R_0 \equiv 8.122~\text{kpc}$, $v_0 \equiv 229.8~\text{km/s}$, $L_{z,0}\sim1866~\text{kpc km/s}$, and $\Omega_0\sim28.3~\text{km/s/kpc}$ in our implementation), re-calculated actions and frequencies, and applied the same strategy to measure pattern speeds at these OLR candidates of $\Omega_\text{bar}\sim1.85~\Omega_0$, $\Omega_\text{bar}\sim1.27\Omega_0$, and $\Omega_\text{bar}\sim1.70\Omega_0$, respectively. 

Only one of these three candidates can be the bar's true OLR and pattern speed. We summarize all measurements in Table \ref{tab:bar_models}.

The \texttt{Hercules} pattern speed $1.85\Omega_0$ is derived from the strongest `red/blue' feature in the data, the transition from the Hercules to the Horn moving groups (see \citetalias{2019MNRAS.484.3291T} for the location of the moving groups in action space). Both potentials give the same result in units of $\Omega_0$, owing to this assumed OLR being close to $L_z/L_{z,0}=1$. With the fixed assumption for the \citet{2010MNRAS.403.1829S} Solar motion, the measurement of this $\Omega_\text{bar}/\Omega_0$ is therefore only weakly dependent on the shape of the rotation curve and quite robust.

The \texttt{Hat} pattern speed $1.2\Omega_0$ is derived from the second strongest `red/blue' feature in the \emph{Gaia} DR2 action space which is located at high $J_R$ around $L_z/L_{z,0}\sim1.3$ and continues---albeit much weaker---down to $J_R=0$. This feature projects to the Hat moving group in the local velocities (at $V\sim40~\text{km/s}$ in \citealt{2018A&A...616A..11G}). The measurements for the two potentials differ by almost $0.1\Omega_0$, indicating that for this pattern speed, derived from an OLR candidate further away from the Sun, the shape of the rotation curve does matter.

The \texttt{Sirius} pattern speed $1.63\Omega_0$ is derived from a third `red/blue' feature, in the \emph{Gaia} action data close to $J_R\sim0$ and $L_z/L_{z,0}\sim1$. In the projection to local velocity space, this corresponds to the transition from the outward-moving Hyades to the inward-moving Sirius group. This OLR candidate feature continues above the Horn, around the ARL at $J_R\sim0.1L_{z,0}$. In Section \ref{sec:discussion_Gaia_vs_sim}, we discuss these pattern speeds in detail and compare to the literature in Section \ref{sec:discussion_literature_comparison}.

\subsection{The imprint in the vertical action due to the OLR} \label{sec:results_vertical_action}

\begin{figure*}
\centering
    \subfigure[Angular momentum histogram of stars in the axisymmetric (orange) and perturbed (pink) disc. \label{fig:Lz_count}]{
        \includegraphics[height=0.265\textwidth,trim=15 0 20 0, clip]{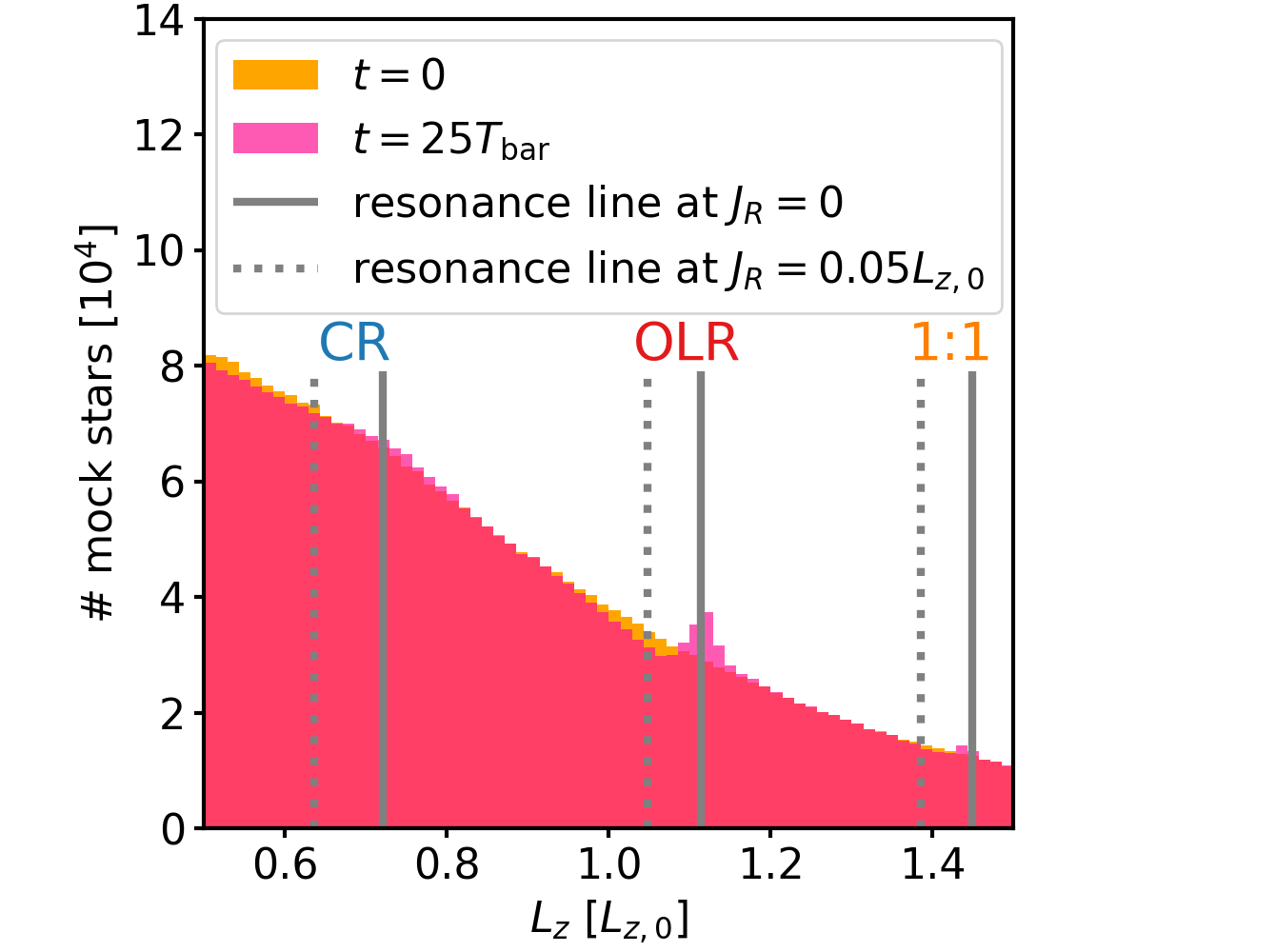}}\hfill
    \subfigure[Mean vertical action for the axisymmetric (green) and perturbed (purple) disc. \label{fig:Lz_Jz}]{
        \includegraphics[height=0.265\textwidth,trim=15 0 20 0, clip]{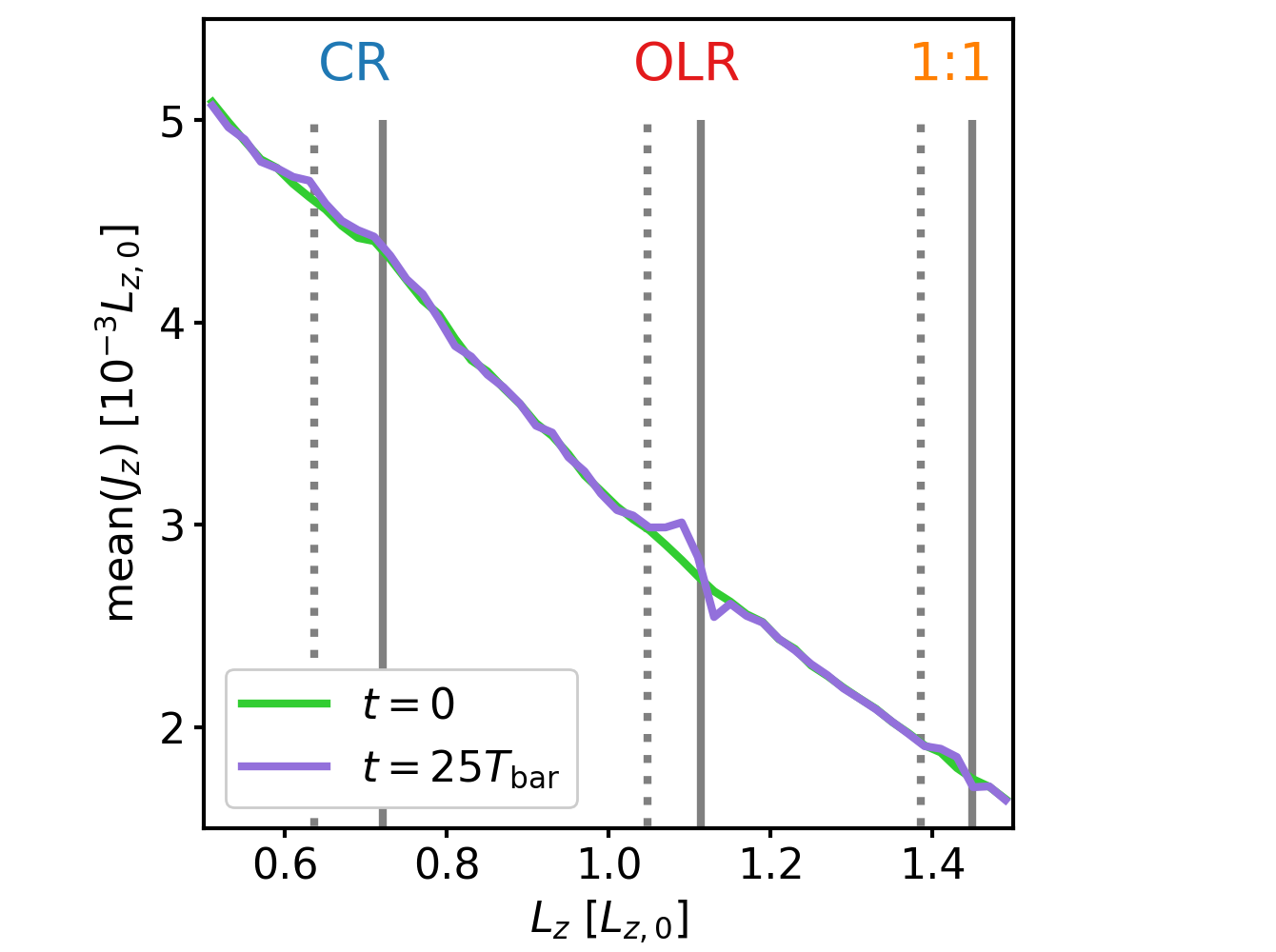}}\hfill
    \subfigure[Mean radial velocity and mean absolute vertical velocity for the `Solar' neighbourhood. \label{fig:vertical_radial_velocities}]{
        \includegraphics[height=0.265\textwidth,trim=10 0 80 0, clip]{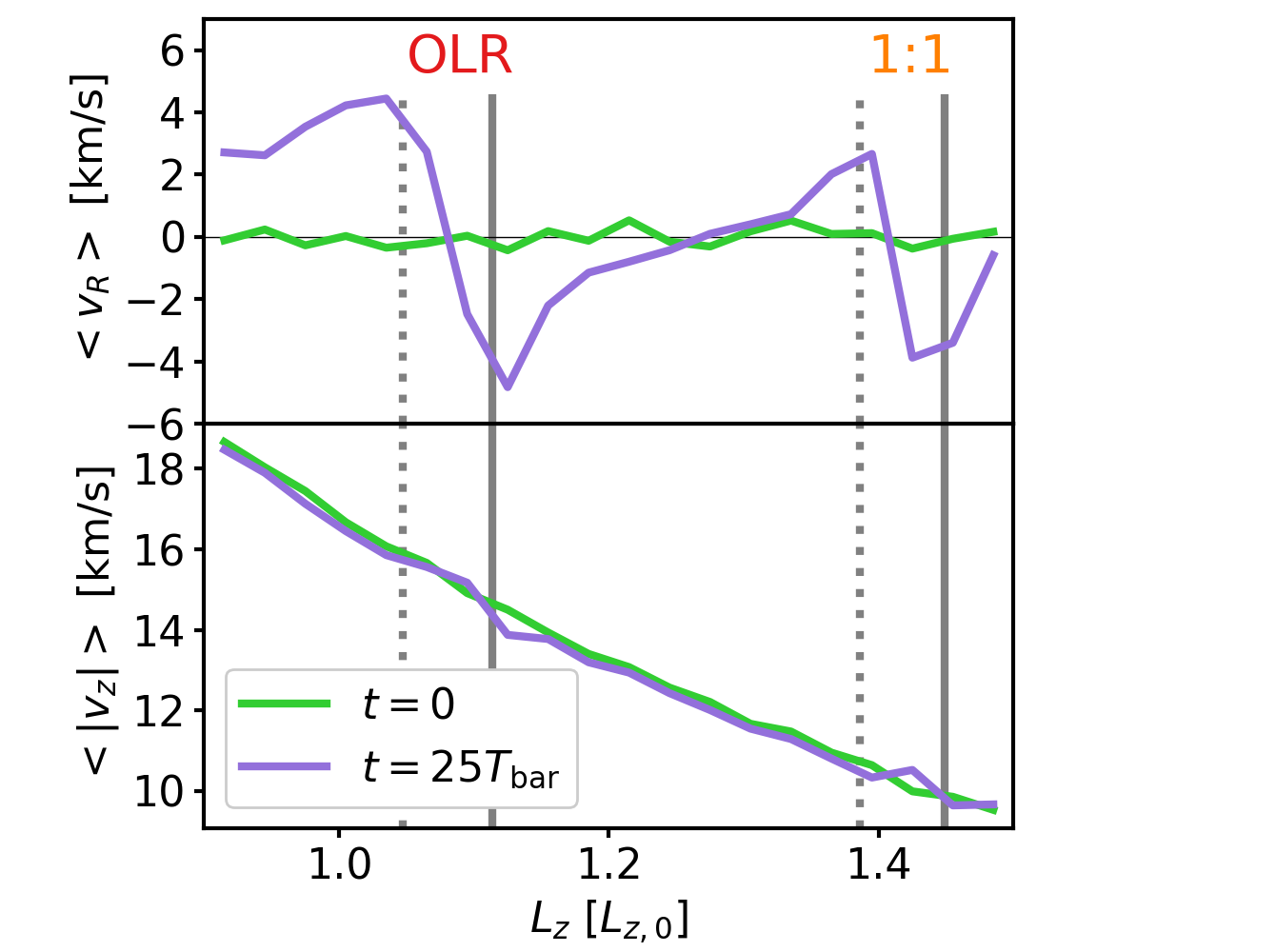}}
    \caption{The vertical action signatures as a function of $L_z$ only. All panels show the resonance signature in the \texttt{Fiducial} bar simulation (at $t=25T_\text{bar}$ within $|z|<500~\text{pc}$) in star counts (Panel \ref{fig:Lz_count}) and mean vertical action $\langle J_z\rangle$ (Panel \ref{fig:Lz_Jz}), comparing them to the axisymmetric disc at $t=0$. The grey vertical lines denote $L_z(J_R\mid\text{ARL})$ evaluated at $J_R=(0,0.05)L_{z,0}$ for CR, OLR, and the 1:1 resonance. Panel \ref{fig:vertical_radial_velocities} shows the mean radial and absolute vertical velocities, $\langle v_R \rangle$ and $\langle |v_z| \rangle$ (for the `Solar' neighbourhood, $R_\odot = 8~\text{kpc}, d < 4~\text{kpc}$), demonstrating that bar resonances could contribute to coupled radial and vertical velocity waves in the Galactic disc. (Note that for this effect still $\langle v_z \rangle = \langle z \rangle = 0$.)}
    \label{fig:Jz_signature_Lz}
\end{figure*}

\emph{$\langle J_z \rangle$ gradient around the OLR.}---In the \emph{Gaia} data, the ridges in $(L_z,J_R)$ are related to signatures in mean vertical action $J_z$ (Figure \ref{fig:Gaia_actions_Solar_neighbourhood}). In Section \ref{sec:method_fiducial_solar_neighbourhood}, we found that our test particle simulations exhibit signatures in mean $J_z$ around the OLR and 1:1 resonance (Figures \ref{fig:Fiducial_bar_model_actions} and \ref{fig:comparison_all_models_data}). This becomes especially obvious in Figures \ref{fig:Lz_count} and \ref{fig:Lz_Jz}, where we show number counts and average $J_z$ for the \texttt{Fiducial} model as a function of $L_z$ only. 

The epicyclic approximation for near-circular disk orbits assumes that vertical and radial motions are decoupled. The in-plane bar resonance should therefore not change the $J_z$ of individual disc stars. And indeed, in the \texttt{Fiducial} model, the average relative change in $J_z$ is only 0.6 percent, as compared to 70 percent in $J_R$ (where $\sigma_R > \sigma_z$). The observed $J_z$-signature at the OLR therefore has to be a \emph{cumulative} effect in the stellar distribution induced by the bar.

\begin{figure*}
    \centering
    \subfigure[Oscillation midpoints for all mock stars, colour-coded by $J_z$. \label{fig:Jmid_Jz_all}]{
        \includegraphics[height=0.265\textwidth,trim=15 0 20 0, clip]{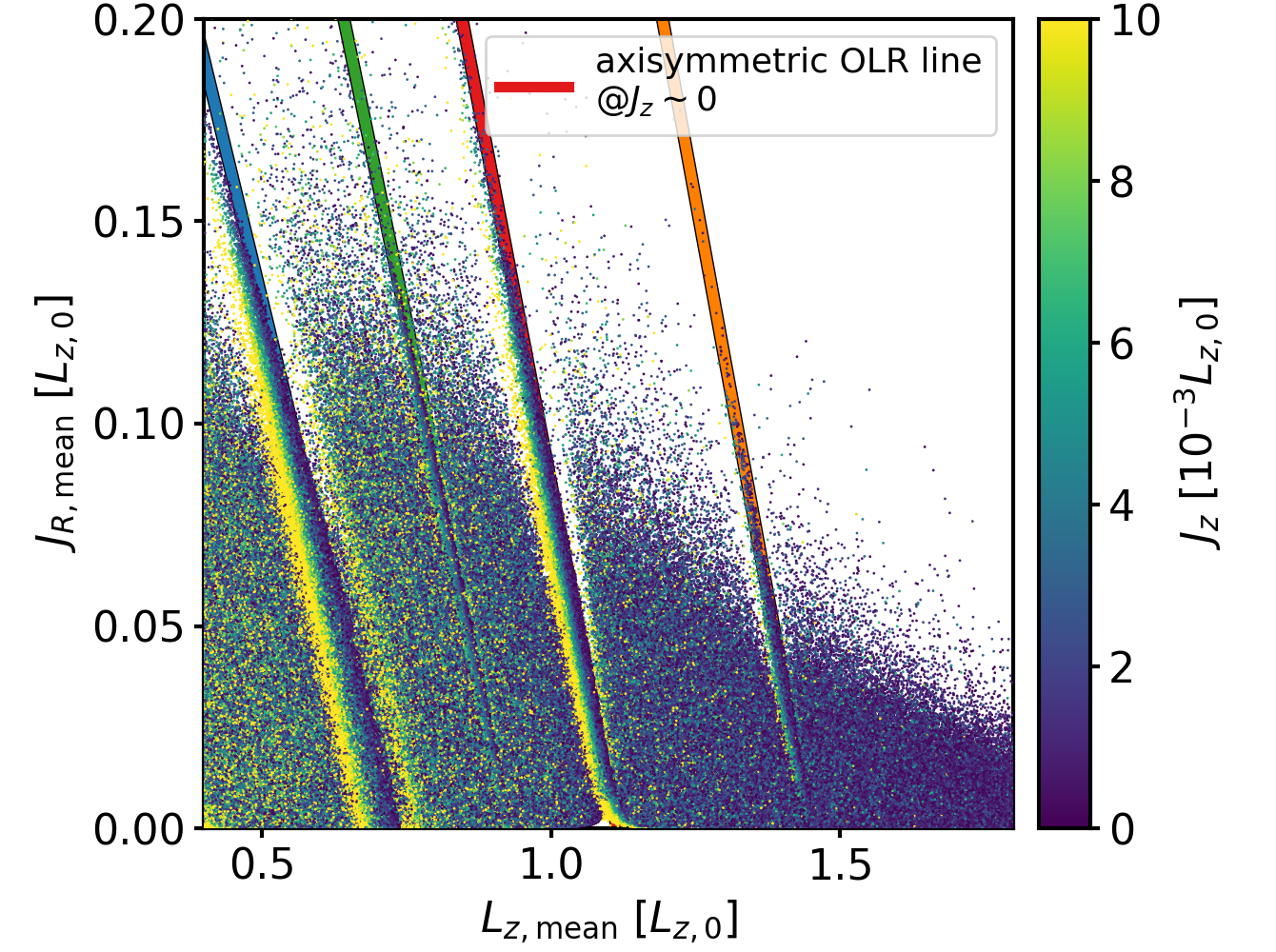}}\hfill
\subfigure[Comparing the mid-location of resonant OLR stars with the ARLs for the corresponding $J_z$. \label{fig:midJ_Jz_OLR}]{
         \includegraphics[height=0.265\textwidth,trim=15 0 20 0, clip]{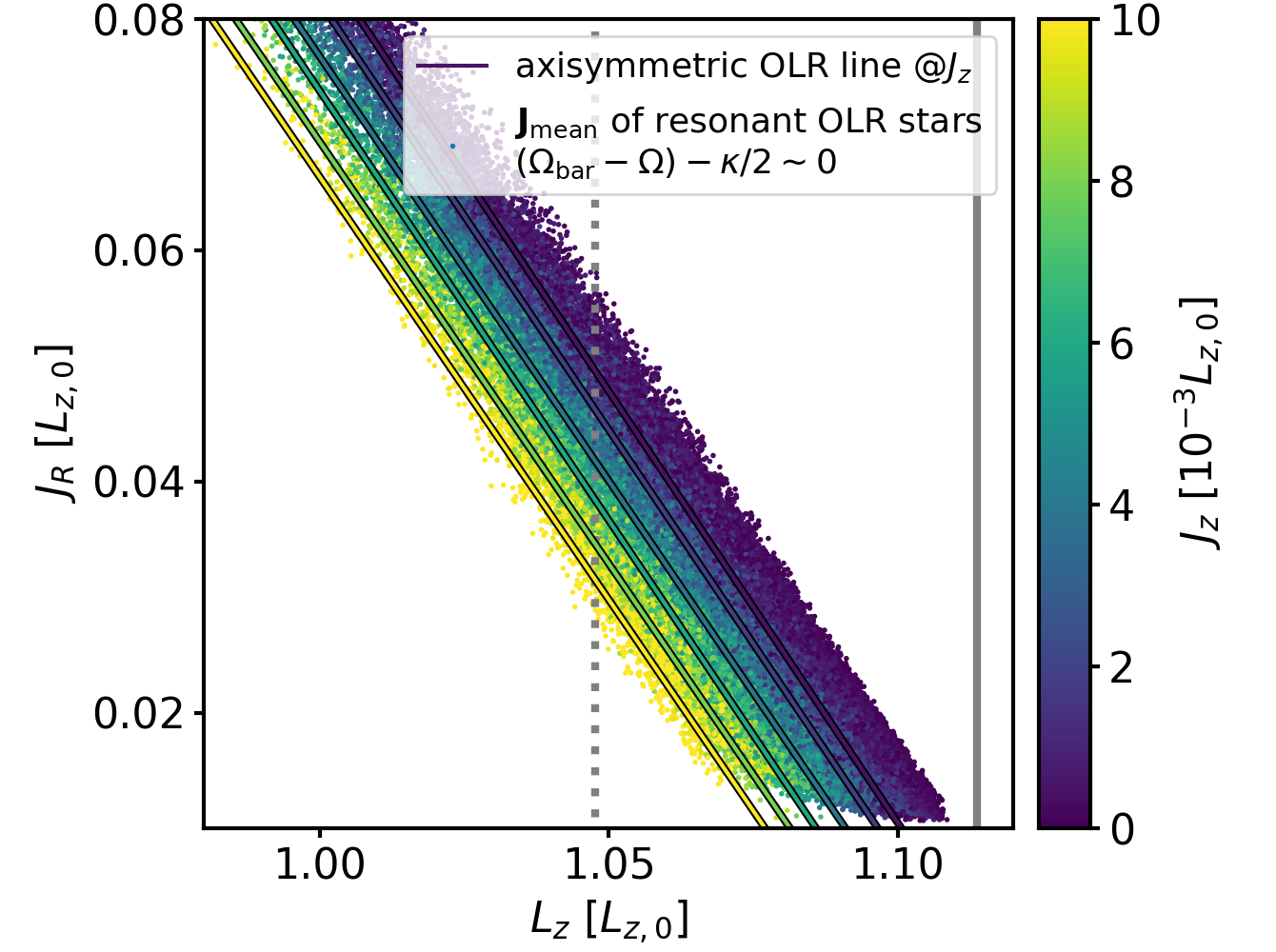}}\hfill
    \subfigure[Oscillation amplitudes (see Equation \eqref{eq:oscillation_amplitude} in Appendix \ref{app:oscillation}) of CR and OLR stars are independent of $J_z$. \label{fig:deltaLzJR_Jz}]{
        \includegraphics[height=0.265\textwidth,trim=10 0 80 0, clip]{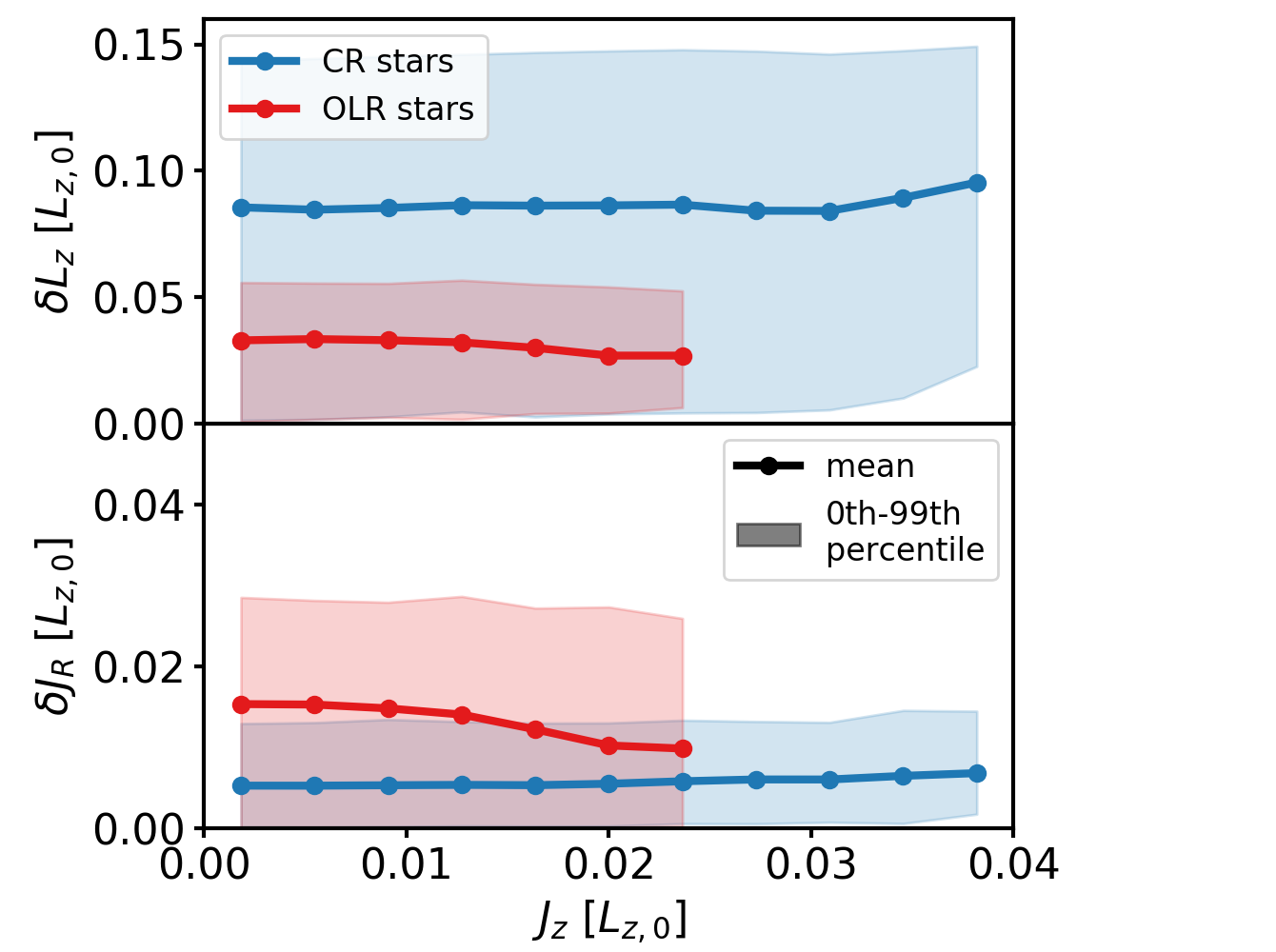}}
    \caption{The underlying reason for the vertical action $J_z$ feature at the resonances in the \texttt{Fiducial} simulation shown in Figures \ref{fig:Fiducial_bar_model_actions} and \ref{fig:Jz_signature_Lz} is (i) that the ARL's location depends on $J_z$ (lines in Panel \ref{fig:midJ_Jz_OLR}), and (ii) the real resonant stars have their orbit midpoints close to the ARL with the same $J_z$ (points in Panel \ref{fig:midJ_Jz_OLR}). This leads to the $J_z$-sorting of (the orbit midpoints of) stars at all resonances (Panel \ref{fig:Jmid_Jz_all}). The $J_z$-independence of the oscillation around the ARLs (Panel \ref{fig:deltaLzJR_Jz}) conserves the $J_z$ ordering during oscillation.}
    \label{fig:my_label}
\end{figure*}

Figure \ref{fig:Jmid_Jz_all} shows again the $\vect{J}_\text{mean}$ distribution from Figure \ref{fig:LzJR_mid}, but colour-coded by the stars' $J_z$. We find that the exact $\vect{J}_\text{mean}$ location at all the resonances shifts with the value of $J_z$: The resonances appear to \emph{sort} stars according to their $J_z$.

This sorting is the consequence of the combination of two different properties:

(i) In axisymmetric Galaxy potentials the ARL depends on $J_z$. 

(ii) In a barred Galaxy potential, a resonant star oscillates around the ARL with the same $J_z$.

\emph{The $J_z$-dependence of the ARL}.---Property (i) follows from the in-plane orbital frequencies being dependent on $J_z$ in galaxy-like potentials---also in the case of axisymmetry. In Figure \ref{fig:midJ_Jz_OLR}, we show the OLR ARL not only for $J_z=0$ as usual, but also for different $J_z > 0$. The lines shift towards smaller $(L_z,J_R)$ with increasing $J_z$. The dependence of $\Omega_{i,\text{axi}}$ on $(L_z,J_R,J_z)$ depends on the exact form of the (axisymmetric) galaxy potential. A general, analytic derivation of this property is non-trivial and beyond the scope of this work. Numerical experiments can however provide some intuition. In an axisymmetric potential, the circular and epicycle frequencies
\begin{eqnarray}
\Omega(R_g\mid\Phi_\text{axi}) &\equiv& \left(\frac1R \frac{\partial \Phi_\text{axi}}{\partial R}\right)^{1/2}_{(R=R_g,z=0)},\\
\kappa(R_g\mid\Phi_\text{axi}) &\equiv& \left(\frac{\partial^2 \Phi_\text{axi}}{\partial R^2} + \frac3R \frac{\partial \Phi_\text{axi}}{\partial R}\right)^{1/2}_{(R=R_g,z=0)}
\end{eqnarray}
(from eq. (3.79) in \citealt{2008gady.book.....B}) can be considered as a property of the potential. For a near-circular orbit in the epicycle approximation, these frequencies evaluated at its $R_g$ are the real orbital frequencies. It can be shown that for an orbit with $J_R \gg 0$ and $J_z \gg 0$ integrated in an axisymmetric potential these $\Omega$ and $\kappa$ are closer to the real\footnote{We have explicitly checked in our simulation that for orbits integrated in the axisymmetric potential, the action frequencies agree with the real frequencies derived from a Fourier-analysis of the orbit, i.e., $\Omega_{\phi,\text{axi}}\equiv\partial \mathscr{H}_\text{axi}/\partial J_\phi = \Omega_{\phi,\text{true}}$ and $\Omega_{R,\text{axi}} \equiv \partial \mathscr{H}_\text{axi} / \partial J_R = \Omega_{R,\text{true}}$.} orbital frequencies $\Omega_R$ and $\Omega_\phi$ when evaluated at the time-averaged radial coordinate $\langle R(t)\rangle_t$ of the orbit rather than at $R_g$. Only in the limit $J_R\longrightarrow 0, J_z\longrightarrow0$ also $\langle R(t)\rangle_t \longrightarrow R_g$. In general, an orbit has a larger $\langle R(t) \rangle_t$ if any of the three actions $(J_R,L_z,J_z)$ is larger. We therefore expect anti-correlations between the actions with the frequencies that satisfy the resonance condition. We observed the $J_z$-dependence of the ARL also for the St\"{a}ckel potential \texttt{KKS-Pot} previously used in \citet{2017ApJ...839...61T}. So even in separable potentials---where the momentum $p_i(x_i)$ is a function of $x_i$ only (with the prolate confocal coordinates $x_i \in [\lambda,\nu,\phi]$) and the actions are $J_i \propto \int_{x_{i,\text{min}}}^{x_{i,\text{max}}} p_i(x_i) \diff x_i$---the frequencies $\Omega_{i,\text{axi}}(\vect{J}) = \partial \mathscr{H}_\text{axi}(\vect{J})/ \partial J_i$ are not independent of $J_z \equiv J_\nu$.

Property (ii) is illustrated in Figure \ref{fig:midJ_Jz_OLR}, where we show in addition the oscillation midpoints for the true resonant OLR stars, demonstrating that the stars oscillate around (or at least close to) their actual, \emph{$J_z$-dependent} ARL, causing therefore the gradual ``sorting'' of the stars by $J_z$ at the resonance.

\emph{The $J_z$-sorting is best visible at the OLR.---}Figure \ref{fig:deltaLzJR_Jz} shows that the oscillation amplitudes of resonant OLR and CR stars are independent of $J_z$. The sorting by $J_z$ of the oscillation midpoints (Figure \ref{fig:Jmid_Jz_all}) remains therefore also visible in the phase-mixed distribution (Figure \ref{fig:Lz_Jz}). The same $(m,l)$-dependence of resonant scattering and oscillation that creates the high-$J_R$ OLR ridge also makes the $J_z$ signature better visible for the OLR than at CR: At the OLR (and also the 1:1 resonance), the asymmetric scattering towards higher $J_R$ and $L_z$ creates a sharply defined ridge dominated by resonant, $J_z$-sorted stars. At CR, the weak $J_R$- and symmetric $L_z$-scattering as well as strong $L_z$-oscillation mixes the resonant with non-resonant stars, diluting the $J_z$ signature.

\emph{Context to other studies.}---The Galactic disc's orbit pattern in $\langle J_z \rangle$ as a function of the $(L_z,J_R)$ plane found by \citetalias{2019MNRAS.484.3291T} is related to the vertical wave-like signatures in projections of $(\vect{x},\vect{v})$: in $\langle v_z \rangle$ vs. $R_g$ \citep{2018MNRAS.478.3809S}, $\langle v_z \rangle$ vs. $(R,v_T)$ \citep{2019MNRAS.485.3134L}, $\langle z \rangle$ and $\langle |z| \rangle$ vs. $(R,v_T)$ \citep{2019MNRAS.489.4962K}. This pattern in the vertical motion is aligned with the overdensities and $v_R$-undulations \citep{2019MNRAS.485.3134L,2019MNRAS.490.5414F,2019MNRAS.489.4962K}. As a cause for this observed coupling of radial and vertical motions, several authors suggested interactions of a satellite galaxy like the Sagittarius dwarf with the Galactic disc \citep{2018MNRAS.478.3809S,2019MNRAS.490..797C,2019MNRAS.489.4962K}.

In the absence of satellite interactions, secular resonance phenomena can also create correlations between radial and vertical motions. \citet{1981MNRAS.196..455B}, for example, showed that coupling between the in-plane and vertical orbital frequencies can excite large vertical motions via instabilities. \citet{1997A&A...318..747M} showed that a spiral wave can, at its OLR, transfer energy to warp waves in the Galactic disk. The $J_z$-sorting mechanism at the OLR presented in this work is different as it leaves the $J_z$ of the orbits unchanged, and neither instabilities nor warps with $\langle z \rangle \neq 0$ occur.

In our simulation, the $\langle J_z \rangle$ vs. $(L_z,J_R)$ signature caused by bar resonances translates into wiggles in the average absolute values of the vertical velocity, $\langle |v_z|\rangle$ vs. $L_z$. This is because a higher $J_z$ describes an orbit that reaches higher above/below the disk and has therefore higher values of $v_z$ when crossing the Galactic plane. As the same amount of stars is moving upwards- and downwards, $\langle v_z \rangle = \langle z \rangle = 0$, but $\langle |v_z| \rangle$ can still be larger. At the OLR, the $\langle |v_z|\rangle$ vs. $L_z$ signature is therefore naturally coupled with the OLRs $\langle v_R \rangle$ wiggle, as shown in Figure \ref{fig:vertical_radial_velocities}. Even though the resonances do not affect $\langle v_z \rangle \sim0$, and the difference in $\langle |v_z|\rangle$ with respect to the axisymmetric disc is no larger than $\sim 1~\text{km/s}$, our work suggests a novel mechanism how in-plane bar resonances \emph{could contribute} to the observed correlation between radial and vertical motions. 

\section{Discussion} \label{sec:discussion}

\subsection{Comparison between the \emph{Gaia} data and the test particle simulation} \label{sec:discussion_Gaia_vs_sim}

The main result of this work is the derivation of three bar OLR and pattern speed candidates from the local \emph{Gaia} DR2 action data in Figure \ref{fig:Gaia_actions_and_OLR_candidates}. In the following, we discuss these pattern speeds in Figure \ref{fig:comparison_all_models_data} by (i) investigating the location of the CR, 1:4, and 1:1 resonance lines in addition to the OLR in the \emph{Gaia} data, and (ii) by comparison to test particle simulations for these pattern speeds (see Table \ref{tab:bar_models} for the model parameters).

The following criteria need to be fulfilled for the pattern speed to be a realistic candidate:
\begin{enumerate}[leftmargin=*,labelwidth=0pt,itemindent=-0.5cm,labelsep=0pt,topsep=1ex,itemsep=1ex,align=left]
    \item[\texttt{[OLR red/blue]}]---The OLR at the Solar azimuth has to exhibit an outward/inward feature around the OLR resonance line (see Section \ref{sec:orbit_orientation_flip_theory}). For our three derived pattern speeds, this is fulfilled by construction. (3rd column in Figure \ref{fig:comparison_all_models_data}.)
    \item[\texttt{[OLR ridge]}]---An underdensity region vs. overdensity ridge in the \emph{Gaia} data associated with this OLR is expected (see Section \ref{sec:scattering_theory}). Our simulation is neither self-consistent nor does it have a cosmological context or spiral arms. \citet{2019MNRAS.488.3324F,2020MNRAS.494.5936F} showed, however, that even in these more realistic cases the OLR ridge is prominent. Moreover, the ridge should have a similar slope in action space as in the simulation. In the discussion, we use the nomenclature for the ridges in the \emph{Gaia} data from \citetalias{2019MNRAS.484.3291T}. (2nd column in Figure \ref{fig:comparison_all_models_data}.)
    \item[\texttt{[OLR Jz]}]---We expect a gradient in $\langle J_z \rangle$ with $L_z$ across the OLR resonance (see Section \ref{sec:results_vertical_action}). (4th column in Figure \ref{fig:comparison_all_models_data}.)
    \item[\texttt{[1:1]}]---Based on the simulations in Figure \ref{fig:comparison_all_models_data}, we expect an outward/inward feature, a scattering ridge, and a $J_z$ signature at the 1:1 resonance. (Resonant orbits at the 1:1 bar resonance have been studied by, e.g., \citet{2000AJ....119..800D}, \citet{1983A&A...127..349A}, and \citet{1989A&ARv...1..261C}.)
    \item[\texttt{[1:4]}]---The 1:4 resonance of a bar with non-zero $m=4$ component might create an overdensity ridge and induce outward/inward features, as suggested by \citet{2018MNRAS.477.3945H}, \citet{2019MNRAS.490.1026H}, and \citet{2019AA...626A..41M}. In this work, we have not included an $m=4$ Fourier component into the bar model to simplify the discussion, even though we have run corresponding simulations (with integration times of more than 20 bar periods). For bar strengths of the order of $|\alpha_{m=4}|\sim0.0005$ no significant 1:4 signatures were observed: Overdensity ridges developed only in the case of large 1:4 scattering in simulations with very strong $|\alpha_{m=4}|\gtrsim 0.001$; $v_R$-asymmetry features did not develop at all as in our simulations only one class of 1:4 orbits got populated by stars---which one depended on the alignment of the $m=2$ and the $m=4$ components with respect to each other. This is in contrast to the above mentioned studies (with backwards orbit integration times for a maximum of 10 bar periods), in which, as \citet{2019MNRAS.490.1026H} demonstrated, two classes of 1:4 orbits are populated, creating `red/blue' features analogously to the OLR in the case of boxy bars ($\alpha_{m=4}<0$), and analogously, 'blue/red' features in the case of pointy bars with ansae ($\alpha_{m=4}>0$). \citet[see their fig. 4]{2001A&A...373..511F} discussed that of the two classes of 1:4 orbits one is stable and the other one is unstable. This could explain the difference to our simulations. Real galaxies might, however, slowly repopulate the unstable 1:4 orbits as more stars get perturbed into the appropriate trapping regions of phase-space by non-axisymmetric structure beyond the bar. Overall, the expected signatures at the 1:4 resonance depend on the strength and orientation of the $m=4$ bar component, and the evolution history how different orbits got populated. As neither of this is well constrained in the MW, we treat the 1:4 resonance only as a weak criterion on the pattern speed for now.
\end{enumerate}

\begin{figure*}
    \centering
    \includegraphics[width=0.84\textwidth]{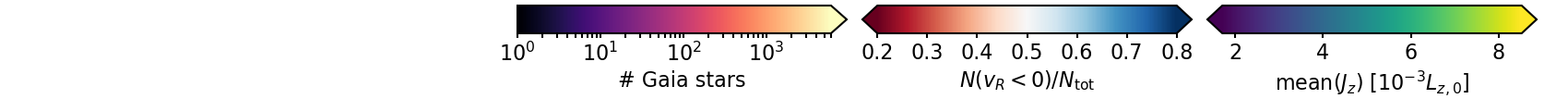}
    \includegraphics[width=0.84\textwidth]{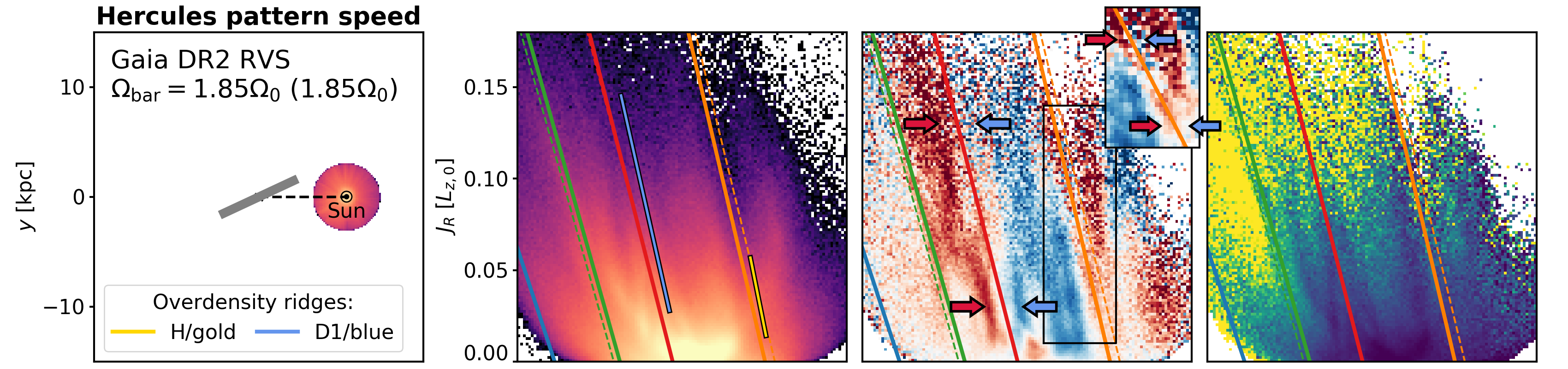}
    \includegraphics[width=0.84\textwidth]{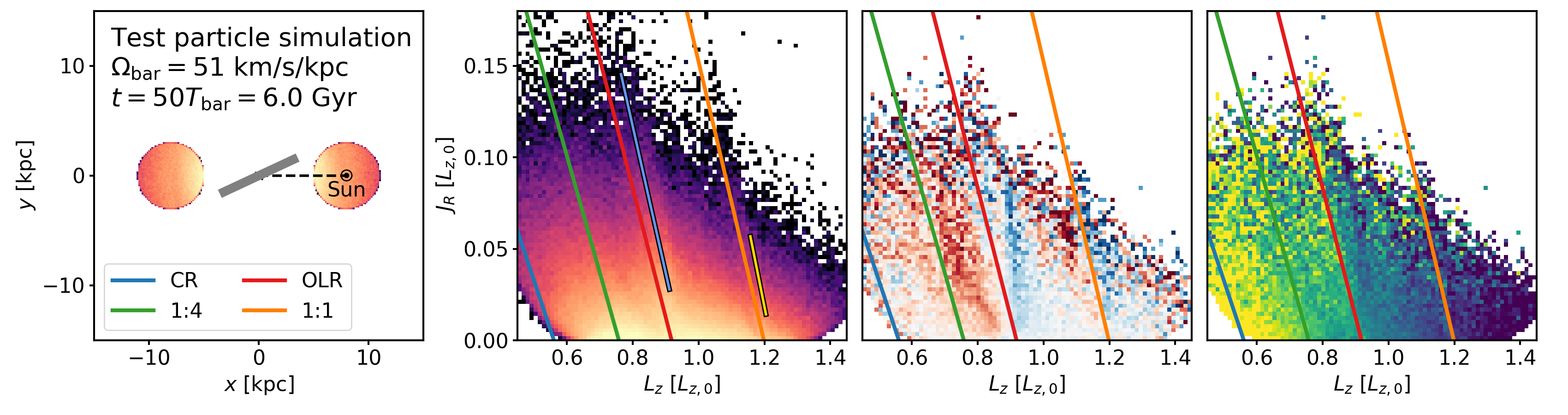}
    \includegraphics[width=0.84\textwidth]{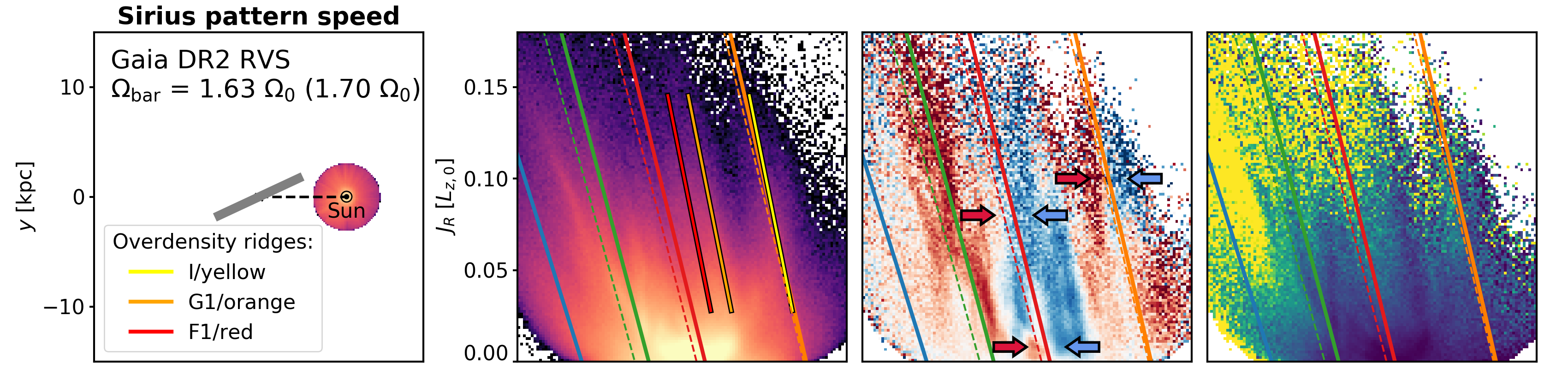}
    \includegraphics[width=0.84\textwidth]{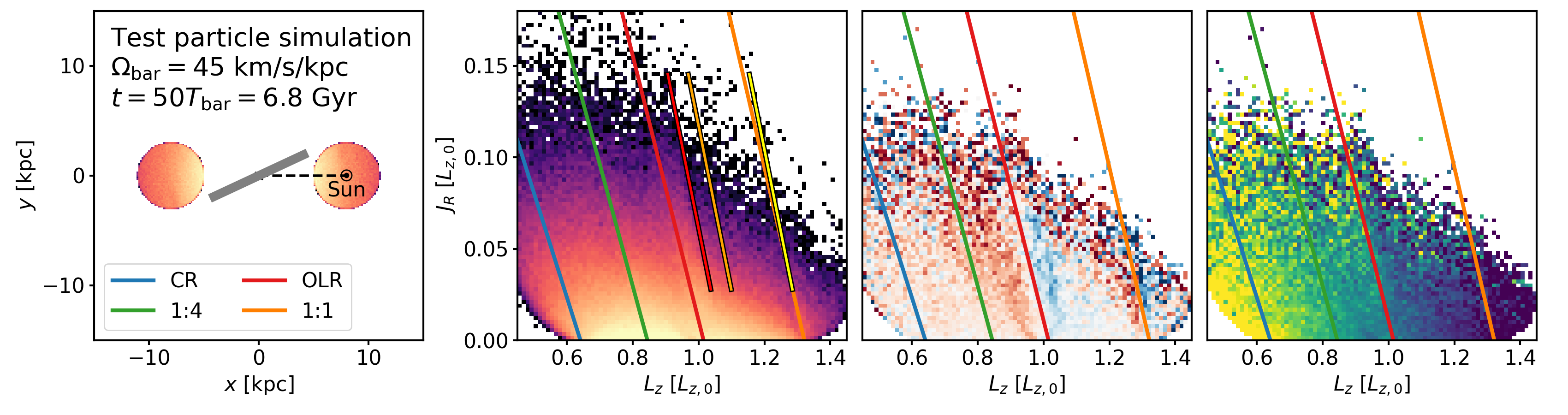}
    \includegraphics[width=0.84\textwidth]{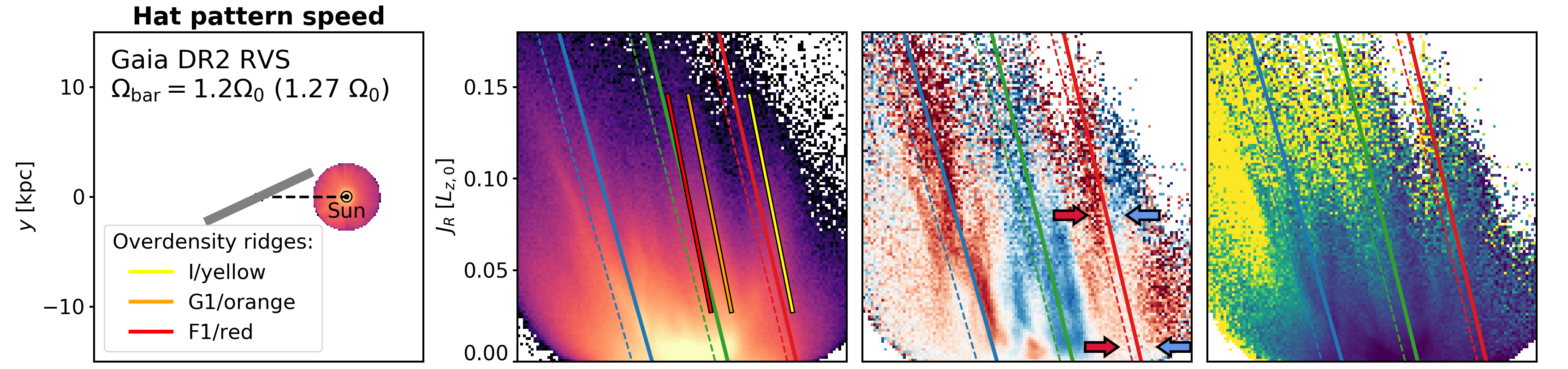}
    \includegraphics[width=0.84\textwidth]{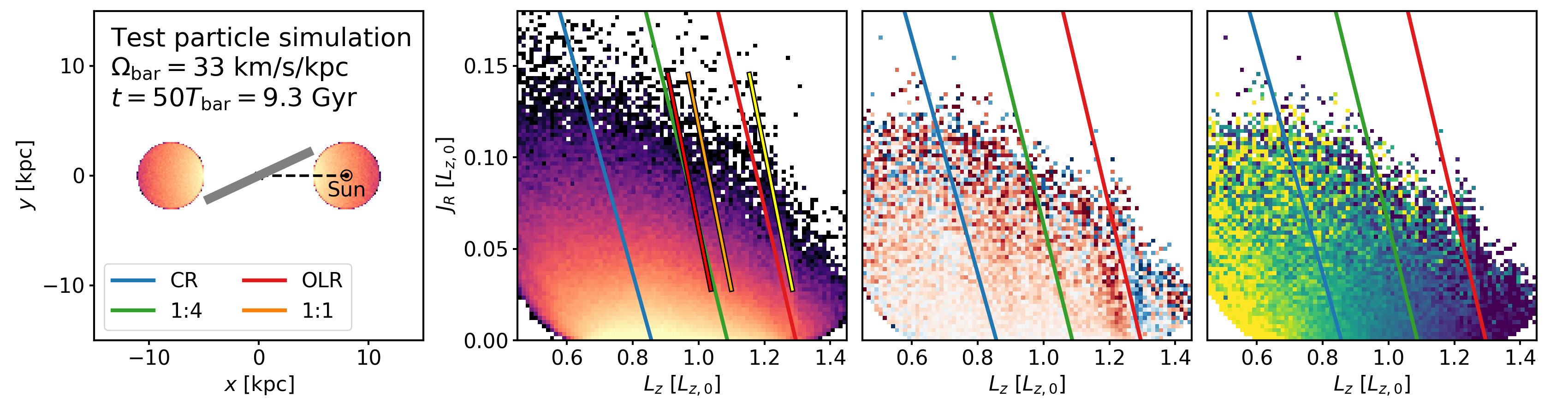}
    \includegraphics[width=0.84\textwidth]{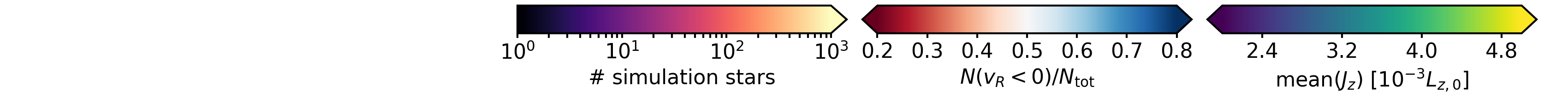}
    \caption{\emph{(Figure caption on the next page.)}}
    \label{fig:comparison_all_models_data}
\end{figure*}

\begin{figure*}
    \centering
    \contcaption{Location of the bar's ARLs in the \emph{Gaia} DR2 RVS action data (upper panels; c.f. Figure \ref{fig:Gaia_actions_Solar_neighbourhood}) and comparison with a bar-only test particle simulation (lower panels) for the three pattern speeds determined in Figure \ref{fig:Gaia_actions_and_OLR_candidates} based on the \texttt{MWPotential2014}. $\Omega_\text{bar}$ is given in units of $\Omega_0=27.5~\text{km/s/kpc}$ and $\text{km/s/kpc}$ in the first column. The dashed lines correspond to the ARLs when assuming the \citet{2019ApJ...871..120E} potential, with the corresponding $\Omega_\text{bar}$ mentioned in brackets in units of $\Omega_0 = 28.3~\text{km/s/kpc}$. All simulation parameters are summarized in Table \ref{tab:bar_models}. The top colourbar refers to the \emph{Gaia} data, the bottom colourbar to the simulation data. The solid lines with black borders denote the high-$J_R$ overdensity features identified in \citetalias{2019MNRAS.484.3291T} which could correspond to the OLR and 1:1 scattering ridges visible in the test particle simulation, and the 1:4 ridge expected for strong $m=4$ bar components (c.f. \citealt{2018MNRAS.477.3945H,2019MNRAS.490.1026H,2019AA...626A..41M}). For the \texttt{Hercules} bar pattern speed in the \emph{Gaia} data in the first row, we show an insert for stars within $d<600~\text{pc}$ only.}
\end{figure*}

\subsubsection{The {\normalfont \texttt{Hercules}} pattern speed}

The first two rows in Figure \ref{fig:comparison_all_models_data} show that the \texttt{Hercules} pattern speed satisfies the \texttt{[OLR red/blue]}, \texttt{[OLR ridge]}, \texttt{[OLR Jz]} and \texttt{[1:1]} criteria.  

\emph{Agreement.---}Around the OLR in action-$v_R$ space, data and simulation are strikingly similar. Because the OLR is closer to the bar, the OLR signature is stronger than in the \texttt{Fiducial} simulation. Because the radial velocity dispersion is higher closer to the Galactic center, the underdensity/overdensity and outward/inward OLR signatures are further apart, as demonstrated by \citet[see their figs. 3 and 4]{2003A&A...401..975M}. The OLR ridge of this proposed model is the action space ridge \texttt{D1/blue} related to the Horn. It even has a similar slope in the test particle simulation. The $\langle J_z \rangle$ data show $J_z$ trends across the OLR. The 1:1 resonance is located in \emph{Gaia}'s action space around $L_z\sim1.2L_{z,0}$, next to the weak \texttt{H/gold} ridge. We show in Figure \ref{fig:comparison_all_models_data} for the region marked by a black box the $v_R$-asymmetry for stars within $d<600~\text{pc}$ as an insert. Locally, a weak and narrow `red/blue' feature separated by the 1:1 resonance line is observed, as predicted in the simulation. 

\emph{Open questions.---}The \texttt{[1:4]} resonance line falls together with a blue/red transition in the \emph{Gaia} data. It is unclear if this supports or contradicts the \texttt{Hercules} model. The \texttt{D1/blue} ridge is not the strongest ridge in the \emph{Gaia} data, contrary to what is expected. Can a different bar strength or bar evolution explain this difference? Or do spiral arms create stronger ridges than the bar? Why is the 1:1 signature so weak and only locally visible?

\emph{Conclusion.---}The action data exhibits just enough agreement with the model prediction to not yet rule-out the \texttt{Hercules} bar pattern speed model. The substructure present in the Hercules region of the \emph{Gaia} data suggests that more than one mechanism might be at work in this region.

\subsubsection{The {\normalfont \texttt{Hat}} pattern speed}

The last two rows in Figure \ref{fig:comparison_all_models_data} show that the \texttt{Hat} pattern speed satisfies the \texttt{[OLR red/blue]}, \texttt{[OLR ridge]}, \texttt{[OLR Jz]} and \texttt{[1:4]} criteria.

\emph{Agreement.---}The prominent `blue' ridge of this proposed OLR feature is called \texttt{I/yellow} in \citetalias{2019MNRAS.484.3291T} and projects to the Hat. The corresponding `red' region is the most prominent underdensity of the \emph{Gaia} DR2 action space. In this model, the 1:4 resonance falls together with the Sirius moving group: If the 1:4 resonance indeed creates an inward-moving ridge, \texttt{G1/orange} or \texttt{F1/red} are good candidates.

\emph{Open questions.---}The 1:1 resonance does not fall within \emph{Gaia}'s survey volume; would it support the \texttt{Hat} pattern speed?

\emph{Conclusion.---}Overall, the \texttt{Hat} pattern speed agrees with the model prediction.

\subsubsection{The {\normalfont \texttt{Sirius}} pattern speed.}

The 3rd and 4th row in Figure \ref{fig:comparison_all_models_data} show that the \texttt{Sirius} pattern speed might satisfy the \texttt{[OLR red/blue]}, \texttt{[OLR ridge]}, \texttt{[OLR Jz]}, \texttt{[1:1]}, and \texttt{[1:4]} criteria.

\emph{Agreement.---} This pattern speed positions \emph{two} ARLs---the OLR and the 1:1 resonance---roughly between `red/blue' features. The two strongest ridges in the \emph{Gaia} data (Sirius \texttt{F1/red} and \texttt{G1/orange} and the Hat \texttt{I/yellow}) are at the same location as in the simulation with clear $J_z$ trends. The 1:4 ARL falls into the outward-moving Hercules region, as in our simulation.

\emph{Open questions.---}The `red' part of this OLR candidate in the \emph{Gaia} data---which is supposed to reach from low to high $J_R$---is partly obscured by the Horn; the `blue' part exhibits a wide double-peaked structure; which mechanisms could explain this? For the 1:1 resonance, the correct `red/blue' flip shows up only for $J_R \gtrsim 0.1L_{z,0}$ as the slope of the 1:1 ARL differs from the slope of the `red/blue' feature in the \emph{Gaia} data; this alignment works even less well in the \citet{2019ApJ...871..120E} potential; could this be resolved with a different potential model?

\emph{Conclusion.---}While being overall the weakest candidate, it is in any case noteworthy that the $m=2$ component of a bar is able to explain \emph{two} ridges and `red/blue' features.

\subsection{Caveats} \label{sec:discussion_caveats}

The only strong assumptions in our ARL-positioning method to measure the pattern speed that we introduced in Section \ref{sec:OLR_signature_and_pattern_speed} are that (a) the spiral arms are weak enough to not wash out the OLR's `red/blue' feature, (b) the MW's bar pattern speed is constant, (c) the axisymmetric potential model, and (d) the assumed Solar motion. 

The ILR of a transient spiral mode can cause an outward/inward signature aligned with the ARL very similar to the OLR of the bar, as presented in fig. 7 by \citet{2019MNRAS.484.3154S}. \citet{2019MNRAS.490.1026H} showed that transient winding spiral arms (that are co-rotating everywhere) can cause a time-dependent pattern of inward and outward moving features. Simulations by \citet{2019MNRAS.482.1983F} and \citet{2019MNRAS.490.1026H} noted that the bar OLR signature can get washed out in some time steps or with some spiral arm models. As we don't know much about the nature and strength of the spiral arms in the Solar vicinity yet, nothing can be done about assumption (a).

\citet{2019arXiv191204304C} investigated the effect of a slowing bar, whose resonances sweep outward in the disk with time. A slowly decelerating bar widens the `blue' part of the OLR signature (their fig. 18). A rapidly decelerating bar slightly shifts the location of the OLR's scattering ridge and `red/blue' feature with respect to the ARL (their fig. 19). As the true deceleration rate of the bar is not known, we cannot estimate how much assumption (b) biases the ARL-positioning method.

The influence of assumption (c) was tested by calculating actions and frequencies also in different potential models. For changes in $v_\text{circ}(R_0)$ of up to $20~\text{km/s}$, and in $R_0$ up to $0.3~\text{kpc}$, as well as experimenting with the slope of the rotation curve, we found deviations from the measured values for $\Omega_\text{bar}$ in Table \ref{tab:bar_models} of up to but no more than $0.1~\Omega_0$, which pushes the pattern speed by $\sim3~\text{km/s/kpc}$. As an explicit test, we stated the pattern speeds derived using the \citet{2019ApJ...871..120E} potential both in Figure \ref{fig:comparison_all_models_data} and Table \ref{tab:bar_models}.

Another source of error is the uncertainty in our knowledge of the Solar motion, assumption (d), which can also shift the distribution across the action plane and therefore $\Omega_\text{bar}$. In this work we used $(U_\odot,V_\odot,W_\odot) = (11.1, 12.24, 7.25)~\text{km/s}$ as measured by \citet{2010MNRAS.403.1829S}. We leave this error source unexplored for now, given the already significant uncertainty of $0.1\Omega_0$ due to the assumed potential model.

\subsection{Comparison to the literature} \label{sec:discussion_literature_comparison}

\subsubsection{The short fast bar model}

Our \texttt{Hercules} pattern speed $\Omega_\text{bar}=51~\text{km/s/kpc}= 1.85 \Omega_0$ re-derived from action space the classic \emph{short fast bar} model. Previous pattern speed measurements based on the Hercules/Horn bimodality include $\Omega_\text{bar}=1.85\pm0.15 \Omega_0$ by \citet{2000AJ....119..800D}, and $\Omega_\text{bar}=1.81\pm0.02 \Omega_0$ for $\phi_\text{bar}=25~\text{deg}$ by \citet{2014AA...563A..60A}. From modeling the effect of the bar on the Oort constants, \citet{2007ApJ...664L..31M} found a pattern speed of $\Omega_\text{bar}=1.87\pm0.04 \Omega_0$. This corresponds to a CR radius of $4.3-4.4~\text{kpc}$ in our two potential models. Independent, older measurements also suggested CR radii in the range $R_\text{CR} \in [3.0-4.5]~\text{kpc}$ from gas dynamics \citep{1999MNRAS.304..512E,1999A&A...345..787F,2003MNRAS.340..949B}.

Substructure in the \emph{Gaia} action-angle data beyond the resonances of a \emph{short fast bar}---in particular the ridges associated with Sirius---could be reproduced by a transient winding spiral, as shown by \citet{2019MNRAS.490.1026H} (their model H).

We also found weak evidence for the corresponding 1:1 resonance in action space. For local stars with $d<200~\text{pc}$, this was first mentioned by \citet{2000AJ....119..800D}: the velocity space above Sirius could look like the 1:1 resonance of a fast bar. We noted, however, that this `red/blue' signature is, however, only visible out to $d\sim600~\text{kpc}$.

\begin{figure*}
    \centering
    \includegraphics[width=0.49\textwidth,trim=332 82 285 0, clip]{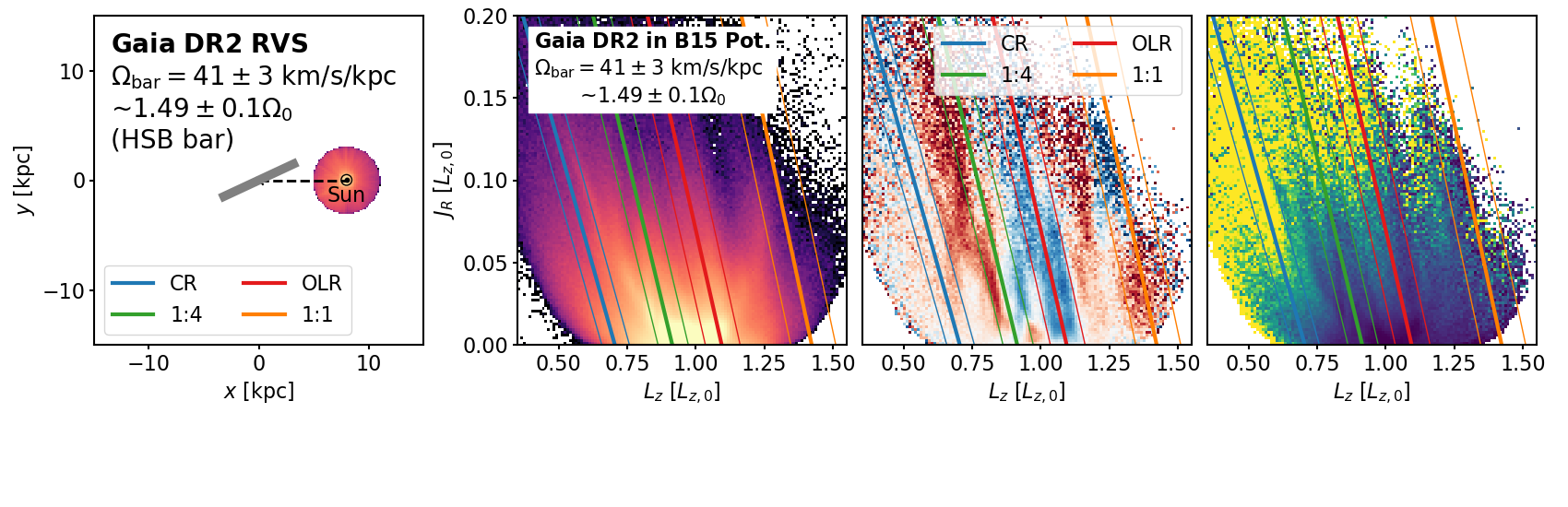}\hfill
    \includegraphics[width=0.49\textwidth,trim=332 82 285 0, clip]{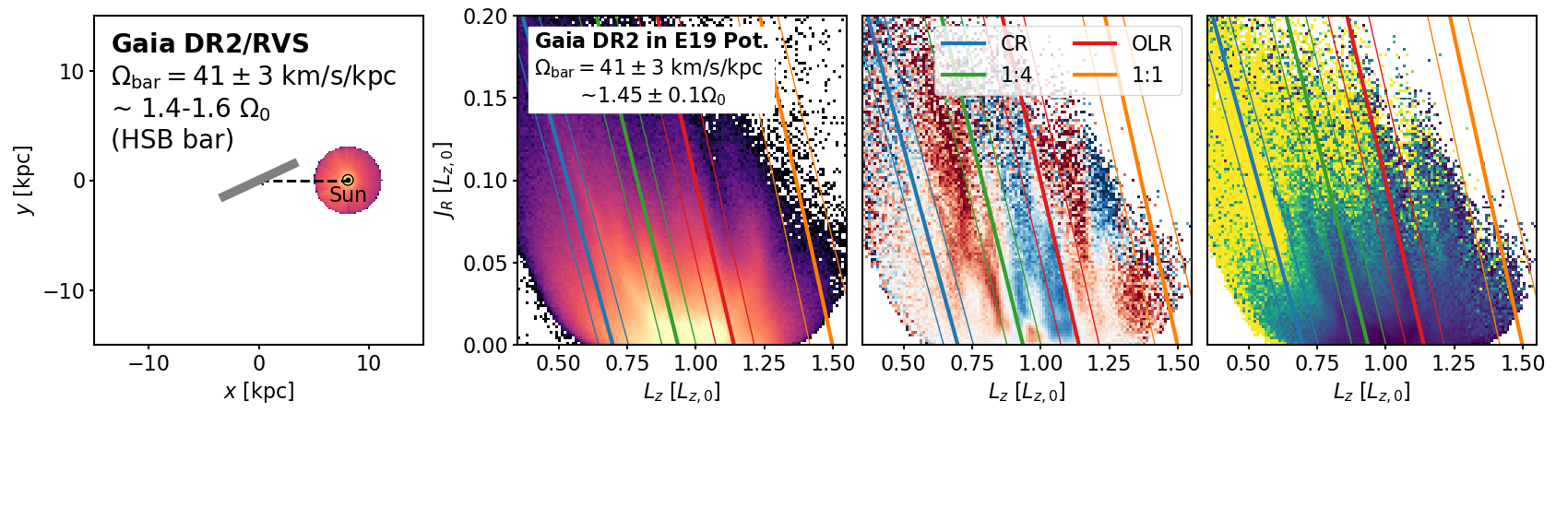}
    \caption{The \emph{Gaia} DR2 RVS action data, overplotted with the ARLs for a pattern speed of $41\pm3~\text{km/s/kpc}$ (corresponding to $1.49\pm0.1\Omega_0$ for the \texttt{MWPotential2014} by \citet{2015ApJS..216...29B}, left panels; $1.45\pm0.1\Omega_0$ for the \citet{2019ApJ...871..120E} potential, right panels), which has recently been measured from data in the inner Galaxy (\citealt{2019MNRAS.488.4552S,2019MNRAS.490.4740B}; \texttt{S19B19} hereafter). (The thin lines correspond to $\Omega_\text{bar}\pm0.1\Omega_0$.) For this pattern speed, the 1:4 resonance line (green) separates the prominent `red/blue' feature close to Hercules. \citet{2018MNRAS.477.3945H} proposed that a bar with a similar pattern speed and a strong $m=4$ Fourier component could be responsible for the Hercules stream. The OLR line (red) for this \emph{slightly faster slow} \texttt{S19B19} bar pattern speed does, however, not separate a prominent outward from an inward-moving (`red/blue') feature as we would expect.}
    \label{fig:Hunt_actions}
\end{figure*}

\subsubsection{The long slow bar model}

Our \texttt{Hat} pattern speed $\Omega_\text{bar}=33-36~\text{km/s/kpc}=1.20-1.27\Omega_0$ is quite close to (albeit slower than) some pre-\emph{Gaia} DR2 measurements from the Galactic center known as the \emph{long slow bar} model: \citet{2016ApJ...824...13L} measured $\Omega_\text{bar}=1.3\Omega_0$ (for $\Omega_0=210/8.3~\text{km/s/kpc}$) from comparing the gas flow in the MW (HI and CO $(l,v)$-diagrams) to $N$-body simulations. \citet{2017MNRAS.465.1621P} found $\Omega_\text{bar}=1.34\pm0.12\Omega_0$ (for $\Omega_0=238/8.2~\text{km/s/kpc}$) from made-to-measure modeling of the bar (3D red clump star density \citep{2013MNRAS.435.1874W} and kinematics from the BRAVA survey \citep{2012AJ....143...57K} and others). A more recent study by \citet{2019MNRAS.489.3519C} found that \emph{Gaia} DR2 and VIRAC \citep{2018MNRAS.474.1826S} proper motions of giant stars in the Galactic bar region are consistent with $\Omega_\text{bar}\sim1.32\Omega_0$ (for $\Omega_0=233/8.2~\text{km/s/kpc}$). 

The CR of the \emph{long slow bar} might explain the Hercules stream \citep{2017ApJ...840L...2P}. This was supported by \citet{2020MNRAS.495..895B} using action-based torus modeling of the Hercules stream in \emph{Gaia} DR2 , finding $\Omega_\text{bar}=1.14-1.25\Omega_0$ (for $\Omega_0=239/8.27~\text{km/s/kpc}$). The \emph{Gaia} DR2 action data and the Hat's `red/blue' feature in this work provide an independent constraint on this pattern speed for the slow \texttt{Hat} bar.

\cite{2019MNRAS.488.3324F,2020MNRAS.494.5936F} showed, however, that CR does not give rise to a Hercules/Horn bimodality and concluded that this strong observed bimodality cannot be a consequence of CR alone in a slow bar scenario, but requires the existence of an additional perturbation mechanism.

Model F in \citet{2019MNRAS.490.1026H} demonstrated that winding spiral arms could be responsible for differences between the \emph{long slow} \texttt{Hat} bar model and the \emph{Gaia} data.

Our pure-quadrupole bar model does not include an $m=4$ component. Some recent studies---which were developed independently and in parallel to this work---did consider the effect of higher-order bar componentssmall but notable discrepancy of other local-only pattern speed measurements from bar resonances---our `OLR = Hat' and the 
`CR = Hercules' by \citet{2020MNRAS.495..895B}---of $\Omega_\text{bar} \sim 1.2\Omega_0$ with respect to Galactic center measurements of $\sim1.32 \Omega_0$.

\subsubsection{An intermediate bar}

\citet{1991dodg.conf..323K} had first pointed out that the outward-moving Hyades and the inward-moving Sirius streams could form together the signature of the bar's OLR. In this work, we showed that this `red/blue' OLR candidate is not just visible in the classical local moving groups at $J_R\lesssim 0.02L_{z,0}$, but continues at $J_R \gtrsim 0.05L_{z,0}$ and out to $\sim 3~\text{kpc}$ from the Sun. That the corresponding \texttt{Sirius} pattern speed $\Omega_\text{bar} \sim 1.63-1.70\Omega_0$ (as measured in this work) has not attracted more attention in the literature is most likely because it cannot provide an explanation for the Hercules/Horn. However and interestingly, the recent study by \citet{2019MNRAS.490.1026H} found that a bar pattern speed of $1.55-1.65\Omega_0$ together with a $m=4$ bar component and a transient winding spiral arm (their Figure 9; Model G) looks quite similar to the \emph{Gaia} data, with the combination of the bar 1:4 resonance and the winding spiral causing the substructure in the Hercules region.

\subsubsection{The slightly faster slow bar}

Recent studies modelled the central bar region using versions of the \citet{1984ApJ...282L...5T} method. They converge on pattern speeds around $\Omega_\text{bar}\sim40~\text{km/s/kpc}$. \citet{2019MNRAS.488.4552S} and \citet{2019MNRAS.490.4740B} both quote $\Omega_\text{bar}=41\pm3~\text{km/s/kpc}$. \citet{2019MNRAS.488.4552S} modelled the transverse proper motions of red giants from \emph{Gaia} DR2 and the VVV survey \citep{2018MNRAS.474.1826S} observed towards the Galactic center. Independently, \citet{2019MNRAS.489.3519C} found that this data agrees with models for a pattern speed of $37.5~\text{km/s/kpc}$. \citet{2019MNRAS.490.4740B} modelled the in-plane velocities $(v_T,v_R)$ of giant stars as derived from \emph{Gaia} DR2 and APOGEE data \citep{2017AJ....154...94M,2019MNRAS.489.2079L} in the bar region ($2 < R/\text{kpc} < 5$). It also agrees with gas dynamics measurements by \citet{1999ApJ...524..112W} and \citet{2015MNRAS.454.1818S}. For the \texttt{MWPotential2014}, $\Omega_\text{bar}=41\pm3~\text{km/s/kpc}$ corresponds to $1.49\pm0.1\Omega_0$, for the \citet{2019ApJ...871..120E} potential to $1.45\pm0.1\Omega_0$. Figure \ref{fig:Hunt_actions} overplots the \emph{Gaia} actions for both potentials with the resonance lines for this pattern speed (including the uncertainty).

To distinguish it from the pattern speeds derived in this work, we refer to this \emph{slightly faster slow bar} pattern speed $\Omega_\text{bar}=41\pm3~\text{km/s/kpc}$ as the \texttt{S19B19} pattern speed, named after \citet{2019MNRAS.488.4552S} and \citet{2019MNRAS.490.4740B}.

In our MW potential models for $\Omega_\text{bar}\sim1.5\Omega_0$, it is the 1:4 resonance that falls right in between the `red/blue' Hercules/Horn feature (see Figure \ref{fig:Hunt_actions}). This agrees with the `Hercules/Horn = 1:4 resonance' explanation by \citet{2018MNRAS.477.3945H} (for a boxy bar with $\alpha_{m=4} < 0$; see their fig. 5). A similar model was revisited in \citet{2019MNRAS.490.1026H} (their model C) in action-angle-frequency space in different potential models, as well as in the $(R,v_T,\langle v_R \rangle)$ plane. They pointed out that this model's OLR ridge lies in the wrong location in action space. And indeed, in our Figure \ref{fig:Hunt_actions}, the inward-moving \texttt{G1/orange} Sirius ridge (not overplotted) around $L_z/L_{z,0}\sim1.1$ could be the scattering ridge of the OLR for $\sim1.5\Omega_0$, but no strong `red/blue' feature exists close-by. In fact, it is close to (and for the \citet{2019ApJ...871..120E} potential exactly on) the most prominent 
`blue/red' transition in the data (i.e. the opposite way around).

The resonance locations of the \texttt{S19B19} pattern speed with respect to the observed action ridges are therefore clearly distinct from those of the \texttt{Hat} and \texttt{Sirius} pattern speeds.

Two side notes on the lower and upper limits of the \texttt{S19B19} pattern speed: Firstly, the upper limit ($\sim1.6\Omega_0$ in the \texttt{MWPotential2014}) would make the \texttt{S19B19} pattern speed agree with our \texttt{Sirius} pattern speed. A potential model with lower circular velocities would have the same effect. Secondly, for the lower limit of the \texttt{S19B19} pattern speed ($\sim 1.4\Omega_0$ in the \texttt{MWPotential2014}), all four shown resonance lines align with transitions from inward- to outward-moving stripes (or vice versa). For the OLR it is the wrong way around, but it is still an interesting coincidence that the spacing of the sign-flips in $v_R$ correspond to the spacings of these resonance lines.

To conclude, we tend to rule out the \texttt{S19B19} pattern speed (for $\Omega_\text{bar}\sim1.4-1.5\Omega_0$) because of the absence of the OLR's characteristic `red/blue' feature---unless we find an explanation, e.g. through obscuration by spiral structure \citep{2019MNRAS.490.1026H,2019MNRAS.482.1983F}, or modifications of $V_\odot$ or $v_\text{circ}(R)$ \citep{2019AA...626A..41M}, or resonance sweeping by a decelerating bar \citep{2019arXiv191204304C}.

Further constraints are required. An obvious place to search is the space of phase-angles, which we investigate in a companion study, Trick et al. (in preparation).

\section{Conclusion}    \label{sec:conclusion}

We illustrated that the axisymmetric actions $\vect{J}=(J_R,L_z,J_z)$ estimated for the real MW stars in an axisymmetric MW potential model are still meaningful in the presence of---and even informative about---perturbations in the Galactic disc. In particular, we used the axisymmetric resonance lines (ARLs), i.e. the line in $(L_z,J_R)$ for a given $(l,m)$ along which $m\left[\Omega_\text{bar}-\Omega_{\phi,\text{axi}}(\vect{J})\right] - l \Omega_{R,\text{axi}}(\vect{J}) = 0$ is satisfied, as a diagnostic tool for resonances of the Galactic bar.

We investigated the behaviour of individual stars in axisymmetric action space as response to an $m=2$ bar model, by means of a test particle simulation that integrates orbits in an analytic barred MW potential. These numerical orbits confirm and illustrate what the field already knows about the characteristics of bar resonances:
\begin{enumerate}[leftmargin=*,topsep=1ex,label=(\roman*)]
    \item All orbits in a bar-affected system \emph{oscillate} in both $L_z$ and $J_R$ direction. Orbits trapped at resonances oscillate around the corresponding $l:m$ ARL. Closed periodic parent orbits swing only between peri- and apocenter. Orbits with the same Jacobi energy $E_J$ librate in addition along lines of $\delta J_R / \delta L_z \sim l/m$, on which $E_J$ is conserved.
    \item Resonances \emph{scatter} stars (when considering their time-averaged oscillation midpoint with respect to their actions in the un-barred system) also along $\Delta J_R / \Delta L_z \sim l/m$. This---together with the disk population's density gradients across the action plane---creates at the Outer Lindblad resonances $l=+1, m\in[1,2]$ a high-$J_R$ scattering ridge on the high-$L_z$ side of the ARL, and an underdensity region on the low-$L_z$ side. This is the action-analogue of the well-studied OLR signature in velocity space.
    \item It has long been known that the shape of the parent orbits flips its orientation with respect to the bar at the principal resonances, leading at the OLR to a sign-flip in radial velocity from outward- to inward-moving at the Solar azimuth. Action space visualizes this especially cleanly, with the flip occurring along the OLR ARL.
\end{enumerate}

Building on these foundations, we have presented in this work two novel findings related to the ARLs and the OLR of the MW's bar:
\begin{enumerate}[leftmargin=*,topsep=1ex,itemsep=1ex,label=(\arabic*)]
 \item We showed that an $l:m$ ARL shifts its location in $(L_z,J_R)$ with vertical action $J_z$, and that a resonant star has its oscillation midpoint (see (i) and (ii) above) close to the ARL that takes into account this star's specific $J_z$. We demonstrated that this causes in the overall disk population at the resonances a gradient in $\langle J_z \rangle$ as a function of $(L_z,J_R)$ which is especially strong at the OLR, with the underdensity region having a higher and the ridge having a lower $\langle J_z \rangle$. This proposes for the first time an additional mechanism that could contribute to the vertical patterns (i.e. in $\langle J_z\rangle$ and $\langle |v_z| \rangle$) observed in the \emph{Gaia} DR2 data.
 \item We proposed a straight-forward strategy to measure all bar pattern speed candidates $\Omega_\text{bar}$ directly and precisely from the \emph{Gaia} DR2 RVS data. When colour-coding the action plane $(L_z,J_R)$ by predominantly outward (`red') and predominantly inward motion (`blue'), three pairs of 'red/blue' transitions are visible that have slopes similar to the OLR ARL. Varying the bar pattern speed $\Omega_\text{bar}$ and positioning the OLR ARL exactly on top of the `red/blue' transitions---the expected signature of the OLR (see (iii) above)---gives precise measurements for $\Omega_\text{bar}$ candidates: $1.85\Omega_0$, $1.20\Omega_0$, and $1.63\Omega_0$, when assuming the \texttt{MWpotential2014} by \citet{2015ApJS..216...29B}. When assuming the potential by \citet{2019ApJ...871..120E}, we measure $1.85\Omega_0$, $1.27\Omega_0$,  and $1.70\Omega_0$. The first measurement is very close to the classic \emph{fast} bar model in the literature with the OLR between the Hercules/horn moving groups. The second is slightly slower than the popular \emph{slow} bar model from Galactic center measurements, with the Hat as the OLR scattering ridge. The last---which we call the \texttt{Sirius} pattern speed---revives an old proposition from the literature and we show that it
`red/blue' moving features in action space with the OLR and with the 1:1 resonance line of the bar.
\end{enumerate}

One of the above three pattern speeds has to be close to the real bar pattern speed, unless (a) our knowledge of the best-fit axisymmetric potential, in particular the rotation curve, of the MW is very wrong, (b) the OLR of the bar does not fall within the \emph{Gaia} survey volume (which, however, is quite unlikely), or (c) spiral arms or other transient perturbations are so strong in the MW disc that the signature of the bar is washed out. We note that the disagreement between Galactic center measurements and our candidates points towards a missing piece in the puzzle of $\Omega_\text{bar}$.

In any case, the signatures in axisymmetric actions space are highly informative about the true nature of perturbers in the Galactic disc.

\section*{Data availability statement}

This work has made use of data from \emph{Gaia} DR2 \citep{2016A&A...595A...1G,2018A&A...616A...1G} available at \url{https://gea.esac.esa.int/archive}. For \emph{Gaia} DR2's radial velocity sample \citep{2019A&A...622A.205K}, stellar distances were taken from \citet{2019MNRAS.487.3568S} and are available at \url{https://zenodo.org/record/2557803}. 

Action estimation and test particle simulations underlying this article were produced with the \texttt{galpy} code by \citet{2015ApJS..216...29B} which is publicly available at \url{http://github.com/jobovy/galpy}. The action data and simulations will be shared on reasonable request to the corresponding author.

\section*{Acknowledgements}

W.H.T.~thanks Jerry Sellwood, Benoit Famaey, Ortwin Gerhard, Jason Sanders, Christophe Pichon, as well as the White research group at MPA for helpful discussions, Irene Abril Cabezas and Adam Wheeler for providing useful comments on the draft, and the anonymous referee for many suggestions to improve this paper.

J.A.S.H.~was supported by a Dunlap Fellowship at the Dunlap Institute for Astronomy \& Astrophysics, funded through an endowment established by the Dunlap family and the University of Toronto. J.A.S.H.~is now supported by a Flatiron Research Fellowship at the Flatiron institute, which is supported by the Simons Foundation.

J.T.M.~acknowledges support from the ERC Consolidator Grant funding scheme (project ASTEROCHRONOMETRY, G.A. n. 772293).

This work has made use of data from the European Space Agency (ESA) mission {\it Gaia} (\url{https://www.cosmos.esa.int/gaia}), processed by the {\it Gaia} Data Processing and Analysis Consortium (DPAC, \url{https://www.cosmos.esa.int/web/gaia/dpac/consortium}). Funding for the DPAC has been provided by national institutions, in particular the institutions participating in the {\it Gaia} Multilateral Agreement.
    
This project was developed in part at the 2018 NYC \emph{Gaia} Sprint, hosted by the Center for Computational Astrophysics of the Flatiron Institute in New York City.

This research was supported in part at KITP by the Heising-Simons Foundation and the National Science Foundation under Grant No. NSF PHY-1748958.


\bibliographystyle{mnras}
\bibliography{bibliography} 

%%%%%%%%%%%%%%%%% APPENDICES %%%%%%%%%%%%%%%%%%%%%

\appendix

\section{Time evolution and azimuthal dependence of the OLR signature} \label{app:OLR_time_evolution}

\begin{figure*}
    \centering
    \includegraphics[width=\textwidth]{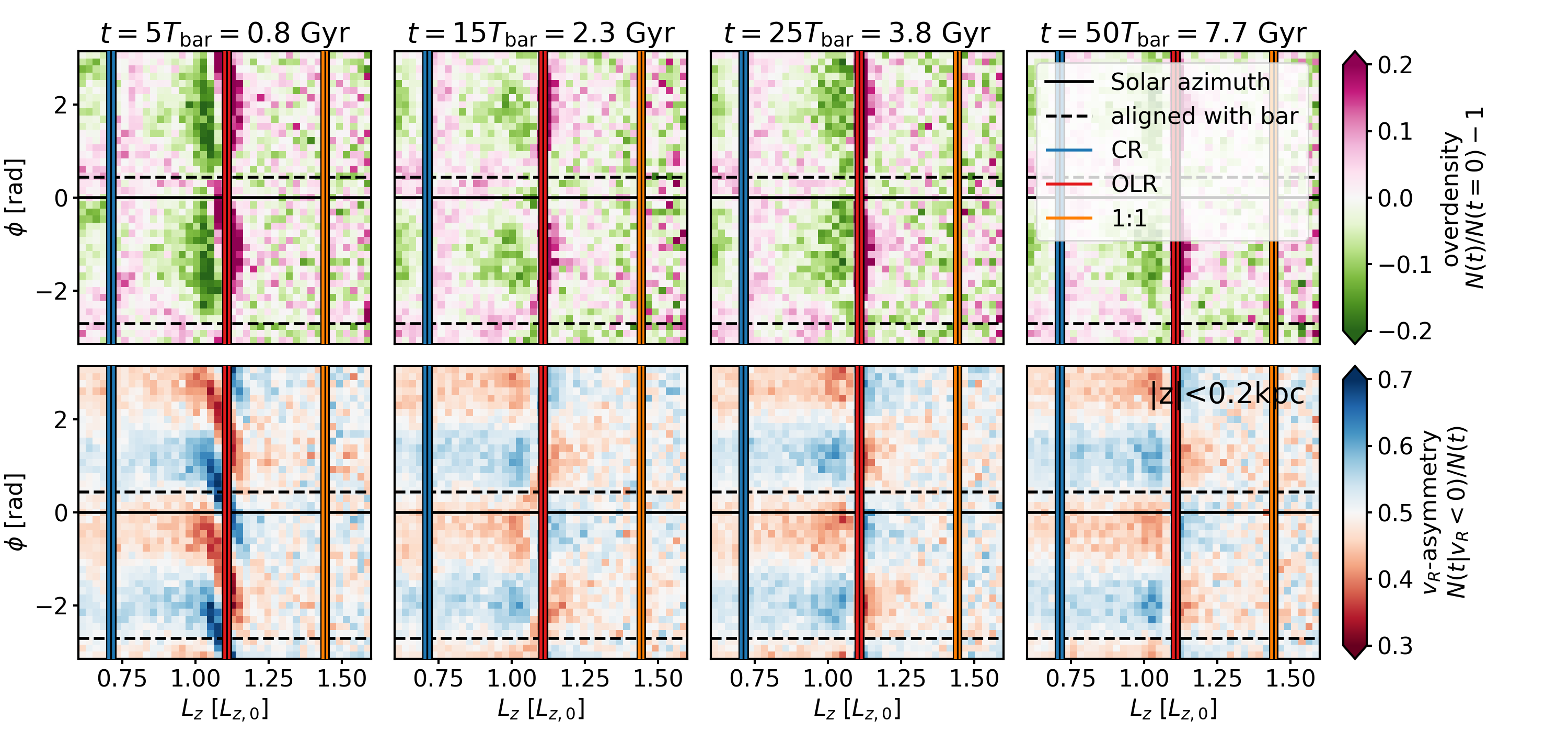}
    \caption{Time evolution of the OLR signature in the \texttt{Fiducial} bar simulation as a function of Galactic azimuth $\phi$. The $x$-axis shows $L_z$, i.e. the guiding center radius, and the vertical lines the location of the OLR (red), CR (blue) and 1:1 (orange) ARLs evaluated at $J_R \in [0,0.01]L_{z,0}$ and $J_z=0$. Around the OLR line, we see the underdensity region and the overdensity ridge and the characteristic asymmetry between inward- and outward moving stars. The first column captures the transition from the axisymmetric to the bar affected state in the simulation. After a few Gyr of orbit integration, the OLR pattern remains stable with respect to the bar.}
    \label{fig:OLR_time_evolution}
\end{figure*} 

Figure \ref{fig:OLR_time_evolution} shows for the \texttt{Fiducial} simulation the time evolution of the OLR signature from Section \ref{sec:OLR_signature_and_pattern_speed}: the underdensity-region/overdensity-ridge signature around the OLR ARL in the upper panels, and the associated outward/inward feature at the Solar azimuth in the lower panels. We show both as a function of Galactic azimuth $\phi$ and $L_z$, which can be considered as the average radius of the star's orbit. The first few Gyr of orbit integration in the barred potential are marked by the (unrealistic) transition phase away from axisymmetry (first column). The characteristic, steady-state OLR signature due to the orbit pattern with its azimuthal $m=2$ symmetry with respect to the bar has been established.

\section{Oscillation amplitude across the action plane} \label{app:oscillation}

\begin{figure*}
    \centering
    \subfigure[Amplitude of oscillation in the action plane in $J_R$-direction. \label{fig:LzJR_JRosc}]{
        \includegraphics[width=\columnwidth]{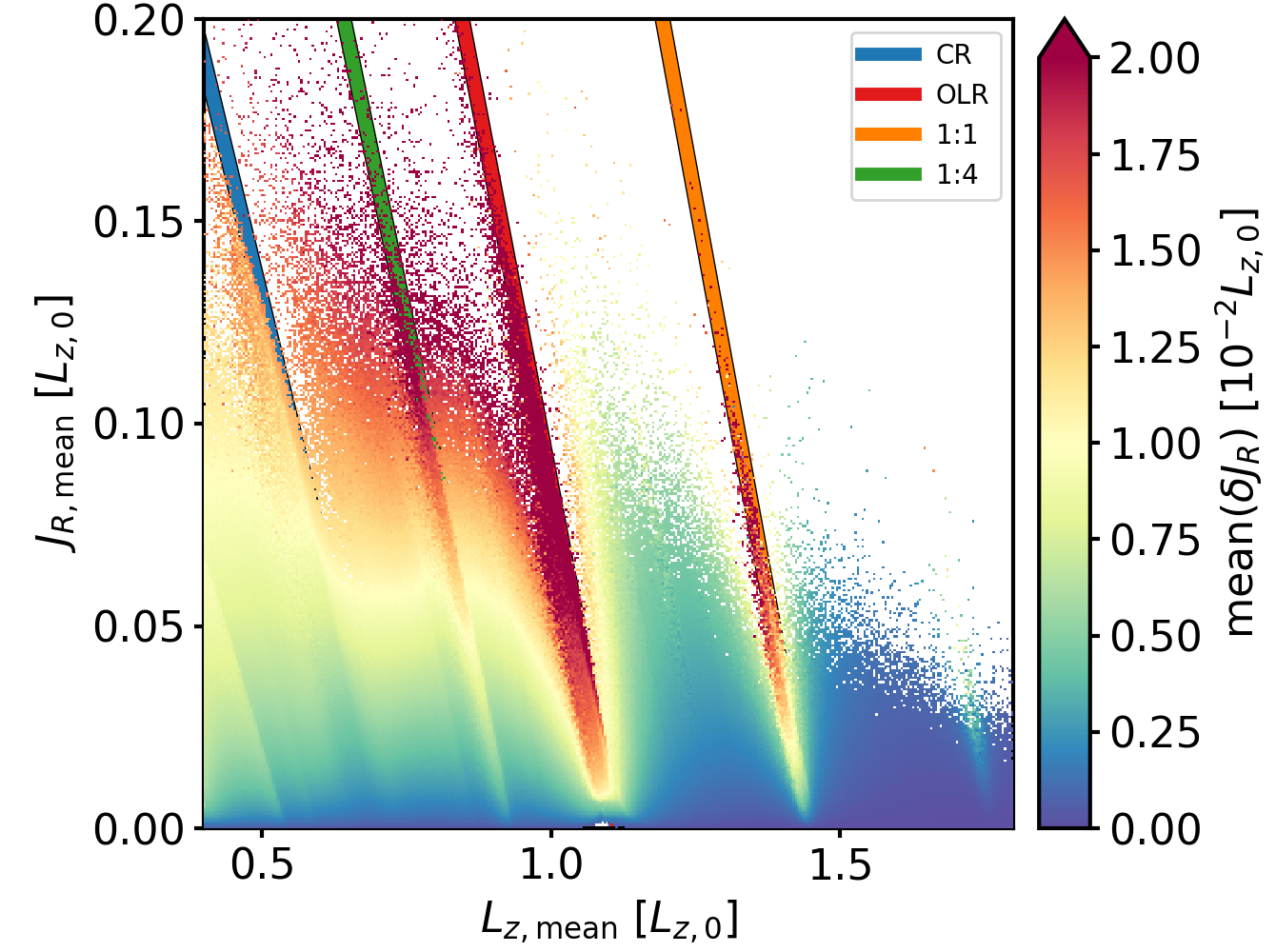}}
    \subfigure[Amplitude of oscillation in $L_z$-direction as a function of the azimuthal position $\phi$ with respect to the bar (at $t=25T_\text{bar}$). \label{fig:Lzphi_Lzosc}]{
        \includegraphics[width=\columnwidth]{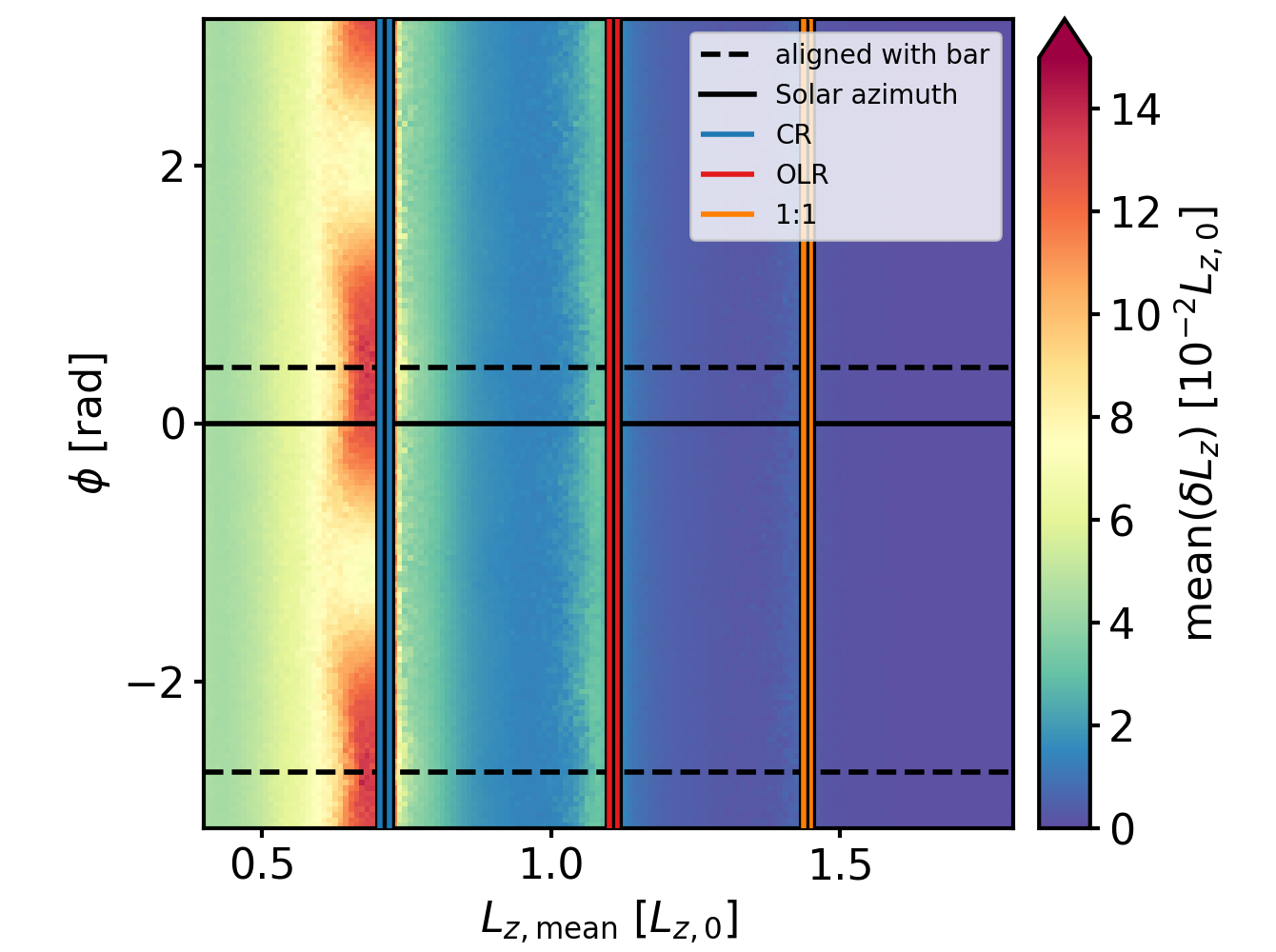}}
    \caption{Oscillation amplitudes of bar affected orbits in $J_R$ (left panel) and $L_z$ (right panel) direction as a function of the oscillation midpoints $\vect{J}_\text{mean}$.}
    \label{fig:oscillation_amplitudes}
\end{figure*}

Orbits oscillate in the space of axisymmetric actions due to the rotating bar potential (Section \ref{sec:oscillation_theory}) around the midpoints $\vect{J}_\text{mean}$ illustrated in Figure \ref{fig:LzJR_mid}. Figure \ref{fig:oscillation_amplitudes} shows for the \texttt{Fiducial} simulation the  \emph{average oscillation amplitudes}, i.e. $\delta L_z$ and $\delta J_R$, which were calculated from the numerically integrated orbits as
\begin{eqnarray}
\delta J_R = \frac12 \left(\text{max}\left[J_R(t)\right] - \text{min}\left[(J_R(t)\right]\right) \label{eq:oscillation_amplitude}
\end{eqnarray}
and equivalently for $\delta L_z$. We show these in the $(L_{z,\text{mean}},J_{R,\text{mean}})$ and the $(L_{z,\text{mean}},\phi)$ planes, respectively, where $\phi$ is the Galactocentric azimuth at $t=25T_\text{bar}$. Figure \ref{fig:oscillation_amplitudes}(a) demonstrates that stars that are on $x_1(1)$ orbits oscillate strongly in $J_R$ (and $L_z$) around the OLR ARL (c.f. Figure \ref{fig:orbit_orientation}, Section \ref{sec:orbit_orientation_flip_theory}, and \citet{2019MNRAS.488.3324F}). Figure \ref{fig:oscillation_amplitudes}(b) shows the strong oscillation $\delta L_z$ at CR. This mixes resonant with non-resonant stars and washes out resonance features, which are therefore less prominent than at the OLR.

\section{Details of the test particle simulation} \label{sec:app_simulation_details}

\begin{figure*}
\centering
\subfigure[Axisymmetric mock data created in a MW potential from a disc DF within the shown annulus around the Galactic center (GC). \label{fig:xy_start}]{\includegraphics[width=0.32\textwidth]{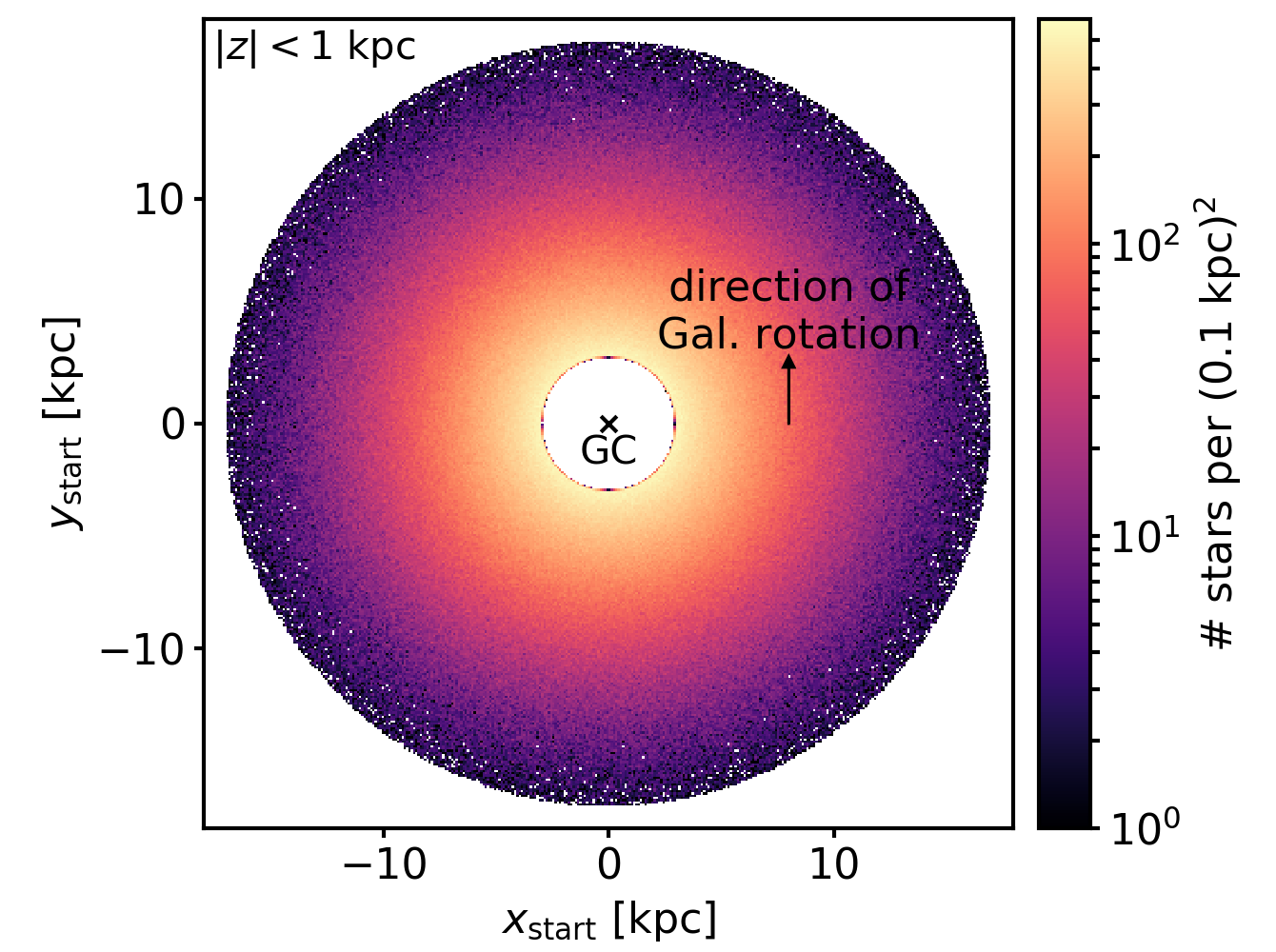}}%
\hspace{0.5cm}%
\subfigure[Actions of the smooth mock data calculated in the axisymmetric MW potential. \label{fig:LzJR_start}]{\includegraphics[width=0.32\textwidth]{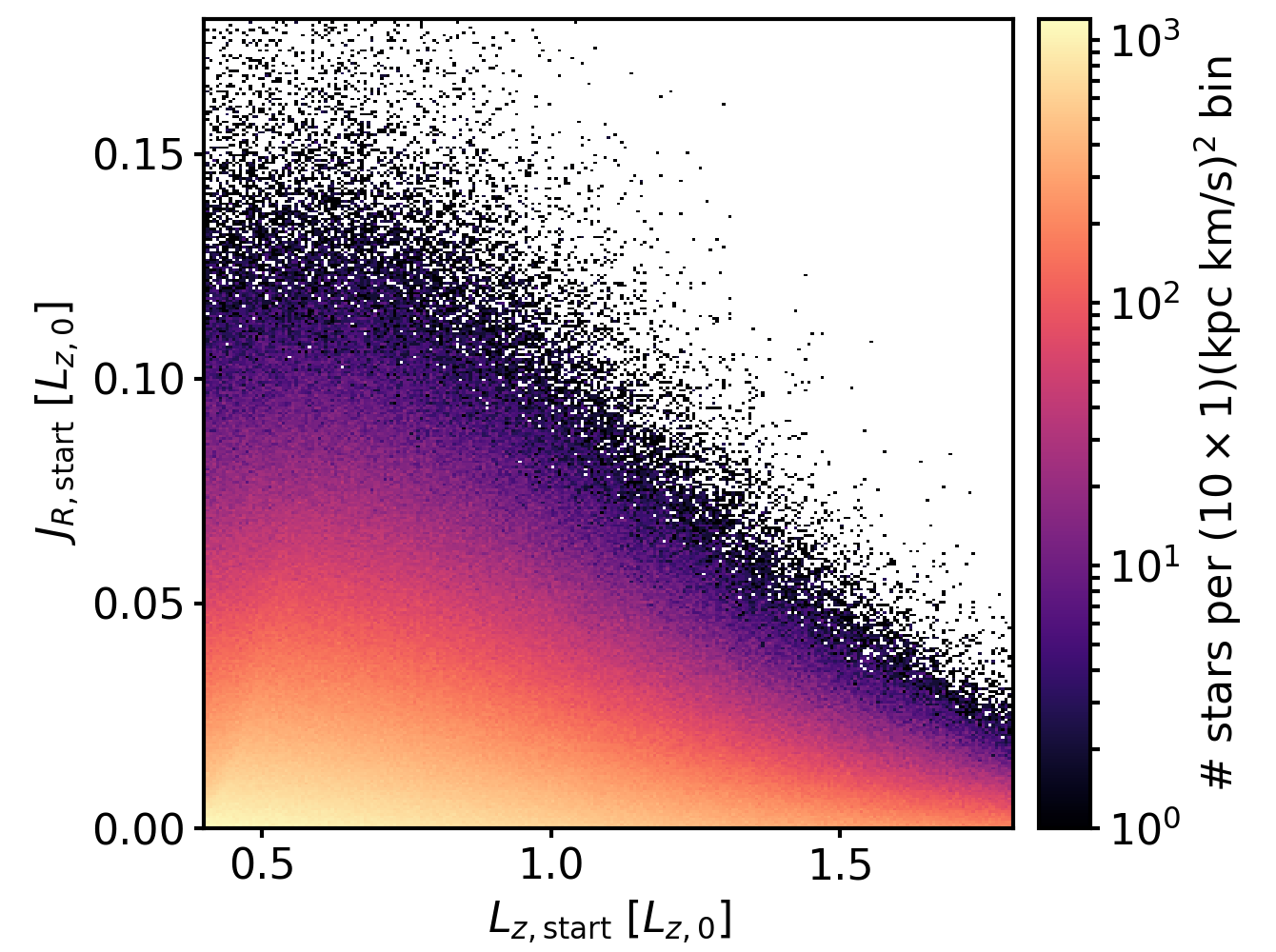}} %
\hfill
\subfigure[Surface density of the ``Dehnen bar'' perturbation which will be superimposed onto the axisymmetric MW potential. \label{fig:bar_surfdens}]{\includegraphics[width=0.32\textwidth]{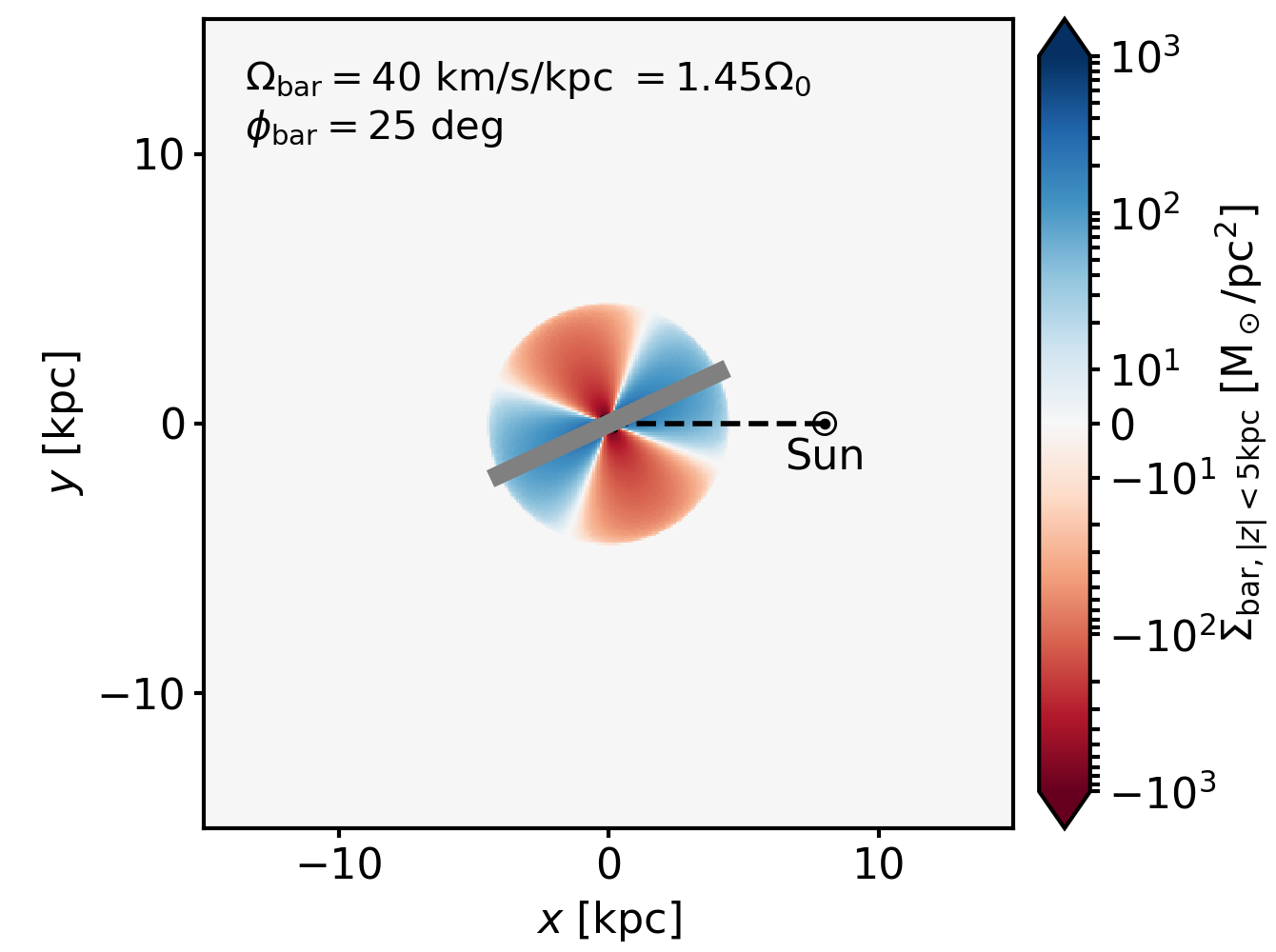}} %
\hfill
\subfigure[Distribution of mock stars after orbit integration for $t=25T_\text{bar}\sim3.8~\text{Gyr}$ in the barred MW potential. \label{fig:xy_end}]{\includegraphics[width=0.32\textwidth]{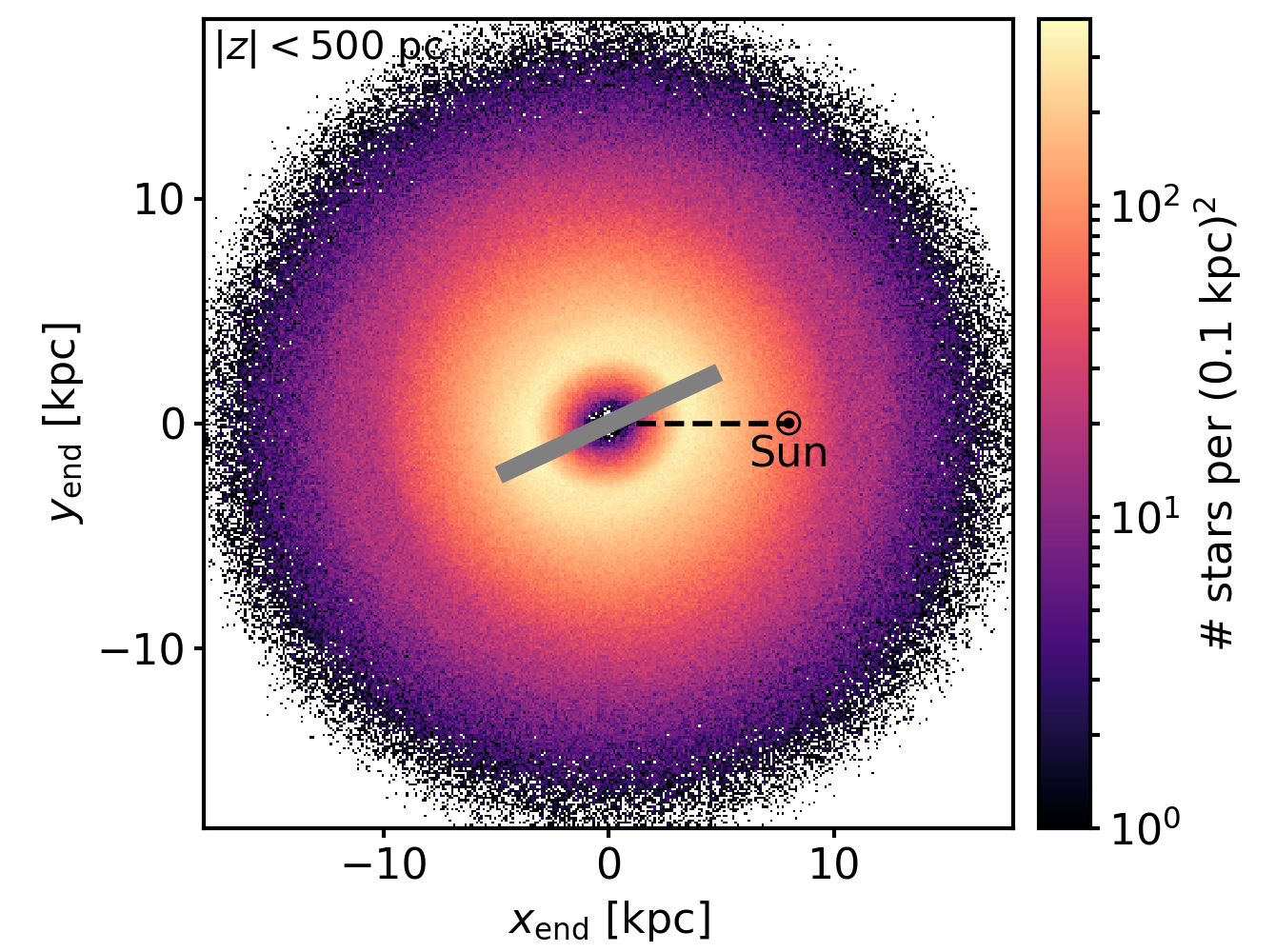}} %
\hspace{0.5cm}%
\subfigure[Actions estimated in the axisymmetric MW potential, after orbit integration. \label{fig:LzJR_end}]{\includegraphics[width=0.32\textwidth]{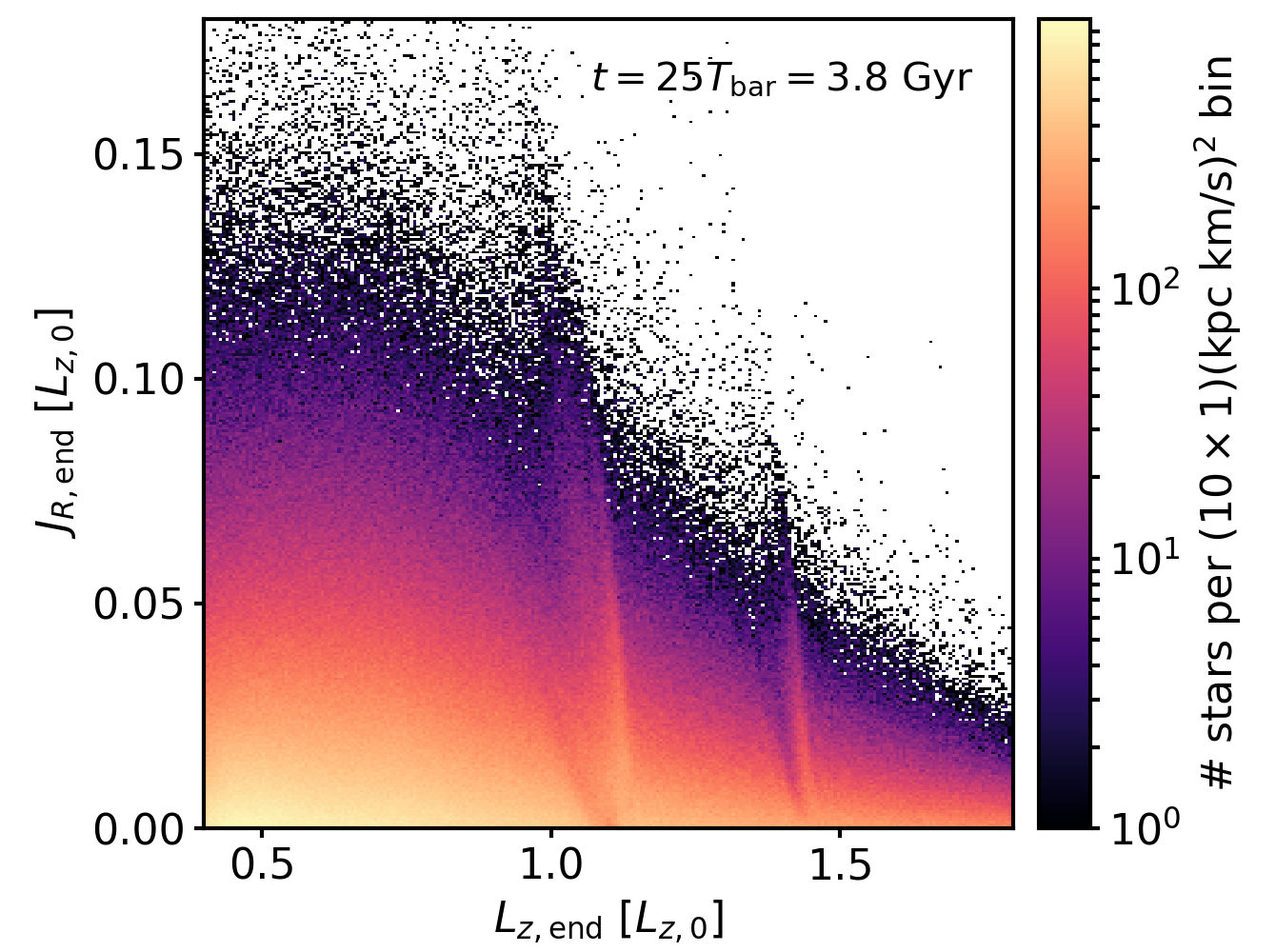}} %
\hfill
\subfigure[Same as Panel (e), but overplotted with the ARL from Equation \eqref{eq:res_condition_axisym_freqs}. \label{fig:LzJR_end_res}]{\includegraphics[width=0.32\textwidth]{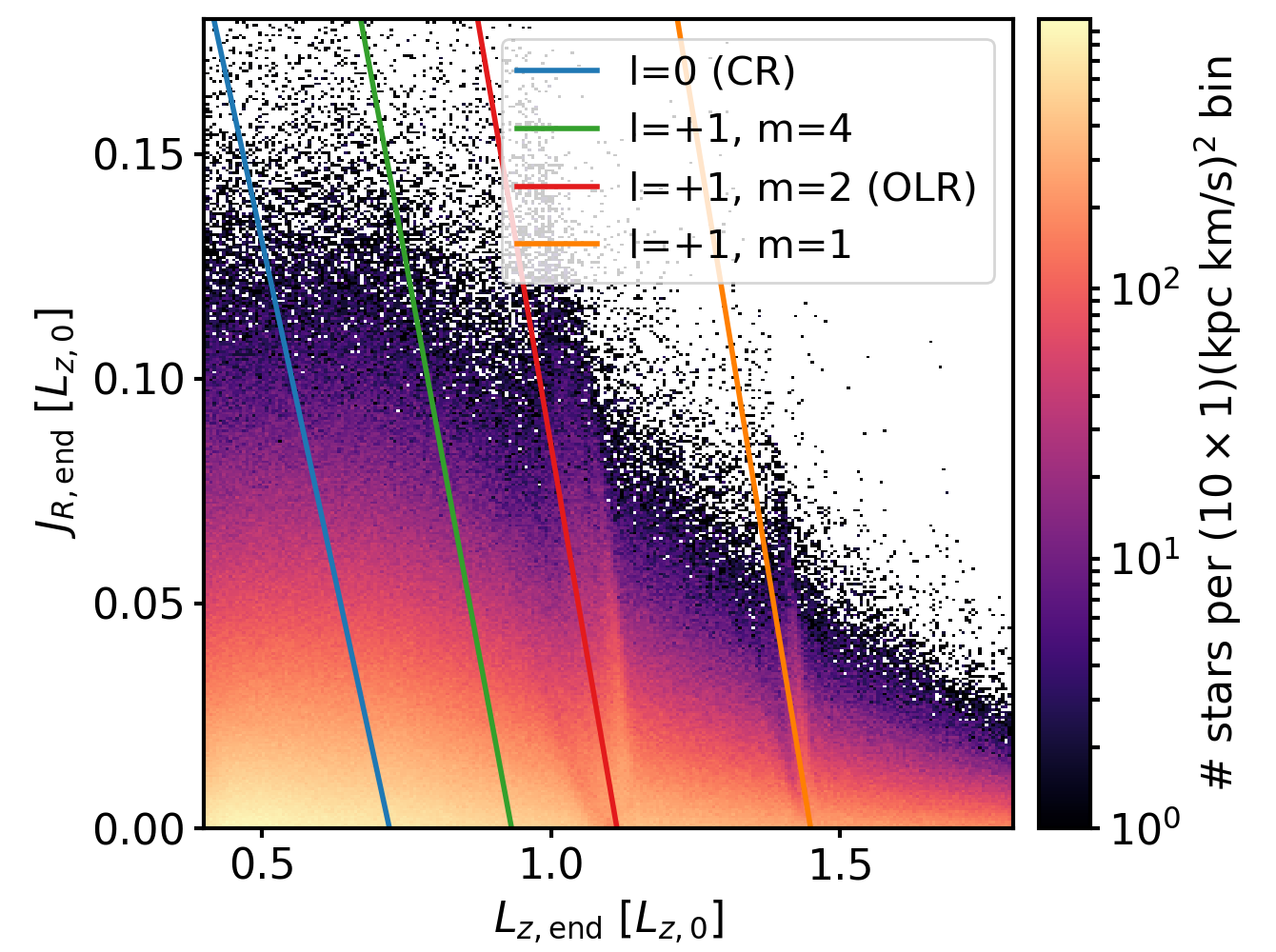}} %
\caption{Illustration of the mock data simulation with test particles in a barred MW potential for the \texttt{Fiducial} bar model in Table \ref{tab:bar_models}. Panels (a)-(b) show the axisymmetric mock data before orbit integration; Panels (d)-(f) the distribution after integration in the barred potential (Panel (c)). Ridges have developed next to the OLR and 1:1 ARLs.}
\label{fig:Fiducial_model_intro}
\end{figure*}

\subsection{The axisymmetric stellar disc model} \label{sec:method_axisymmetric}

The test particle simulation is set up in the analytic axisymmetric \texttt{MWPotential2014} by \citet{2015ApJS..216...29B}. We mimic an axisymmetric, exponential stellar disc by sampling the action-based quasi-isothermal DF by \citet{2011MNRAS.413.1889B}. This qDF requires the potential as input and ensures that the collisionless Boltzmann equation is satisfied. The resulting stellar distribution is phase-mixed by construction. We use the same qdf parameters as for the mock data in \citetalias{2019MNRAS.484.3291T}, which roughly reproduce the velocity dispersion of \emph{Gaia} DR2 in the Solar neighbourhood, $\sigma_{z,0}^\text{qdf}=20~\text{km/s}$ and $\sigma_{R,0}^\text{qdf}=37~\text{km/s}$. The vertical velocity dispersion is exponentially decreasing with radius with a scale length of $h^\text{qdf}_{\sigma,z} = 7~\text{kpc}$ \citep{2012ApJ...755..115B}.

We restrict the sampling of the mock data to the large annulus around the galactic center illustrated in Figure \ref{fig:Fiducial_model_intro}(a): $R \in [3,17]~\text{kpc}$, $|z| < 1~\text{kpc}$, $\phi \in [-\pi,\pi]~\text{rad}$. The exact mock data generation procedure is described in appendix A of \citet{2016ApJ...830...97T}.

Figure \ref{fig:Fiducial_model_intro}(b) shows the distribution of the axisymmetric mock data in the action plane $(L_z,J_R)$. The sharp unrealistic radial edges of the annulus cause phase artifacts in the distribution. We therefore show the action data only for the range $L_z \in [0.5,1.8]L_{z,0}$ within which the mock data are fully phase-mixed, as required. $(L_z,J_R)$ are in this setup the ``real'' actions of the orbits, i.e. they are true integrals of motion and stay constant along the orbit if integrated in the axisymmetric potential. Their distribution is smooth and nicely illustrates the realistic property of the qdf that stars in the inner galaxy (at smaller $L_z$) are more numerous and on ``hotter'' orbits (i.e. have larger $J_R$ on average).

\subsection{Orbit integration in the barred galaxy potential}  \label{app:method_bar}

In a second step, we in the test particle simulation the quadrupole bar by \citet{2000AJ....119..800D}, generalized to 3D by \citet{2016MNRAS.461.3835M}, implemented in \texttt{galpy}. Its strength $\alpha_{m=2} = 0.01$ is the ratio of the maximum radial force at $R_0$ due to the ($m=2$) bar potential alone to the axisymmetric background potential. This bar strength is similar to those used by, e.g., \citet{2017MNRAS.465.1443M} and \citet{2019MNRAS.490.1026H}. In the surface density of the stellar component of the potential, the bar imposes outside of $R\sim1~\text{kpc}$ a maximum perturbation of $A_2=0.26$ (following the definition of $A_2$ by \citealt{2013MNRAS.429.1949A}). The parameters for the \texttt{Fiducial} bar model are summarized in Table \ref{tab:bar_models}. The total mass of the bar is zero. When imposing it onto the \texttt{MWPotential2014} it can be considered as a redistribution of the matter (see Figure \ref{fig:Fiducial_model_intro}(c)). Averaged over $\phi$, the circular velocity curve is the same as for the purely axisymmetric galaxy model.

The bar strength is instantaneously switched on from zero to its full value at time $t=0$. Similar studies usually grow the bar adiabatically from zero to its full strength to avoid a shock to the system (e.g. \citealt{2003A&A...401..975M,2010MNRAS.407.2122M,2019MNRAS.490.1026H}). We have run additional test simulations that grow the bar over $N=15$ bar periods. The qualitative bar signatures were very similar. The reasons are: (i) The particles are massless. The disc distribution is therefore not modified by self-gravity and the wake that the bar induces. A star's orbit depends only on its current $(\vect{x},\vect{v})$ and the analytic potential model without knowledge of its past orbital evolution. No shocks are therefore induced. (ii) The qdf we used to set-up the system generates a stellar population in steady-state. As long as the bar is weak, the system remains in almost the same steady-state.

We integrate particle orbits in the barred potential using the \texttt{RK4} method provided by \texttt{galpy}. At integration times of $t=0$ and $t=n \times T_\text{bar}\equiv2\pi n/\Omega_\text{bar}, n\in\mathbb{N}$, the bar is orientated at an angle of 25 degrees with respect to the $x=0$ line (see Figure \ref{fig:Fiducial_model_intro}(c) and (d)), similar to the angle between the Galactic bar and the line-of-sight line between the Galactic center and the Sun \citep{2019MNRAS.490.4740B}. The lower panels in Figure \ref{fig:Fiducial_model_intro} show the distribution after 25 bar periods, which corresponds to $3.8~\text{Gyr}$. This time was chosen to be well past the initial transition from the axisymmetric to the bar-affected system (see Figure \ref{fig:OLR_time_evolution} and fig. 2 in \citealt{2003A&A...401..975M}). We applied an additional cut of $|z| < 500~\text{pc}$ to reduce artifacts in the vertical phase and action due to the vertical cut in the initial conditions.

The bar-affected action distribution of all particles in Figure \ref{fig:Fiducial_model_intro}(e) is overplotted by the resonance ARLs. As our simulation is restricted to the action distribution outside of $L_z = 0.4L_{z,0}$, we cannot pick up any resonant signatures inside the CR.

%%%%%%%%%%%%%%%%%%%%%%%%%%%%%%%%%%%%%%%%%%%%%%%%%%

% Don't change these lines
\bsp	% typesetting comment
\label{lastpage}
\end{document}